\documentstyle[seceqn,psfig,amsfonts]{elsart}

\newcommand{\ep}{\epsilon}
\newcommand{\al}{\alpha}

\renewcommand{\i}{{\mathrm i}}

\newsymbol\lesssim 132E
\newsymbol\gtrsim 1326

\begin{document}

\setcounter{tocdepth}{3}

\begin{frontmatter} 

\title{\bf Spike autosolitons in the Gray-Scott model}

\author{C. B. Muratov}

\address{Department of Mathematical Sciences, New Jersey Institute of
Technology, \\ University Heights, Newark, NJ 07102}

\author{V. V. Osipov} 

\address{CSIC, Laboratorio de Fisica de Sistemas Peque\~nos y
Nanotecnologia, \\ Calle Serrano, 144, 28006, Madrid, Spain}


\begin{abstract}
   We performed a comprehensive study of the spike autosolitons:
   self-sustained solitary inhomogeneous states, in the classical
   reaction-diffusion system --- the Gray-Scott model. We developed
   singular perturbation techniques based on the strong separation of
   the length scales to construct asymptotically the solutions in the
   form of a one-dimensional static autosolitons, higher-dimensional
   radially-symmetric static autosolitons, and two types of traveling
   autosolitons. We studied the stability of the static autosolitons
   in one and three dimensions and analyzed the properties of the
   static and the traveling autosolitons.
\end{abstract}

\begin{keyword}
Pattern formation; Self-organization; Reaction-diffusion systems;
Singular perturbation theory
\end{keyword}

\end{frontmatter}

\tableofcontents 

\newpage

\section{Introduction} \label{s:intro}

Self-organization and pattern formation in nonequilibrium systems are
among the most fascinating phenomena in nonlinear physics
\cite{Prig,Vas,Osc,Murr,Cross,Mik,Kapral,KO86,KO89,KO90,KO94}.
Pattern formation is observed in various physical systems including
aero- and hydrodynamic systems; gas and electron-hole plasmas;
various semiconductor, superconductor and gas-discharge structures;
some ferroelectric, magnetic and optical media; combustion systems
(see, for example, \cite {Cross,KO89,KO90,KO94,Semi,Bode,Gor}), as
well as in many chemical and biological systems (see, for example,
\cite{Prig,Vas,Osc,Murr,Cross,Mik,Kapral,Lee}).

Self-organization is often associated with the destabilization of the
uniform state of the system \cite{Prig,Vas,Cross,KO90,KO94}. At the
same time, when the uniform state of the system is stable, by applying
a sufficiently strong perturbation one can excite large-amplitude
patterns, including {\em autosolitons} (ASs) --- self-sustained
solitary inhomogeneous states
\cite{KO86,KO89,KO90,KO94,KO78,KO79,KO80,Koga}. Autosolitons are the
elementary objects in open dissipative systems away from
equilibrium. They share the properties of both solitons and traveling
waves (or autowaves, as they are also referred to
\cite{Vas,Mik}). They are similar to solitons since they are localized
objects whose existence is due to the nonlinearities of the system. On
the other hand, from the physical point of view they are essentially
different from solitons in that they are {\em dissipative structures},
that is, they are self-sustained objects which form in strongly
dissipative systems as a result of the balance between the dissipation
and pumping of energy or matter. This is the reason why, in contrast
to solitons, their properties are independent of the initial
conditions and are determined primarily by the nonlinearities of the
system \cite{KO86,KO89,KO90,KO94}. ASs can be static, pulsating, or
traveling. As a result of their various instabilities, these simplest
localized patterns can spontaneously transform into complex
space-filling static or dynamic patterns, including complex pulsating
and traveling patterns, or spatio-temporal chaos
\cite{KO86,KO89,KO90,KO94,Semi,Bode,Gor,KO80,Koga,Gol,Mur96a,Mur96b,%
Hag,thesis,KO85,Pear,Ross,KO82,KO83,Kuzn,Mik94,Gaf95,Os96}. Thus, it
is the destabilization of the ASs that is the main source of
self-organization in nonequilibrium systems with the stable
homogeneous state.

Real physical, chemical, and biological systems exhibiting pattern
formation and self-organization are extremely complicated, so
simplified models are used to describe these phenomena. A prototype
model of this kind is a pair of reaction-diffusion equations of the
activator-inhibitor type
\begin{eqnarray}
  \tau _\theta \frac{\partial \theta }{\partial t} & = & l^2\Delta
  \theta - q \left( \theta, \eta, A \right), \label{gen:act} \\
  \tau_\eta \frac{\partial \eta }{\partial t} & = & L^2 \Delta \eta -
  Q \left( \theta , \eta , A \right), \label{gen:inh}
\end{eqnarray}
where $\theta$ is the activator, $\eta$ is the inhibitor,
$\tau_\theta$, $l$ and $\tau _\eta $, $L$ are the time and the length
scales of the activator and the inhibitor, respectively; $A$ is the
control (bifurcation) parameter; $q$ and $Q$ are certain nonlinear
functions representing the activation and the inhibition processes.
Examples of these equations for various physical systems are given in
\cite{KO89,KO90,KO94,Semi,Bode,KO85} where the physical meaning of the
variables $\theta $ and $\eta $ and the nature of the activation and
the inhibition processes are discussed. The well-known Brusselator
\cite{Prig} and the Gray-Scott \cite{GS} models of autocatalytic
chemical reactions, the classical Gierer-Meinhardt model of
morphogenesis \cite{GM}, the FitzHugh-Nagumo \cite{FHN} and the
piecewise-linear Rinzel-Keller model \cite{RK} for the propagation of
pulses in the nerve fibers are all special cases of
Eqs. (\ref{gen:act}) and (\ref{gen:inh}).

The fact that $\theta$ is the activator means that for certain
parameters the uniform fluctuations of $\theta$ will grow when the
value of $\eta$ is fixed. From the mathematical point of view, this is
given by the condition \cite{KO89,KO90,KO94,KO80}
\begin{eqnarray} \label{act:cond}
q'_\theta < 0
\end{eqnarray}
for certain values of $\theta$ and $\eta$. On the other hand, the fact
that $\eta$ is the inhibitor means that its own fluctuations decay and
that it damps the fluctuations of the activator. Mathematically, these
conditions are expressed by \cite{KO89,KO90,KO94,KO80}
\begin{eqnarray}
\label{inh:cond}
Q'_\eta > 0, ~~~ q'_\eta Q'_\theta < 0
\end{eqnarray}
for all values of $\theta$ and $\eta$, provided that the derivatives
in Eq. (\ref{inh:cond}) do not change sign.

\begin{figure}
\centerline{\psfig{figure=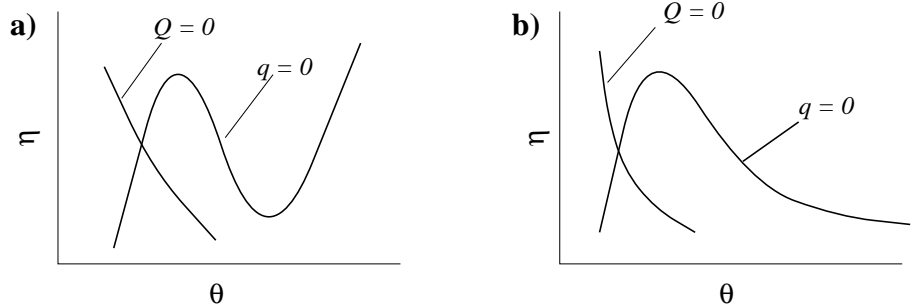,angle=0,width=12cm}}
\begin{center}
\caption{Two qualitatively different types of the nullclines of
Eqs. (\ref{gen:act}) and (\ref{gen:inh}): N-systems (a) and
$\Lambda$-systems (b). }
\label{nullgen}
\end{center}
\end{figure}

Kerner and Osipov showed
\cite{KO86,KO89,KO90,KO94,KO78,KO79,KO80,KO85} that the properties of
the patterns and the self-organization scenarios in systems described
by Eqs. (\ref{gen:act}) and (\ref{gen:inh}) are chiefly determined by
the parameters $\epsilon \equiv l/L$ and $\alpha \equiv \tau _\theta
/\tau _\eta $ and the shape of the nullcline of the equation for the
activator, that is, the dependence $\eta (\theta )$ given by the
equation $q(\theta ,\eta ,A)=0$ for $A={\mathrm const}$. They
demonstrated that depending on the shape of the activator nullcline
all systems involved can be divided into two fundamentally different
classes: N-systems, for which the nullcline is N- or inverted N-shaped
and, $\Lambda$- or V-systems, for which the nullcline is $\Lambda $-
or V-shaped, respectively (see Fig. \ref{nullgen}).

Most works devoted to the description of pattern formation on the
basis of Eqs. (\ref{gen:act}) and (\ref{gen:inh}) deal with
N-systems. In N-systems the equation $q(\theta ,\eta ,A)=0$ has three
roots: $\theta_1,\theta_2,$ and $\theta_3$, for given values of $A$
and $\eta $. The roots $\theta _1$ and $\theta_3$ correspond to the
stable states and $\theta _2$ corresponds to the unstable state in the
system with $\eta = {\mathrm const}$. It is easy to see that the
FitzHugh-Nagumo and the piecewise-linear models belong to
N-systems. For these models it was proved \cite{Ross,RK} that Eqs.
(\ref{gen:act}) and (\ref{gen:inh}) with $L=0$ and $\alpha =\tau
_\theta /\tau _\eta \ll 1$ have solutions in the form of the traveling
waves (also called autowaves \cite{Vas,Mik}, or traveling ASs \cite
{KO86,KO89,KO90,KO94}). In \cite{KO78,KO79,KO80,Koga} it was shown
that in another limit $L \gg l$ (or, more precisely, when $\epsilon =
l/L \ll 1$ and $\alpha \gtrsim 1$) Eqs. (\ref{gen:act}) and
(\ref{gen:inh}) admit solutions in the form of the stable static
patterns including ASs (see also \cite{KO89,KO90,KO94}).  Furthermore,
it was shown that in systems with $\epsilon \ll 1$ and $\alpha \ll 1$
one can excite static, pulsating, and traveling patterns \cite
{KO86,KO89,KO90,KO94,KO80,KO85,KO82,KO83,Kuzn,Mik94,Gaf95,Os96}. The
characteristic velocity of the traveling patterns in N-systems does
not exceed the value of order $l/\tau _\theta$
\cite{KO89,KO94,KO80,Ross}.

It is important to emphasize that $\ep$ or $\al$ are the natural small
parameters in these system.  Their smallness is in fact a necessary
condition for the feasibility of any patterns
\cite{KO89,KO90,KO94}. Indeed, if the inverse were true, that is, if
both the characteristic time and length scales of the variation of the
inhibitor were much smaller than those of the activator, the inhibitor
would easily damp all the deviations of the activator from the
homogeneous steady state, making the formation of any kinds of
persistent patterns impossible. On the other hand, the fact that we
must have either $\ep \lesssim 1$ or $\al \lesssim 1$ implies that it
is advantageous to consider the extreme cases of $\ep \ll 1$ or $\al
\ll 1$. These conditions, in turn, will result in a significant
simplification of the original highly nonlinear problem. Recently,
this kind of approach has been successfully applied to a variety of
problems (see, for example, \cite{Fife}).

In N-systems with $\epsilon \ll 1$ the static ASs and other patterns
are essentially the domains of high and low values of the activator
separated by the interfaces (walls) where $\theta $ varies sharply
over a distance of order $l$ from one stable state $\theta _1$ to the
other $\theta _3$. The characteristic size ${\mathcal L}_s$ of these
domain patterns lies in the range $l \ll {\mathcal L}_s \lesssim L$
and their amplitude (the value $\theta _1-$ $\theta _3$) is determined
by the form of $q$ and $Q$ and becomes independent of $\epsilon $ as
$\epsilon \rightarrow 0$
\cite{KO89,KO90,KO94,KO79,KO80,Koga,Gol,Mur96a,Mur96b}. The properties
of these patterns are essentially determined by the dynamics of their
interfaces and the interaction between them.  So, the majority of the
theoretical work devoted to the description of the complex domain
patterns in N-systems in fact developed an interfacial dynamics
approach
\cite{KO89,KO90,KO94,KO79,KO80,Koga,Gol,Mur96a,Mur96b,Hag,thesis,KO85,%
KO83,Kuzn,Mik94,Gaf95,Os96,Ohta,m:pre96,m1:pre97}. On the basis of
this approach, Kerner and Osipov developed a theory of instabilities
of the domain patterns in the general N-systems in one dimension
\cite{KO86,KO89,KO90,KO94}. More recently, we extended this theory for
arbitrary domain patterns in higher-dimensional systems and
extensively studied various pattern formation scenarios in these
systems \cite{Mur96a,Mur96b,m:pre96,m1:pre97}.

At the same time, there are many physical, chemical and biological
systems for which the activator nullcline is $\Lambda$- or V-shaped
[Fig. \ref{nullgen}(b)]. In this case the equation $q(\theta ,\eta
,A)=0$ for given $A$ and $\eta $ has only two roots: $\theta_1$
corresponding to the stable state, and $\theta _2$ corresponding to
the unstable state in the system with $\eta = {\mathrm const}$
\cite{KO89,KO90,KO94,KO78}. Among $\Lambda$-systems are many
semiconductor and gas discharge structures, electron-hole and gas
plasmas, radiation heated gas mixtures (see, for example, \cite
{KO89,KO90,KO94,Semi,KO78,KO85}).  It is not difficult to see that the
Brusselator and the Gray-Scott models are $\Lambda$-systems, and the
Gierer-Meinhardt model is a V-system.

Kerner and Osipov qualitatively showed that in $\Lambda$-systems the
so-called spike ASs and more complex spike patterns can be excited
\cite {KO89,KO94,KO78,Dub,Os93}. They were the first to analyze the
static spike ASs and strata in the Brusselator, the Gierer-Meinhardt
model, and the electron-hole plasma \cite{KO78,Dub}. They found that
when $\epsilon \ll 1$ and $\alpha \gtrsim 1$, the one-dimensional
static spike AS can have small size of order $l$ and huge amplitude
which goes to infinity as $\epsilon \rightarrow 0$.  Dubitskii,
Kerner, and Osipov formulated the asymptotic procedure for finding the
stationary solutions in $\Lambda$-systems for sufficiently small $\ep$
\cite{KO94,Dub}. Recently, we showed that in another limiting case
$\alpha \ll 1$ and $\epsilon \gg 1$ one can excite the one-dimensional
traveling spike AS which also has small size and whose amplitude goes
to infinity as $\alpha \rightarrow 0$ \cite{Mur95}. We also showed
that, in contrast to the traveling patterns in N-systems, the velocity
of this one-dimensional traveling spike AS can have huge values ($c
\gg l/ \tau_\theta$) and that the inhibitor distribution varies
stepwise in the front of the spike. Thus, one can see that the
properties of the spike patterns forming in $\Lambda $-systems differ
fundamentally from those of the domain patterns forming in
N-systems. In particular, since the interface connecting the two
stable states at $\eta = {\mathrm const}$ does not exist in $\Lambda
$- systems, the size of the spike should be of the order of the
smallest system length scale. For this reason, the concept of the
interfacial dynamics developed for the domain patterns in N-systems is
generally inapplicable to the description of spike patterns. Some
properties of the one-dimensional static spike ASs and the main types
of their instabilities in the simplified version of the Gray-Scott
model have recently been studied by Osipov and Severtsev \cite{Sev}.

Spike patterns including the spike ASs are observed experimentally in
the nerve tissue \cite{Katz}, chemical reactions \cite{Cross,Wood},
electron-hole plasma \cite{Vin}, gas-discharge structures \cite{Purw},
as well as numerically in the simulations of the Brusselator, the
Gierer-Meinhardt, and the Gray-Scott models
\cite{Prig,Lee,Pear,GM,Reyn}. At the same time, there is a only a
limited number of theoretical studies of these patterns. Moreover,
many aspects of these patterns including the stability of the spike
ASs and their properties in higher dimensions as well as the
spontaneous transitions between the static, the pulsating and the
traveling spike ASs have not been studied at all. So far, there have
been no general methods for dealing with spike patterns.

In the present paper we develop asymptotic methods for the description
of the spike patterns and study their major properties in arbitrary
dimensions as well as their shape and stability. We find the
conditions of the spontaneous transitions between different types of
the spike ASs and study the scenarios of the formation of the spike
ASs and more complex spike patterns in one- and two-dimensional
systems. To be specific, we consider the Gray-Scott model of an
autocatalytic chemical reaction, which possesses a number of
advantages. First, it is one of the rarest models for which in many
cases one can obtain exact results. Second, a lot of the numerical
studies of this model were performed recently
\cite{Lee,Pear,Reyn,kaper}. Finally, the Gray-Scott model possesses a
particularly simple set of nonlinearities, so one can expect a certain
degree of universality in the pattern formation scenarios exhibited by
it.

The outline of our paper is as follows. In Sec. \ref{s:mod} we
introduce the model we will study, in Sec. \ref{s:stat} we
asymptotically construct the one-dimensional static AS, and the two-
and the three-dimensional radially-symmetric ASs, in
Sec. {\ref{trav:s1} we asymptotically construct the solutions in the
form of the two types of traveling spike ASs, in Sec. \ref{s:stab} we
analyze the stability of the one-dimensional and the
higher-dimensional radially-symmetric static ASs and show the
existence of various instabilities, in Sec. \ref{s:pf1d} we compare
our results with the numerical simulations of the one-dimensional
system, and in Sec. \ref{s:pf2d} we do that for the two-dimensional
system, in Sec. \ref{s:dis} we discuss the works of other authors on
the Gray-Scott model in light of our results, and in Sec. \ref{s:conc}
we give the summary of our work and draw conclusions.

\section{The model} \label{s:mod}

The Gray-Scott model describes the kinetics of a simple autocatalytic
reaction in an unstirred flow reactor. The reactor is a narrow space
between two porous walls. Substance $Y$ whose concentration is kept
fixed outside of the reactor is supplied through the walls into the
reactor with the rate $k_0$ and the products of the reaction are
removed from the reactor with the same rate. Inside the reactor $Y$
undergoes the reaction involving an intermediate species $X$:
\begin{eqnarray}
2 X + Y & \stackrel{k_1\ }{\rightarrow} & 3 X, \label{reacta} \\ X &
\stackrel{k_2\ }{\rightarrow} & {\mathrm inert}. \label{reactb}
\end{eqnarray}
The first reaction is a cubic autocatalytic reaction resulting in the
self-production of species $X$; therefore, $X$ is the activator
species. On the other hand, the production of $X$ is controlled by
species $Y$, so $Y$ is the inhibitor species. The equations of
chemical kinetics which describe the spatiotemporal variations of the
concentrations of $X$ and $Y$ in the reactor and take into account the
supply and the removal of the substances through the porous walls take
the following form \cite{GS}:
\begin{eqnarray}
{\partial X \over \partial t} & = & - (k_0 + k_2) X + k_1 X^2 Y + D_X
\Delta X, \label{X} \\ {\partial Y \over \partial t} & = & k_0 (Y_0 -
Y) - k_1 X^2 Y + D_Y \Delta Y, \label{Y}
\end{eqnarray}
where now $X$ and $Y$ are the concentrations of the activator and the
inhibitor species, respectively, $Y_0$ is the concentration of $Y$ in
the reservoir, $\Delta$ is the two-dimensional Laplacian, and $D_X$
and $D_Y$ are the diffusion coefficients of $X$ and $Y$.

In order to be able to understand various pattern formation phenomena
in a system of this kind, it is crucial to introduce the variables and
the time and length scales that truly represent the physical processes
acting in the system. The first and the most important is the choice
of the characteristic time scales. These are primarily dictated by the
time constants of the dissipation processes. For $Y$ this is the
supply and the removal with the rate $k_0$, whereas for $X$ this is
the removal from the system and the decay via the second reaction with
the total rate $k_0 + k_2$. The natural way to introduce the
dimensionless inhibitor concentration is to scale it with $Y_0$. Since
we want to fix the time scale of the variation of the inhibitor (with
the fixed activator), we will rescale $X$ in such a way that the
reaction term in Eq. (\ref{Y}) will generate the same time scale as
the dissipative term. This leads to the following dimensionless
quantities:
\begin{eqnarray}
\theta = X / X_0, ~~~\eta = Y / Y_0, ~~~X_0 = \left( {k_0 \over k_1}
\right)^{1/2}.
\end{eqnarray}
The characteristic time and length scales for these quantities are
\begin{eqnarray}
\tau_\theta = \left(k_0 + k_2\right)^{-1}, & ~ & \tau_\eta = k_0^{-1}
\label{tau}, \\ l = \left( D_X \tau_\theta \right)^{1/2}, ~ & & L =
\left( D_Y \tau_\eta \right)^{1/2}.
\end{eqnarray}
Naturally, one should require the positivity of $\theta$ and $\eta$.

If we now write Eqs. (\ref{X}) and (\ref{Y}) in the dimensionless
form, we will arrive at the following set of equations:
\begin{eqnarray}
\tau_\theta {\partial \theta \over \partial t} & = & l^2 \Delta \theta
+ A \theta^2 \eta - \theta, \label{act0} \\ \tau_\eta {\partial \eta
\over \partial t} & = & L^2 \Delta \eta - \theta^2 \eta + 1 - \eta,
\label{inh0}
\end{eqnarray}
where we introduced a dimensionless parameter
\begin{eqnarray} \label{y0}
A = {Y_0 k_0^{1/2} k_1^{1/2} \over (k_0 + k_2)}.
\end{eqnarray}
One can see from Eqs. (\ref{act0}) and (\ref{inh0}) that $\tau_\theta$
and $\tau_\eta$ are in fact the characteristic time scales, and $l$
and $L$ the characteristic length scales of the variation of small
deviations of $\theta$ and $\eta$ from the stationary homogeneous
state $\theta = \theta_h$ and $\eta = \eta_h$:
\begin{eqnarray} \label{hom} 
\theta_h = 0, ~~~ \eta_h = 1.
\end{eqnarray}

Thus, the system is characterized by only three dimensionless
parameters: $\al \equiv \tau_\theta / \tau_\eta$, $\ep \equiv l / L$,
and $A$. As can be seen from Eq. (\ref{act0}), the parameter $A$ is
the dimensionless strength of the activation process, that is, it
describes the degree of deviation of the system from thermal
equilibrium. With all this, Eqs. (\ref{act0}) and (\ref{inh0}) are
reduced to the form of Eqs. (\ref{gen:act}) and
(\ref{gen:inh}). Notice that the system given by Eqs. (\ref{act0}) and
(\ref{inh0}) is indeed a system of the activator-inhibitor type: the
condition in Eq. (\ref{act:cond}) is satisfied for $\theta >
\frac{1}{2 A \eta}$, and the conditions in Eq. (\ref{inh:cond}) are
satisfied with $q'_\eta < 0$ and $Q'_\theta > 0$ for all $\theta > 0$
and $\eta > 0$.

The nullclines of Eqs. (\ref{act0}) and (\ref{inh0}) are shown in
Fig. \ref{null}. 
\begin{figure}
\centerline{\psfig{figure=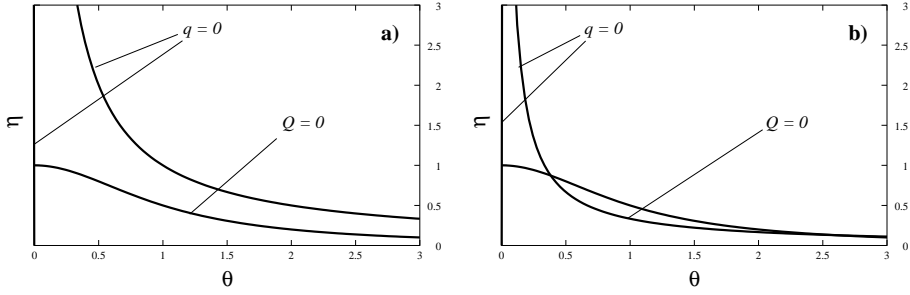,angle=-90,width=12cm}}
\begin{center}
\caption{The nullclines of Eqs. (\ref{act0}) and (\ref{inh0}) for $A =
1$ (a) and $A = 3$ (b).} \label{null}
\end{center}
\end{figure}
From this figure one can see that the nullcline of the equation for
the activator has degenerate $\Lambda$-form. It consists of two
separate branches: $\theta = 0$ and $\theta = {1 \over A \eta}$. One
can easily check that for $0 < A < 2$ there is only one stationary
homogeneous state given by Eq. (\ref{hom}), whereas for $A > 2$ two
extra stationary homogeneous states exist
\begin{eqnarray} \label{hom2}
\theta_{h2,3} = {A \mp \sqrt{A^2 - 4} \over 2}, ~~~\eta_{h2,3} = {A
\pm \sqrt{A^2 - 4} \over 2 A}.
\end{eqnarray}
The stability analysis of these homogeneous states shows that for $\ep
\ll 1$ or $\al \ll 1$ the homogeneous state $\theta = \theta_{h2}$,
$\eta = \eta_{h2}$ is always unstable. For $\ep \ll 1$ the homogeneous
state $\theta = \theta_{h3}$, $\eta = \eta_{h3}$ is unstable with
respect to the Turing instability if $A < 0.41 \ep^{-1}$. For $\al \ll
1$ it is unstable with respect to the homogeneous oscillations (Hopf
bifurcation) if $0.41 \al^{-1/2} < A < \al^{-1/2}$, or it is an
unstable node if $A < 0.41 \al^{-1/2}$. On the other hand, the
homogeneous state $\theta = \theta_h$, $\eta = \eta_h$ is stable for
all values of the system's parameters. The latter is simple to
understand: in order for the reaction to begin there has to be at
least some amount of the activator put in at the start. Equivalently,
the fact that the homogeneous state in Eq. (\ref{hom}) is stable for
all values of the parameter $A$ (for an arbitrary deviation from
thermal equilibrium) is the consequence of the degeneracy of the
nullcline of Eq. (\ref{act0}). Thus, self-organization associated with
the Turing instability of the homogeneous state $\theta_h = 0$ and
$\eta_h = 1$ is not realized in the Gray-Scott model. In such a stable
homogeneous system any inhomogeneous pattern, including the ASs, can
only be excited by a sufficiently strong localized stimulus. In turn,
self-organization will occur as a result of the instabilities of the
large-amplitude patterns already present in the system.

Note that in the opposite case $\ep \gg 1$ and $\al \gg 1$ the
dynamics of the system becomes dramatically simpler. Indeed, if we put
both $L$ and $\tau_\eta$ to zero, from Eq. (\ref{inh0}) we get a local
relationship $\eta = {1 \over 1 + \theta^2}$. Substituting this back
to Eq. (\ref{act0}), we obtain
\begin{eqnarray}
\tau_\theta {\partial \theta \over \partial t} = l^2 \Delta \theta +
{A \theta^2 \over 1 + \theta^2} - \theta.
\end{eqnarray}
This equation possesses a simple variational structure
\begin{eqnarray}
\tau_\theta {\partial \theta \over \partial t} = - {\delta {\mathcal
F} \over \delta \theta}, ~~~~{\mathcal F} = \int d^d x \left( {l^2
(\nabla \theta)^2 \over 2} - A \theta + A \arctan \theta + {\theta^2
\over 2} \right).
\end{eqnarray}
For $A < 2$ the functional $\mathcal F$ has a unique global minimum at
$\theta = \theta_h = 0$, so any initial condition will relax to the
homogeneous state $\theta_h$. For $A > 2$ there are two stable
homogeneous states $\theta = \theta_h$ and $\theta = \theta_{h3}$ (see
above), so it is possible to have the waves of switching from one
homogeneous state to the other \cite{Cross}. It is easily checked that
for $2 < A < 2.18$ the dominant homogeneous state is $\theta_h$, while
for $A > 2.18$ the dominant homogeneous state is $\theta_{h3}$.

In the case of $\ep \ll 1$ and $\al \ll 1$ the largest length scale in
the system is $L$ and the longest time scale is $\tau_\eta$, so it is
natural to scale length and time with $L$ and $\tau_\eta$,
respectively. In these units Eqs. (\ref{act0}) and (\ref{inh0}) will
take the following form:
\begin{eqnarray}
\al {\partial \theta \over \partial t} = \ep^2 \Delta \theta + A
\theta^2 \eta - \theta, \label{act} \\ {\partial \eta \over \partial
t} = \Delta \eta - \theta^2 \eta + 1 - \eta. \label{inh}
\end{eqnarray}
We will assume that the problem is defined on the sufficiently large
domain with neutral boundary conditions. Notice that the kinetic model
used to arrive at Eqs. (\ref{act}) and (\ref{inh}) imposes a
restriction $\al \leq 1$ [see Eq. (\ref{tau})]. Also, in the
derivation we assumed that the system is essentially
two-dimensional. For the sake of generality, in the following we will
allow $\al$ to take arbitrary values and will work with the
arbitrarily-dimensional Gray-Scott model.

\section{Static spike autosolitons} \label{s:stat}

Let us now study the simplest possible stationary pattern in the
Gray-Scott model --- the static spike AS. According to the general
qualitative theory, these ASs form in $\Lambda$-systems when $\ep \ll
1$ \cite{KO89,KO90,KO94,KO78}. The condition $\epsilon \ll 1$ will
therefore be assumed throughout this section.

\subsection{One-dimensional static spike autosoliton} \label{stat:1d}

We begin with the analysis of the one-dimensional static spike AS.  In
the Gray-Scott model it is described by the following equations
\begin{eqnarray}
\ep^2 {d^2 \theta \over d x^2} + A \theta^2 \eta - \theta = 0, 
\label{act:1d} \\ {d^2 \eta \over d x^2} - \theta^2 \eta + 1 -
\eta = 0. && \label{inh:1d}
\end{eqnarray}

Since $\ep \ll 1$, there is a strong separation of the length scales
in the AS \cite{KO89,KO90,KO94,KO78}. One can separate the spike
region where the distribution of $\theta$ varies on the length scale
of $\ep$, and the periphery of the AS where $\eta$ decays into the
homogeneous state $\eta_h = 1$ on the length scale of order 1. One can
use this separation of the length scales to construct a singular
perturbation theory which describes the distributions in the form of
the static one-dimensional spike AS \cite{Dub}. But before we do that,
it is instructive to use a more qualitative approach which will give
us an idea about the scaling of the main parameters of the AS and its
qualitative shape. As will be seen below, this approach works when
$\ep^{1/2} \lesssim A \ll 1$.

\subsubsection{Case $A \sim \ep^{1/2}$: autosoliton collapse}
\label{stat:ss1} 

According to this approach \cite{KO89,KO90,KO94,KO78}, one assumes
that the value of $\eta$ inside the spike (on the length scale of
$\ep$) is close to a constant. This is a reasonable assumption as long
as $\eta \gg \ep$ in the spike since the characteristic length scale
of the variation of $\eta$ is 1. Let us denote this constant value of
$\eta$ as $\eta_s$. Then, Eq. (\ref{act:1d}) with $\eta = \eta_s$ can
be solved exactly. Its solution has the form
\begin{eqnarray} \label{th1dsh0}
\theta(x) = \theta_m \cosh^{-2} \left( {x \over 2 \ep} \right) {\mathrm 
~with~} \theta_m = \frac{3}{2 A \eta_s}.
\end{eqnarray}

On the other hand, the distribution of $\theta$ given by
Eq. (\ref{th1dsh0}) acts in Eq. (\ref{inh:1d}) as a $\delta$-function,
so away from the spike the distribution of $\eta$ is given by
\begin{eqnarray} \label{et1dsm0}
\eta(x) = 1 - {3 \ep \over \eta_s A^2 } \e^{- |x| }.
\end{eqnarray}
Now, matching this solution for $\eta(x)$ with the condition that
$\eta(0) = \eta_s$, we obtain the following expressions
\begin{eqnarray} \label{pm1d}
\theta_m = \frac{3 A}{A_b^2} \left[ 1 \pm \sqrt{1 - {A_b^2 \over A^2}
} ~\right], ~~ \eta_s = \frac{A_b^2}{2 A^2} \left[ 1 \pm \sqrt{1 -
{A_b^2 \over A^2} } ~\right]^{-1},
\end{eqnarray}
where 
\begin{eqnarray} \label{Ab1d}
A_b = \sqrt{12 \ep}.
\end{eqnarray}
Note that these results were also obtained in \cite{kaper} by applying
Melnikov analysis to Eqs (\ref{act:1d}) and (\ref{inh:1d}). Similar
results for the simplified version of the Gray-Scott model were
obtained in \cite{Sev}.

From Eq. (\ref{Ab1d}) one can see that at $A < A_b$ the solution in
the form of the spike AS does not exist. When $A > A_b$ there are two
solutions: the one corresponding to the plus sign has larger amplitude
and the one corresponding to the minus sign has smaller amplitude. As
was shown by Kerner and Osipov, the solutions that have smaller
amplitude are always unstable \cite{KO89,KO90,KO94}, so the only
interesting solution corresponds to the plus sign in
Eq. (\ref{pm1d}). This solution is precisely the static spike AS. The
numerical simulations of Eqs. (\ref{act}) and (\ref{inh}) show that if
the value of $A$ is lowered, at $A = A_b$ a stable one-dimensional
static spike AS collapses into the homogeneous state (see
Sec. \ref{s:pf1d}).

Let us look more closely at the parameters of the static spike AS and
the conditions of validity of the approximations made in the preceding
paragraphs. As can be seen from Eqs. (\ref{th1dsh0}) and
(\ref{et1dsm0}), the distribution of the activator indeed has a form
of the spike whose characteristic width is of order $\ep$, and the
distribution of the inhibitor varies on the much larger length scale
of order 1. Also, according to Eq. (\ref{pm1d}), the amplitude of the
spike at $A$ close to $A_b$ is of order $\ep^{-1/2} \gg 1$ (and can in
fact have huge values as $\ep$ gets smaller) and grows as the value of
$A$ increases. At $A \sim 1$ the amplitude $\theta_m \sim
\ep^{-1}$. These features fundamentally differ the AS forming in
$\Lambda$-systems from the AS in N-systems.

Recall that in the derivation we neglected the variation of the
inhibitor inside the spike. Since the characteristic length of
the variation of $\eta$ is of order 1, this means that the value of $\eta
= \eta_s$ in the center of the AS must be much greater than
$\ep$. According to Eq. (\ref{pm1d}), this is indeed the case as long
as $A \ll 1$, so the solution obtained above is a good approximation
to the actual solution in this case. Also, in this case one can easily
calculate the distribution of $\eta$ in the spike. To do this, we note
that, according to Eq. (\ref{pm1d}), for $A \ll 1$ we have $\theta^2
\eta \gg 1$ in the spike, so the last two terms in Eq. (\ref{inh:1d})
can be neglected. Since the variation of $\eta$ in the spike is much
smaller compared to $\eta_s$, we can put $\eta = \eta_s$ in the
right-hand side of Eq. (\ref{inh:1d}). Then, substituting $\theta$
from Eq. (\ref{th1dsh0}) into this equation, after simple integration
we obtain an expression for $\eta$ in the spike region
\begin{eqnarray} \label{esh1d0}
\eta(x) = \eta_s + {4 \ep^2 \theta_m^2 \eta_s \over 3} \left( 2 \ln
\cosh {x \over 2 \ep} + \frac{1}{2} \tanh^2 {x \over 2 \ep} \right).
\end{eqnarray}

\subsubsection{Case $A \sim 1$: local breakdown}
\label{stat:ss2} 

On the other hand, according to Eqs. (\ref{pm1d}), when $A \sim 1$, we
have
\begin{eqnarray} \label{minmax1d}
\theta_{\mathrm max} \sim \ep^{-1}, ~~~~\eta_{\mathrm min} \sim \ep,
\end{eqnarray}
and the approximation used by us ceases to be valid. However, it is
clear that qualitatively the character of the solution should not
change even for these values of $A$. Therefore, we can still assume
that the spike of the AS has the width of order $\ep$ and that the
values of the activator and the inhibitor scale the same way as those
in Eq. (\ref{minmax1d}). With all this in mind, we are now able to
introduce singular perturbation expansion and separate the ``sharp''
distributions (inner solutions) that vary on the length scale of $\ep$
and the ``smooth'' distributions (outer solutions) that vary on the
length scale of order 1.

At distances much greater than $\ep$ away from the spike (in the outer
region) the value of $\theta$ is exponentially zero,\footnote{This
follows from the fact that in the region of the smooth distributions
$\theta$ and $\eta$ are related locally through the equation
$q(\theta, \eta) = 0$ and therefore must lie on the stable branch of
the nullcline of the equation for the activator \cite{KO89,KO90,KO94},
which in the case of the Gray-Scott model gives especially simple
relation: $\theta = 0$} so the equation for the smooth distributions
becomes
\begin{eqnarray} \label{sm1d}
{d^2 \eta \over d x^2} + 1 - \eta = 0,
\end{eqnarray}
with the boundary condition in the spike $\eta(0) = 0$ (to order
$\ep$) and $\eta = \eta_h$ at infinity. This immediately
gives us the smooth distribution of $\eta$
\begin{eqnarray} \label{sm1dsol}
\eta(x) = 1 - \e^{-|x|}. 
\end{eqnarray}

Let us scale the activator and the inhibitor according to
Eq. (\ref{minmax1d}) and introduce the stretched variable $\xi$:
\begin{eqnarray} \label{scale1d}
\tilde\theta = \ep \theta, ~~\tilde\eta = \ep ^{-1} \eta, ~~\xi =
\frac{x}{\ep}.
\end{eqnarray}
Using these variables, after a little algebra we can write
Eqs. (\ref{act:1d}) and (\ref{inh:1d}) as
\begin{eqnarray}
\tilde\theta_{\xi\xi}+ A \tilde\theta^2 \tilde\eta - \tilde\theta = 0,
\label{sh1dscaleda} \\ \tilde\eta_{\xi\xi} = A^{-1} (\tilde\theta -
\tilde\theta_{\xi\xi} ), \label{sh1dscaledb}
\end{eqnarray}
where we kept only the leading terms. In this equation
$\tilde\theta_{\xi\xi}$ denotes the second derivative with respect to
$\xi$. Thus, the solution of Eqs. (\ref{sh1dscaleda}) and
(\ref{sh1dscaledb}) properly matched with the smooth distribution,
given by Eq. (\ref{sm1dsol}), will give the sharp distributions of the
activator and the inhibitor in the spike.

The matching of the sharp and the smooth distributions is performed by
noting that, according to Eq. (\ref{scale1d}), to order $\ep$ we have
$\eta = 0$ and $\eta_x \sim 1$ for $\ep \ll |x| \ll 1$. Therefore, it
is the derivative of $\eta$ obtained from the sharp distribution at
$|\xi| \gg 1$ that must coincide with that of the smooth distribution
for $|x| \ll 1$. This condition is obtained by imposing the boundary
condition $\tilde\eta_\xi (\pm \infty) = \pm 1$ in
Eq. (\ref{sh1dscaledb}) [see Eq. (\ref{sm1dsol})]. One can obtain an
integral representation of this boundary condition by integrating
Eq. (\ref{sh1dscaledb}) over $\xi$. Let us introduce the variables
\begin{eqnarray} \label{bar1d}
\bar\theta = {\tilde\theta \over A}, ~~~ \bar\eta = \tilde\eta +
{\tilde\theta \over A}.
\end{eqnarray}
Then, this integral condition takes the form
\begin{eqnarray} \label{match1d}
\int_{-\infty}^{+\infty} \bar\theta d \xi = 2 |\lambda|^{-1/2},
\end{eqnarray}
where $\lambda$ is a constant that should be equal to $-1$ (the reason
for introducing this coefficient will be explained in the following
paragraph).  In terms of the new variables Eq. (\ref{sh1dscaledb})
becomes especially simple
\begin{eqnarray} \label{sh1deta}
\bar\eta_{\xi\xi} = |\lambda| \bar\theta,
\end{eqnarray}
where $\lambda$ is the same constant. Notice that Eq. (\ref{sh1deta})
has an obvious symmetry which allows us to add an arbitrary constant
to $\bar\eta$, so we can replace $\bar\eta \rightarrow \bar\eta +
\bar\eta_s$, where $\bar\eta$ satisfies the condition $\bar\eta(0) = 
\bar\eta_\xi (0) = 0$ and thus is uniquely determined by $\bar\theta$,
and $\bar\eta_s$ is an arbitrary constant.

To analyze Eqs. (\ref{sh1dscaleda}) and (\ref{sh1dscaledb}), it is
convenient to rewrite them as a nonlinear eigenvalue problem
\begin{eqnarray}
\left[ - {d^2 \over d \xi^2} + V(\xi) \right] \bar\theta = \lambda
\bar\theta, \label{neigen1da} \\ V(\xi) = - A^2 \bar\theta
(\bar\eta_s + \bar\eta - \bar\theta),
\label{neigen1db} 
\end{eqnarray}
where $\bar\eta$ is in turn related to $\bar\theta$ via
Eq. (\ref{sh1deta}). Then, since $\bar\theta$ is positive for all
$\xi$ and therefore has no nodes, the solution in the form of the
static spike AS will correspond to the lowest bound state of the
operator in Eq. (\ref{neigen1da}) with $\lambda = -1$. The latter is
achieved by adjusting the value of $\bar\eta_s$. 

The nonlinear eigenvalue problem given by Eqs. (\ref{neigen1da}) and
(\ref{neigen1db}) together with Eqs. (\ref{match1d}) and
(\ref{sh1deta}) with fixed $\bar\eta_s$ can be solved
iteratively. Indeed, for a given potential well $V$ there is a unique
eigenvalue $\lambda$ and a unique eigenfunction $\bar\theta$ (up to
normalization) that correspond to the lowest bound state of the
Schr\"odinger operator in Eq. (\ref{neigen1da}). Equation
(\ref{match1d}) gives a unique normalization for $\bar\theta$, which
then uniquely determines $\bar\eta$ through
Eq. (\ref{sh1deta}). Knowing the distributions $\bar\theta$ and
$\bar\eta$, one can then reconstruct the potential $V'$, thus defining
an iterative map. It is convenient to think of the solutions of the
nonlinear eigenvalue problem as fixed points of this iterative map.

Observe that the nonlinear eigenvalue problem is invariant with
respect to the following transformation
\begin{eqnarray} \label{sym1d}
\xi \rightarrow \frac{\xi}{b}, ~~~\lambda \rightarrow b^2 \lambda,
~~~A \rightarrow b A,
\end{eqnarray}
where $b$ is an arbitrary positive constant. It is clear that if one
knows a solution of the nonlinear eigenvalue problem with certain
$\lambda$, one can obtain a solution of Eqs. (\ref{sh1dscaleda}) and
(\ref{sh1dscaledb}) by simply using the symmetry transformation in
Eq. (\ref{sym1d}) with $b = |\lambda|^{-1/2}$, so there is in fact a
one-to-one correspondence between the solutions of the nonlinear
eigenvalue problem with arbitrary $\lambda$ and its solution with
$\lambda = -1$ which corresponds to the sharp distributions.

Since we are interested in the lowest bound state whose eigenvalue is
equal to $-1$, the characteristic length scale of the variation of
$\bar\theta$ and, according to Eq. (\ref{sh1deta}), of $\bar\eta$ and
$V$ as well, is of order 1. Equation (\ref{match1d}) fixes the
normalization of $\bar\theta$, so we must have $\bar\theta \sim
1$. Notice that in view of Eq. (\ref{sh1deta}) and the fact that
$\bar\theta \sim 1$, we must always have $\bar\eta \sim 1$.

Let us write the potential $V$ in Eq. (\ref{neigen1da}) and
(\ref{neigen1db}) as a sum of two parts: $V = V_0 + V_1$, where
\begin{eqnarray} \label{V01d}
V_0 = - A^2 \bar\theta \bar\eta_s, ~~~V_1 = - A^2 \bar\theta
(\bar\eta - \bar\theta).
\end{eqnarray}
From the qualitative form of $\tilde\eta(\xi)$ [see, for example,
Eq. (\ref{esh1d0})] it is easy to see that the potential $V_0$ has the
form of a simple potential well, while $V_1$ has the form of a double
well (see Fig. \ref{v1d}).
\begin{figure}
\centerline{\psfig{figure=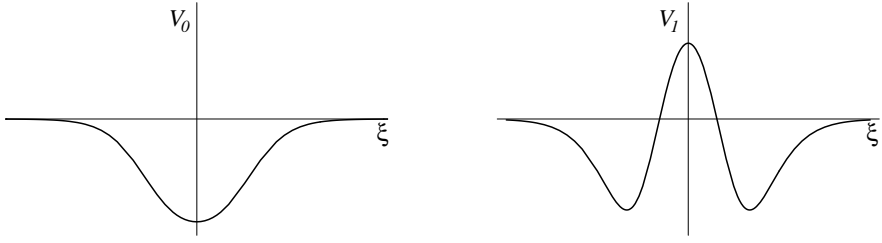,angle=-90,width=12cm}}
\begin{center}
\caption{Qualitative form of the potentials $V_0$ and $V_1$ from
Eqs. (\ref{V01d}). }
\label{v1d}
\end{center}
\end{figure}

In order for the operator in Eq. (\ref{neigen1da}) to have the lowest
bound state with $\lambda = -1$ the potential $V$ must have the depth
of order 1.  When $A \ll 1$, the function $\tilde\eta(\xi)$ must be
chosen in such a way that it compensates for the small factor of $A$
in Eq. (\ref{V01d}). However, since $\bar\eta \sim 1$, this can only
be achieved by choosing $\bar\eta_s \sim A^{-2} \gg 1$. This means
that we will have $V_0 \gg V_1$. If one neglects $V_1$ compared to
$V_0$, one can solve the nonlinear eigenvalue problem exactly. This
solution will be
\begin{eqnarray} \label{sh1d000}
\bar\theta(\xi) = {1 \over 2} \cosh^{-2} \left( {\xi \over 2} \right),
~~~\bar\eta_s = {3 \over A^2}.
\end{eqnarray}
The potential $V_1$ can then be treated as a perturbation which will
give corrections to $\bar\theta$ and $\bar\eta$, so one should not
expect any qualitative changes in the behavior of the solution for $A
\lesssim 1$.

In the other limiting case $A \gg 1$ the nonlinear eigenvalue problem
will not have solutions with $\lambda = -1$. Indeed, in this case the
potential $V$ is always deep with the depth $\gtrsim A^2$ regardless
of the choice of $\bar\eta_s$, so the lowest eigenvalue of the
operator in Eq. (\ref{neigen1da}) will be $|\lambda| \sim A^2 \gg 1$
(assuming that $V$ varies on the length scale of order 1). This means
that the solution in the form of the static spike AS exists only when
$A \lesssim 1$. Note that this result was also obtained in
\cite{kaper} by the method of topological shooting. 

The numerical solution of the nonlinear eigenvalue problem shows that for
$A = 1$ and arbitrary $\lambda$ [recall that the solutions for all
other values of $A$ can be obtained using the symmetry transformation
given by Eq. (\ref{sym1d})] there exists a unique  stable
solution for all $\bar\eta_s$ greater than some critical value
$\bar\eta_s^*$ (see Fig. \ref{lam1d}).
\begin{figure}
\centerline{\psfig{figure=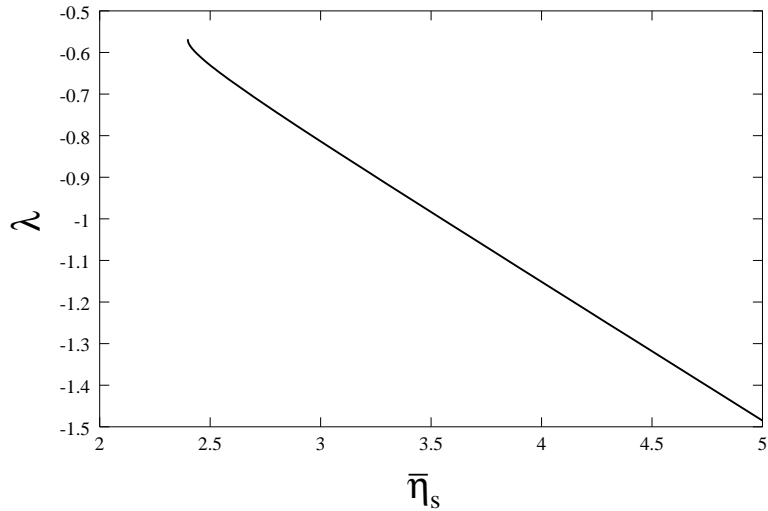,width=10cm}}
\begin{center}
\caption{Dependence $\lambda(\bar\eta_s)$ for the one-dimensional
static AS. Results of the numerical solution of the nonlinear
eigenvalue problem with $A = 1$.}
\label{lam1d}
\end{center}
\end{figure}
This can be explained in the following way. For large enough values of
$\bar\eta_s$ the potential $V_0$, which can always produce a localized
state, dominates in the total potential $V$. As the value of
$\bar\eta_s$ decreases, the effect of the potential $V_1$ becomes more
and more pronounced, so $V$ gradually transforms from a single-well to
a double-well potential. This means that with the decrease of
$\bar\eta_s$ the function $\bar\theta$ will tend to localize in the
minima of $V_1$ instead of the minimum of $V_0$ (see
Fig. \ref{v1d}). On the other hand, the localization of $\bar\theta$
in the minima of $V_1$ will in turn increase $V_1$, since the latter
is self-consistently determined by $\bar\theta$ [see
Eq. (\ref{sh1deta})]. If one constructs the solution of the nonlinear
eigenvalue problem iteratively, for small enough values of
$\bar\eta_s$ one will find that at each step of the iterations the
potential $V$ is such that at the next step the distribution of
$\bar\theta$ will become localized further and further away from the
origin. On the other hand, it is easy to show that there are no
solution of the nonlinear eigenvalue problem in the form of a pair of
spikes some distance $L \gtrsim 1$ apart. Suppose that we have a
solution in the form of two spikes centered at $\xi = \pm L/2$, with
$L \gg 1$. Let us multiply Eq. (\ref{neigen1da}) by $\bar\theta_\xi$
and integrate it over positive $\xi$. As a result, using the fact that
for $L \gg 1$ with exponential accuracy $\bar\theta(0) = 0$, we obtain
the relation $A^2 \int_0^\infty \bar\theta^3 \bar\eta_\xi d \xi =
0$. This relation, however, cannot be satisfied since the function
$\bar\eta_\xi$ is positive definite for all positive $\xi$, so the
solution of the assumed form does not exist. So, for $\bar\eta_s <
\bar\eta_s^*$ the iterative procedure will not converge and at
$\bar\eta_s = \bar\eta_s^*$ the solution of the nonlinear eigenvalue
problem abruptly disappears.

From Fig. \ref{lam1d} one can see that $\lambda$ is a monotonically
decreasing function of $\bar\eta_s$, with its maximum attained at
$\bar\eta_s = \bar\eta_s^* \simeq 2.40$ for $A = 1$ (see
Fig. \ref{lam1d}). According to Eq. (\ref{sym1d}), we have
$|\lambda_{\mathrm max}| = |\lambda(\bar\eta_s^*)| = {\mathrm const}/
A^2$. Therefore, at some $A = A_d \sim 1$ we will have
$\lambda_{\mathrm max} = -1$, so that for $A > A_d$ there will be no
solutions corresponding to the one-dimensional static spike AS. Thus,
at $A = A_d$ there is a bifurcation of the static spike AS which
results in the local breakdown and leads to its splitting and
self-replication (see Sec. \ref{stab:1d} and \ref{s:pf1d}).

The numerical solution of Eqs. (\ref{sh1dscaleda}) and (\ref{sh1dscaledb})
together with Eq. (\ref{match1d}) confirm our conclusions about the
behavior of the sharp distributions as the value of $A$ is
varied. Figure \ref{ett1d} shows the dependences of the values of
$\tilde\theta$ and $\tilde\eta$ at $\xi = 0$ on $A$ obtained from the
numerical solution of Eqs. (\ref{sh1dscaleda}) and
(\ref{sh1dscaledb}). 
\begin{figure}
\centerline{\psfig{figure=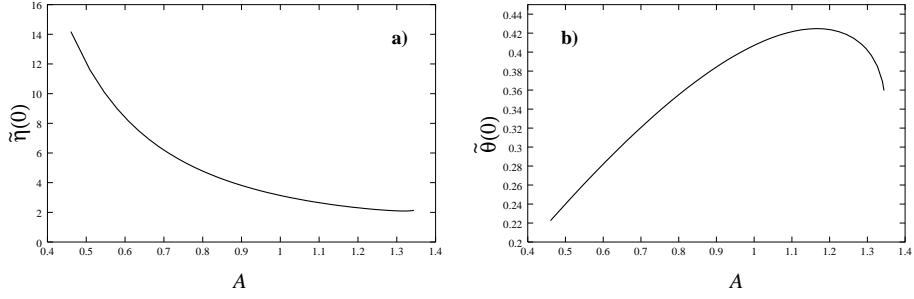,angle=-90,width=12cm}}
\begin{center}
\caption{The values of $\tilde\eta(0)$ (a) and $\tilde\theta(0)$ (b)
as functions of $A$ for the one-dimensional AS obtained from the
numerical solution of Eqs. (\ref{sh1dscaleda}) and (\ref{sh1dscaledb}).}
\label{ett1d}
\end{center}
\end{figure}
From this figure one can see that the solution indeed disappears at $A
= A_d$ with the value of $A_d$ found to be $A_d = 1.35$. Figure
\ref{sh1df} shows the distributions of $\tilde\theta$ and $\tilde\eta$
in the spike obtained from the numerical solution of
Eqs. (\ref{sh1dscaleda}) and (\ref{sh1dscaledb}) for a particular
value of $A$.
\begin{figure}
\centerline{\psfig{figure=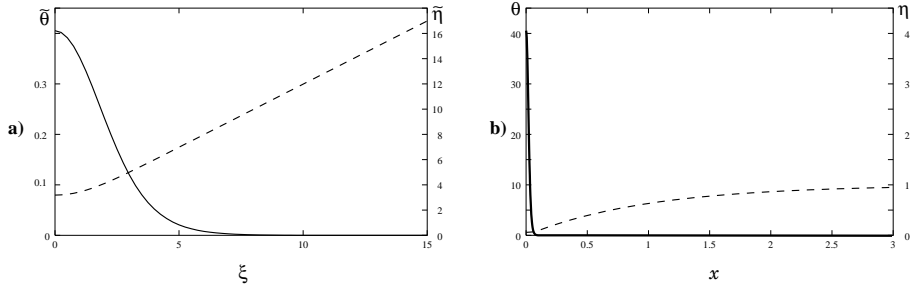,angle=-90,width=12cm}}
\begin{center}
\caption{Sharp distributions $\tilde\theta(\xi)$ (solid line) and
$\tilde\eta(\xi)$ (dashed line) in a one-dimensional static AS,
obtained from the numerical solution of Eqs. (\ref{sh1dscaleda}) and
(\ref{sh1dscaledb}) with $A = 1$; (b) The full solution for $\theta$
(solid line) and $\eta$ (dashed line) at $A = 1$ and $\ep = 0.01$. }
\label{sh1df}
\end{center}
\end{figure}
It also shows the entire solution obtained by matching the sharp and
the smooth distributions for the particular values of $A$ and
$\ep$. Notice that the distributions given by Eqs. (\ref{sh1d000})
give a very good approximation to the actual solution whenever $A$ is
not in the immediate vicinity of $A_d$ (for example, at $A < 0.8$
these distributions give the solutions with the accuracy better than
10\%).

\subsubsection{Case $\ep^{1/2} \ll A \ll 1$} \label{stat:ss3}

Let us now consider the intermediate case $\ep^{1/2} \ll A \ll 1$. In
this case the results of Sec. \ref{stat:ss1} and \ref{stat:ss2} both
predict for $\theta$ and $\eta$ in the spike
\begin{eqnarray} \label{sh1d00}
\theta(x) = {A \over 2 \ep} \cosh^{-2} \left(x \over 2 \ep \right),
~~~\eta_s = {3 \ep \over A^2},
\end{eqnarray}
and the correction to $\eta$ to be given by Eq. (\ref{esh1d0}) with
$4 \ep^2 \theta_m^2 \eta_s / 3 = \ep$ [see Eqs. (\ref{pm1d}) for $A
\gg \ep^{1/2}$]. Note that Eqs. (\ref{sh1d00}) were also obtained in
\cite{kaper}. 

Observe that the procedure presented in Sec. \ref{stat:ss2} is valid
with accuracy $\ep$ when $A \sim 1$. For these values of $A$ it is
justified to assume in the matching condition that with this accuracy
$\eta(0) = 0$. As the value of $A$ decreases, the actual value of
$\eta(0) \sim \ep/A^2$ grows, so the accuracy of the above mentioned
approximation decreases, and at $A \sim \ep^{1/2} \sim A_b$ this
approximation becomes invalid. On the other hand, in the procedure
discussed in Sec. \ref{stat:ss1} the matching condition uses the true
value of $\eta(0) \sim 1$ for $A \sim A_b$, but neglects the variation
of $\eta$ in the spike. The latter gives the corrections of order
$A^2$ to the solution given by Eq. (\ref{th1dsh0}), which are of order
$\ep$ when $A \sim A_b$ and grow as $A$ increases. For $A \sim
\ep^{1/4}$ both these procedures give the same solutions with the
accuracy $\ep^{1/2}$, so in fact for $\ep \ll 1$ one can construct the
solution in the form of the static spike AS asymptotically for all
values of $A$. When $A \lesssim \ep^{1/4}$, one should use the
procedure described in Sec. \ref{stat:ss1} and when $A \gtrsim
\ep^{1/4}$ one should use that of Sec. \ref{stat:ss2}.

Above we presented the ways to construct asymptotically the solution
in the form of the one-dimensional static spike AS. As such, these
procedures should be good only for sufficiently small values of $\ep
\ll 1$. According to the analysis above, this AS exists in a wide
range $A_b < A < A_d$ with $A_b \sim \ep^{1/2} \ll 1$ and $A_d \sim
1$. This implies that in order for the whole asymptotic procedure to
be in quantitative agreement with the actual solution we must have
$A_b \ll A_d$. In view of Eq. (\ref{Ab1d}), this will be the case when
$\ep \lesssim 0.03$.

\subsection{Three-dimensional radially-symmetric static spike
autosoliton} \label{stat:3d}

Let us now study the higher dimensional static spike ASs. These are
radially symmetric spikes of large amplitude and the size of order
$\ep$ \cite{KO89,KO90,KO94}. As we will show below, in the Gray-Scott
model the properties of the solutions in the form of the
higher-dimensional radially-symmetric spike ASs turn out to be
different from those of the static spike AS in one dimension.

Let us consider a three-dimensional AS first. The distributions of
$\theta$ and $\eta$ in the form of the AS will be determined by
Eqs. (\ref{act}) and (\ref{inh}) in which the time derivatives are put
to zero and only the radially symmetric part of the Laplacian is
retained, with the spike centered at zero. When $\ep \ll 1$, one can
once again use singular perturbation theory and separate the sharp
distributions (inner solutions) in the spike from the smooth
distributions (outer solutions) away from the spike.

As in the case of the one-dimensional AS, away from the spike the
activator and the inhibitor become decoupled, so that $\theta = 0$
there and the smooth distribution of $\eta$ is given by
\begin{eqnarray} \label{sm3d}
\eta(r) = 1 - \frac{a \e^{-r}}{r},
\end{eqnarray}
where $r$ is the radial coordinate and $a$ is a certain constant [see
Eq. (\ref{inh})]. The constant $a$ is determined by the strength of
the $\delta$-function like source term at $r = 0$. Integrating
Eqs. (\ref{act}) and (\ref{inh}) over the spike region, we obtain that
\begin{eqnarray} \label{c3d}
a = \frac{1}{A} \int_0^\infty r^2 \theta(r) d r,
\end{eqnarray}
where the integration was extended to the whole space since $\theta =
0$ away from the spike. One can see an important difference between
Eq. (\ref{sm3d}) and Eq. (\ref{sm1dsol}) for the one-dimensional AS:
in the case of the three-dimensional AS the derivative of $\eta$ with
respect to $r$ in the smooth distribution becomes singular at $r =
0$. This means that the scaling of $\theta$ and $\eta$ in the spike
will be different from that of the one-dimensional AS. Indeed, for $r
\sim \ep$ the value of $\eta$ must be positive, so we must have $a
\sim \ep$. According to Eq. (\ref{c3d}), this implies that in the
spike $\theta \sim A \ep^{-2}$. Also, from Eq. (\ref{sm3d}) one can
see that near the spike $\eta$ varies by values of order 1 on the
length scale of $\ep$. Since we must have $A \theta^2 \eta \sim
\theta$ in Eq. (\ref{act}) in the spike region to have a solution, the
scaling for the variables and the parameter $A$ will be the following
\begin{eqnarray} \label{minmax3d}
\theta_{\mathrm max} \sim \ep^{-1}, ~~\eta_{\mathrm min} \sim 1, ~~A \sim
\ep. 
\end{eqnarray}

Introducing the scaled quantities and the stretched variable
\begin{eqnarray} \label{scale3d}
\tilde\theta = \ep \theta, ~~\tilde\eta = \eta, ~~\tilde A = \ep^{-1}
A, ~~\xi = \frac{r}{\ep},
\end{eqnarray}
and retaining only the leading terms in Eqs. (\ref{act}) and
(\ref{inh}), we can write the equations describing the sharp
distributions (inner solutions) of the activator and the inhibitor in
the spike as
\begin{eqnarray}
{d^2 \tilde\theta \over d\xi^2} + {d - 1 \over \xi} {d \tilde\theta \over
d \xi} + \tilde A \tilde\theta^2 \tilde\eta - \tilde\theta & = & 0,
\label{sh3da} \\ {d^2 \tilde\eta \over d\xi^2} + {d - 1 \over \xi} {d
\tilde\eta \over d\xi} - \tilde\theta^2 \tilde\eta & = & 0,
\label{sh3db}
\end{eqnarray}
with $d = 3$ [for generality we put an arbitrary dimensionality of
space $d$ in Eqs. (\ref{sh3da}) and (\ref{sh3db})]. The boundary
conditions are neutral at $\xi = 0$, and zero for $\tilde\theta$ at
infinity. The precise boundary condition for $\tilde\eta$ at infinity
which ensures the proper matching between the sharp and the smooth
distributions has to be specified. To do this, we note that, according
to Eq. (\ref{sm3d}), to order $\ep$ we have $\eta = 1$ at $r \gg \ep$
(or $\xi \gg 1$). This means that the boundary condition for
$\tilde\eta$ in Eq. (\ref{sh3db}) must be taken to be
$\tilde\eta(\infty) = 1$.

It is convenient to perform the following change of variables
\begin{eqnarray} \label{new3d}
\tilde\theta = \frac{\tilde A \bar\theta}{\xi}, ~~\tilde\eta =
\bar\eta_s + \frac{\bar\eta - \bar\theta}{\xi},
\end{eqnarray}
where we define $\bar\eta(0) = \bar\eta_\xi(0) = 0$. Obviously,
$\bar\eta_s$ must satisfy $0 < \bar\eta_s < 1$. In these variables, we
can write Eq. (\ref{sh3da}) in the form of the nonlinear eigenvalue
problem
\begin{eqnarray}
\left[ - {d^2 \over d \xi^2} + V(\xi) \right] \bar\theta = \lambda
\bar\theta, \\ V(\xi) = - {\tilde A^2 \bar\theta \over \xi} \left(
\bar\eta_s + {\bar\eta - \bar\theta \over \xi} \right),
\end{eqnarray}
and Eq. (\ref{sh3db}) as
\begin{eqnarray} \label{sh3deta}
\bar\eta_{\xi\xi} = |\lambda| \bar\theta.
\end{eqnarray}
The problem now has a one-dimensional form similar to the one in
Sec. \ref{stat:1d}, but is defined for $\xi > 0$, with zero boundary
condition for $\bar\theta$ at $\xi = 0$. As in the one-dimensional
case, the solution that corresponds to the AS must be the lowest bound
state and have $\lambda = -1$. The latter is achieved by adjusting the
value of $\bar\eta_s$. Also, according to the definition of
$\bar\eta$, the matching condition $\tilde\eta(\infty) = 1$
corresponds to the boundary condition $\bar\eta_\xi(\infty) = 1 -
\bar\eta_s$ for Eq. (\ref{sh3deta}). Integrating Eq. (\ref{sh3deta})
with $\lambda = -1$ over $\xi$, we transform it into an integral
condition
\begin{eqnarray} \label{match3d}
\int_0^\infty \bar\theta d \xi = 1 - \bar\eta_s.
\end{eqnarray}
This condition fixes the normalization of $\bar\theta$. 

It is possible to show that the nonlinear eigenvalue problem possesses
a continuous symmetry generated by
\begin{eqnarray}
{d \xi \over d b} & = & - \xi, \nonumber \\ {d \lambda \over d b} & =
& 2 \lambda, \nonumber \\ {d \bar\eta_s \over d b} & = & 2 \bar\eta_s
( 1 - \bar\eta_s), \\ {d \bar\theta \over d b} & = & \bar\theta ( 1 -
2 \bar\eta_s), \nonumber \\ {d \bar\eta \over d b} & = & \bar\eta ( 1
- 2 \bar\eta_s), \nonumber \\ {d \tilde A \over d b} & = & - \tilde A
(1 - 2 \bar\eta_s). \nonumber
\end{eqnarray}
From these equations one can see that if there is a solution of the
nonlinear eigenvalue problem with certain $\bar\eta_s$, $\lambda$, and
$A$, there is also a solution with
\begin{eqnarray}
\lambda' = \lambda {\bar\eta_s' ( 1 - \bar\eta_s) \over \bar\eta_s ( 1
- \bar\eta_s') }, ~~~\tilde A' = \tilde A \left[ {\bar\eta_s ( 1 -
\bar\eta_s) \over \bar\eta_s' ( 1 - \bar\eta_s') } \right]^{1/2},
\end{eqnarray}
with $\bar\eta_s'$ arbitrary. Since $\lambda'(\bar\eta_s')$ is a
monotone function of $\bar\eta_s'$ that goes from 0 to infinity as
$\bar\eta_s'$ changes from 0 to 1, for any $\bar\eta_s$ it is always
possible to choose a unique value of $\bar\eta_s'$ for which $\lambda'
= -1$. So, as in the one-dimensional case, there is a one-to-one
correspondence between the solutions of the nonlinear eigenvalue
problem with arbitrary $\lambda$ and the sharp distributions.

The potential $V$ can be written as a sum of two parts: $V = V_0 +
V_1$, where
\begin{eqnarray} \label{v03d}
V_0 = - {\tilde A^2 \bar\theta \bar\eta_s \over \xi}, ~~~ V_1 = -
{\tilde A^2 \bar\theta (\bar\eta - \bar\theta) \over \xi^2}.
\end{eqnarray}
The qualitative form of these potentials for sufficiently small values
of $\tilde A$ coincides with the one shown in Fig. \ref{v1d} for $\xi
> 0$. According to Eqs. (\ref{sh3deta}) and (\ref{match3d}), at these
values of $\tilde A$ we have the following estimates for $V_0$ and
$V_1$ when $\bar\eta_s$ is close to either 1 or 0
\begin{eqnarray} \label{estv3d}
V_0 \sim - \bar\eta_s (1 - \bar\eta_s), ~~~V_1 \sim - (1 -
\bar\eta_s)^2.
\end{eqnarray}
In writing these estimates, we used the fact that the characteristic
length scale of the variation of $\bar\theta$ and $\bar\eta$ is 1.

\subsubsection{Case $A \sim \ep$: autosoliton collapse} \label{stat:ss4}

For $\tilde A \sim 1$, one can analyze the solutions of the nonlinear
eigenvalue problem in the following way. First of all, for
sufficiently small values of $\tilde A$ the potential $V$ will be so
shallow that there will be no bound states in the eigenvalue problem
at all. When the value of $\tilde A$ is increased, at some $\tilde A =
\tilde A_0$ the potential $V$ will become capable to localize a state
at $\bar\eta_s \sim \frac{1}{2}$ [see Eq. (\ref{estv3d})]. When the
value of $\tilde A$ is further increased, the minimum value of
$\lambda = \lambda_{\mathrm min}$ will decrease, so that at some
$\tilde A = \tilde A_b$ we will have $\lambda_{\mathrm min} = -1$ (see
Fig. \ref{lam3d} in which the numerical solution of the nonlinear
eigenvalue problem for several values of $\tilde A$ is shown).
\begin{figure}
\centerline{\psfig{figure=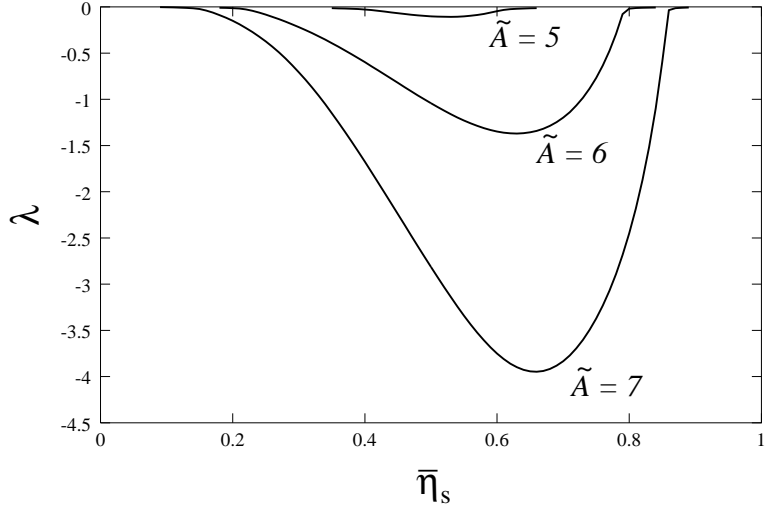,width=10cm}}
\begin{center}
\caption{Dependence $\lambda(\bar\eta_s)$ for the three-dimensional
radially-symmetric AS. Results of the numerical solution of the
nonlinear eigenvalue problem for different values of $\tilde A$.}
\label{lam3d}
\end{center}
\end{figure}
For $\tilde A > \tilde A_b$ we will have $\lambda_{\mathrm min} < -1$,
so there will be two values of $\bar\eta_s$ at which $\lambda = -1$
(Fig. \ref{lam3d}). Therefore, the solutions of the nonlinear
eigenvalue problem with these values of $\bar\eta_s$ will correspond
to the sharp distributions we are looking for. Thus, the solution in
the form of the three-dimensional static spike AS exists only for
$\tilde A > \tilde A_b$. From the numerical solution of
Eqs. (\ref{sh3da}) and (\ref{sh3db}) we obtain that in this case
$\tilde A_b = 5.8$.

It is not difficult to see that the solution corresponding to the
largest value of $\bar\eta_s$ will be unstable. Indeed, if the value
of $\bar\eta_s$ is decreased, we will have an increase in $V_1$ and a
decrease in $V_0$. Since $V_1$ can localize $\bar\theta$ easier than
$V_0$, a small decrease of $\bar\eta_s$ will produce such a deviation
of $\bar\theta$ that will [through Eqs. (\ref{sh3deta}) and
(\ref{v03d})] further distort the potential $V$ in the same manner. In
other words, if we construct an iterative map that takes $V$,
calculates the solution $\bar\theta$ of the eigenvalue problem, and
then generates the new $V$ by solving for $\bar\eta$, it will take us
from the unstable solution with greater $\bar\eta_s$ to the stable
solution with lower $\bar\eta_s$ if $\bar\eta_s$ is decreased at the
start, or to the trivial solution $\tilde\theta = 0$ and $\tilde\eta =
1$, which is obviously stable, if the value of $\bar\eta_s$ is
increased. Thus, the solution corresponding to the stable
radially-symmetric static AS should be unique.

\subsubsection{Case $A \gg \ep$: annulus} \label{stat:ss5}

The numerical solution of Eqs. (\ref{sh3da}) and (\ref{sh3db}) shows that
for $\tilde A$ not far from $\tilde A_b$ the distribution of $\theta$
in the AS has the form of a spike centered at zero. As the value of
$\tilde A$ increases, the shape of the AS changes. To see how the AS
behaves as the value of $\tilde A$ is increased, a special treatment
of the case $\tilde A \gg 1$ is needed. When $\tilde A$ becomes large,
the potential $V$ contains a large factor of $\tilde A$. This factor
can be compensated by choosing, for example, $1 - \bar\eta_s \sim
\tilde A^{-2} \ll 1$, which will correspond to the unstable solution
with larger $\bar\eta_s$. Alternatively, one could have the potential
$V$ shifted along $\xi$, so that it is centered around $\xi = R \gg
1$. In that case the main contribution to $V$ will be given by $V_0
\sim - \tilde A^2 / R$ for $\bar\eta_s$ not close to either 0 or
1. Since $\bar\theta$ exponentially decays at large distances from
$\xi = R$, for $R \gg 1$, the boundary conditions at $\xi = 0$ become
inessential and can be moved to minus infinity. In this case, if one
neglects the terms of order $1/R$ in the potential, one can solve the
nonlinear eigenvalue problem exactly. The solution will be given by
\begin{eqnarray} \label{ann3d}
\bar\theta = {1 - \bar\eta_s \over 4} \cosh^{-2} \left( {\xi - R \over
2} \right),
\end{eqnarray}
with 
\begin{eqnarray} \label{R3d}
R = \frac{A^2 \bar\eta_s (1 - \bar\eta_s)}{6}.
\end{eqnarray}

So far, we obtained a continuous family of solutions parameterized by
$\bar\eta_s$ in the case $\tilde A \gg 1$. This result may seem
surprising, since for $\tilde A \sim 1$ we showed that there should be
only one stable solution to the nonlinear eigenvalue problem. However,
as we will show below, all the solutions found in the preceding
paragraph except for a single solution are in fact structurally
unstable, so the stable solution is indeed unique for $\tilde A \gg
1$.

The reason for the structural instability of the solutions with $R \gg
1$ is that in the limit $R \rightarrow \infty$ the problem possesses
translational symmetry. As a result, for $R \gg 1$ there is a
degenerate mode corresponding to the translations of the spike as a
whole along $\xi$. One should therefore study the stability of the
solution with respect to that mode.

To analyze the  stability of the solutions with $R \gg 1$,
we need to calculate the $1/R$ correction to the solution obtained
above. Let us write Eqs. (\ref{sh3da}) and (\ref{sh3db}) in the form
that is valid to order $1/R$
\begin{eqnarray}
\tilde\theta_{\xi\xi} + c \tilde\theta_\xi + {2 \tilde\theta_\xi \over
R} + \tilde A \tilde \theta^2 \tilde \eta - \tilde\theta & = & 0,
\label{sh3dc} \\
\tilde\eta_{\xi\xi} - \bar\eta_s \tilde\theta^2 & = & 0, \label{sh3dd}
\end{eqnarray}
where for the solution sought we must have $c = 0$. We wrote
Eq. (\ref{sh3dc}) so that it reminisces an equation describing the
solution traveling with constant speed $c$.

We can use the expression for $\bar\theta$ obtained above to calculate
the variation of $\tilde\eta$ in the spike. Integrating
Eq. (\ref{sh3dd}) with $\tilde\theta$ from Eq. (\ref{ann3d}) and using
the boundary condition $\tilde\eta_\xi(-\infty) = 0$, we obtain
\begin{eqnarray} \label{etaann3d}
\tilde\eta = \bar\eta_s + \frac{1 - \bar\eta_s}{2 R} && \left( 2 \ln
\cosh {\xi - R \over 2} \right. \nonumber \\ && \left. + \frac{1}{2}
\tanh^2 {\xi - R \over 2} + \xi - R \right).
\end{eqnarray}
Substituting this expression for $\tilde\eta$ into Eq. (\ref{sh3dc}),
multiplying it by $\tilde\theta_\xi$, and integrating over $\xi$, for
$\bar\eta_s < 1/2$ we obtain the following expression for $c$
\begin{eqnarray} \label{cc3d}
c = \frac{1}{\tilde A^2} \left[ - {2 \tilde A^2 \over R} + {12 \over 1
- {12 R \over \tilde A^2} - \sqrt{1 - \frac{24 R}{\tilde A^2}} }
\right].
\end{eqnarray}
In writing this equation we assumed the relationship between
$\bar\eta_s$ and $R$ from Eq. (\ref{R3d}).

The analysis of Eq. (\ref{cc3d}) shows that we have $c = 0$ only for
$R = R^*$ (and $\bar\eta_s = \bar\eta_s^*$), where
\begin{eqnarray} \label{fp3d}
R^* = {\tilde A^2 \over 27}, ~~~\bar\eta_s^* = \frac{1}{3}.
\end{eqnarray}
Thus, the corrections of order $1/R$ destroy the solutions with $R
\not= R^*$.

Consider the flow generated by the equation $dR/dt = c(R)$. The
behavior of this flow near the fixed point $R = R^*$ determines the
stability of the solution. According to Eq. (\ref{cc3d}), we have $c >
0$ for $R < R^*$ and $c < 0$ for $R > R^*$, so the flow is into the
fixed point. Therefore, the solution with $R = R^*$ is stable.  One
can also write an equation similar to Eq. (\ref{cc3d}) in the case
$\bar\eta_s > 1/2$, which corresponds to another branch of the
solutions obtained above. The analysis of this equation shows that for
those solutions $c < 0$ for all $\bar\eta_s$, so the flow transforms
the solution into the trivial solution $\tilde\theta = 0$. Thus, the
solution that corresponds to the radially-symmetric static AS is
indeed unique even for $\tilde A \gg 1$.

\subsubsection{Comparison of the two cases} \label{stat:ss6}

From the arguments given above it is clear that when the value of
$\tilde A$ is increased from $\tilde A_b \sim 1$ to $\tilde A \gg 1$,
the solution in the form of the AS should gradually transform from the
spike to the annulus of large radius $R \sim \tilde A^2$. This is what
we see from the numerical solution of Eqs. (\ref{sh3da}) and
(\ref{sh3db}). Figure \ref{ett3d} shows the dependence $\tilde\eta(0)$
and $\tilde\theta(0)$ as a function of $\tilde{A}$ in the
radially-symmetric AS obtained from this solution.
\begin{figure}
\centerline{\psfig{figure=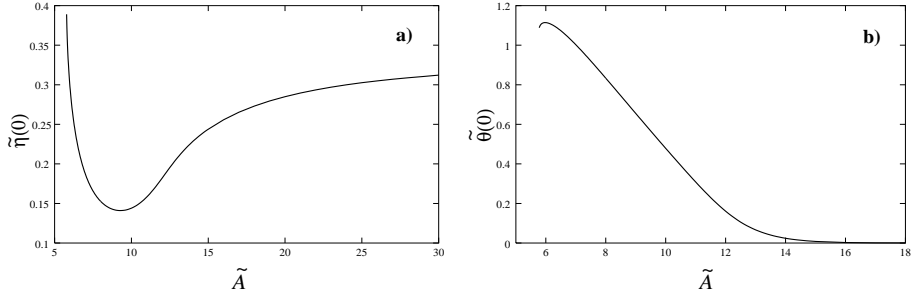,angle=-90,width=12cm}}
\begin{center}
\caption{The values of $\tilde\eta(0)$ (a) and $\tilde\theta(0)$ (b)
as functions of $\tilde{A}$ for the three-dimensional
radially-symmetric AS obtained from the numerical solution of
Eqs. (\ref{sh3da}) and (\ref{sh3db}).}  \label{ett3d}
\end{center}
\end{figure}
From this figure one can see that $\tilde\eta(0)$ indeed approaches
$\bar\eta_s^* = 1/3$ as $\tilde A$ increases. The dependence of the
radius of the annulus was also found to be in good agreement with
Eq. (\ref{fp3d}) for large values of $\tilde A$.

Figure \ref{sh3df} shows the distributions of $\tilde\theta$ and
$\tilde\eta$ in the radially-symmetric three-dimensional AS for a
particular value of $\tilde{A}$ which is intermediate between the
spike and the annulus. 
\begin{figure}
\centerline{\psfig{figure=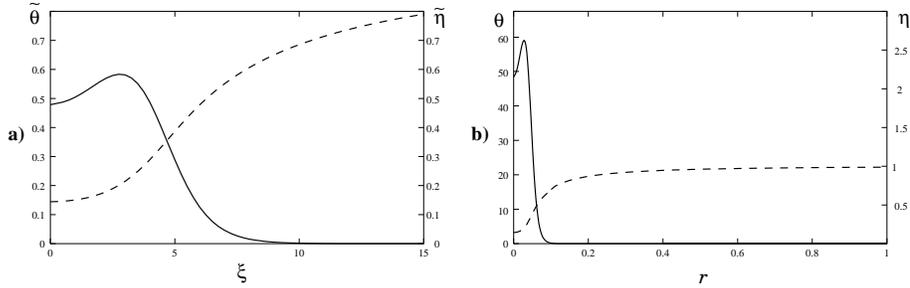,angle=-90,width=12cm}}
\begin{center}
\caption{(a) Sharp distributions $\tilde\theta(\xi)$ (solid line) and
$\tilde\eta(\xi)$ (dashed line) in a three-dimensional
radially-symmetric static AS, obtained from the numerical solution of
Eqs. (\ref{sh3da}) and (\ref{sh3db}) with $\tilde A = 10$; (b)
Construction of the full solution for $\theta$ (solid line) and $\eta$
(dashed line) at $A = 0.1$ and $\ep = 0.01$. } \label{sh3df}
\end{center}
\end{figure}
This figure also shows the entire solution in the form of an AS
obtained by matching the sharp and the smooth distributions for the
particular values of $A$ and $\ep$.

So far we studied the situations when $R \gg 1$ but still smaller than
the inhibitor length, which in these units is of order
$\ep^{-1}$. When $R$ reaches this value, Eqs. (\ref{sh3da}) and
(\ref{sh3db}) will no longer be justified for the description of the
distributions of $\theta$ and $\eta$ in the annulus. In the unscaled
variables this will happen when $A \sim \ep^{1/2} \sim A_b^{(1)}$ [see
Eq. (\ref{fp3d})], where $A_b^{(1)}$ is the minimum value of $A$ at
which the one-dimensional static AS exists (see
Sec. \ref{stat:ss1}). At these values of $A$ the radially-symmetric AS
can be effectively considered as a one-dimensional AS, so when $A$
reaches some value $A_d \sim \ep^{1/2}$, the solution in the form of
an annulus will transform into a quasi one-dimensional AS of infinite
radius. Note, however, that this bifurcation point is essentially
different from the bifurcation at $A = A_d$ of the one-dimensional AS
(Sec. \ref{stat:ss2}) in that it does not involve local breakdown or
splitting of the AS.

\subsection{Two-dimensional static spike autosoliton} \label{stat:2d}

Let us now turn to the two-dimensional case. The analysis of the
two-dimensional radially-symmetric static AS turns out to be analogous
to that of the three-dimensional AS, so we will not go into much
detail here, but will only give the main results.

As in the case of the three-dimensional AS, away from the spike
$\theta = 0$ and the distribution of $\eta$ is given by
\begin{eqnarray} \label{sm2d}
\eta(r) = 1 - a K_0 (r),
\end{eqnarray}
where $K_0$ is the modified Bessel function, $r$ is the radial
coordinate, and $a$ is a certain constant. The value of $a$ is
determined by integrating Eqs. (\ref{act}) and (\ref{inh}) with the
time derivatives set to zero over the spike region. In two dimensions
$a$ is given by
\begin{eqnarray} \label{c2d}
a = \frac{1}{A} \int r \theta(r) dr.
\end{eqnarray}
According to Eq. (\ref{sm2d}), when $r$ becomes of order $\ep$,
we have 
\begin{eqnarray} \label{sm2dest}
\eta \simeq 1 - a \ln \ep^{-1} + a \ln (r/\ep). 
\end{eqnarray}
In order for $\eta$ to remain positive, we must have $a \sim 1/\ln
\ep^{-1}$. Then, on the length scales of $\ep$ the variation of $\eta$
in the spike will be of order $a \ll 1$. In the same way as in the
case of the three-dimensional AS, this results in the following
scaling for the main parameters of the AS
\begin{eqnarray} \label{scale2d}
\theta_{\mathrm max} \sim \ep^{-1}, ~~~\eta_{\mathrm min} \sim {1
\over \ln \ep^{-1}}, ~~~A \sim \ep \ln \ep^{-1}.
\end{eqnarray}

In the scaled variables
\begin{eqnarray} \label{tilde2d}
\tilde\theta = \ep \theta, ~~\tilde \eta = \eta \ln \ep^{-1}, ~~\tilde
A = {A \over \ep \ln \ep^{-1} }, ~~~\xi = {r \over \ep},
\end{eqnarray}
the equations for the sharp distributions in the spike will take the
form of Eqs. (\ref{sh3da}) and (\ref{sh3db}) with $d = 2$. Since the
variation of $\eta$ in the spike is of order $1/\ln \ep^{-1}$,
according to Eq. (\ref{sm2dest}) we must have $a = 1/\ln \ep^{-1}$ in
the limit $\ep \rightarrow 0$. This gives us the condition for
matching the sharp and the smooth distributions.  Let us introduce the
variables
\begin{eqnarray}
\bar\theta = \tilde A^{-1} \tilde\theta, ~~~ \bar\eta = \tilde\eta +
\tilde A^{-1} \tilde\theta - \bar\eta_s,
\end{eqnarray}
where $\bar\eta_s$ is a constant that is chosen so that $\bar\eta(0) =
0$. Then, the matching condition can be written in the integral form
as
\begin{eqnarray} \label{match2d}
\int_0^\infty \bar\theta \xi d \xi = |\lambda|^{-1},
\end{eqnarray}
where $\lambda$ is a constant that must be equal to --1. Also, in
terms of the new variables the equation for the sharp distribution of
$\bar\eta$ can be written as
\begin{eqnarray} \label{sh2deta}
{1 \over \xi} {d \over d \xi} \left( \xi { d \bar\eta \over d \xi}
\right) = |\lambda| \bar\theta
\end{eqnarray}
with $\bar\eta_\xi (0) = 0$.

As with the three-dimensional AS, in two dimensions the problem of
finding the sharp distributions can be written as a nonlinear
eigenvalue problem
\begin{eqnarray}
- {1 \over \xi} {d \over d \xi} \left( \xi {d \bar\theta \over d \xi}
\right) + V(\xi) \bar\theta = \lambda \bar\theta,
\label{neigen2da} \\ V(\xi) = - \tilde A^2 \bar\theta ( \bar\eta_s +
\bar\eta - \bar\theta), \label{neigen2db}
\end{eqnarray}
with the potential $V$ that can be separated as $V = V_0 + V_1$, where
\begin{eqnarray} \label{v02d}
V_0 = - \tilde A^2 \bar\theta \bar\eta_s, ~~~ V_1 = - \tilde A^2
\bar\theta (\bar\eta - \bar\theta).
\end{eqnarray}
These potentials have the form shown in Fig. \ref{v1d} for $\xi >
0$. The lowest bound state with $\lambda = -1$ will give us the
solution we are looking for. This condition is achieved by adjusting
the value of $\bar\eta_s$. The nonlinear eigenvalue problem is
invariant with respect to the transformation given by
Eq. (\ref{sym1d}).

When $\tilde A \ll 1$, the potential $V$ acquires a small factor (as
in the case of the one-dimensional AS), which can be compensated only
by choosing $\bar\eta_s \sim \tilde A^{-2} \gg 1$. Therefore, the
potential $V$ will be dominated by $V_0$ which can always localize a
bound state with $\lambda = -1$. Notice, however, that in order for
the approximations made to derive the equations for the sharp
distributions to remain valid, we must have $\bar\eta_s \lesssim \ln
\ep^{-1}$, so in fact this argument is valid only down to $\tilde A
\sim (\ln \ep^{-1})^{-1/2}$. It is easy to show that, similarly to the
one- and three-dimensional cases, the solution in the form of the
two-dimensional AS will disappear at $A < A_b \sim \ep (\ln
\ep^{-1})^{1/2}$.

At $\tilde A$ sufficiently small, the AS looks like a spike with
the maximum value of $\tilde\theta$ centered at $\xi = 0$. As in the
case of the three-dimensional AS, when the value of $A$ is increased,
the AS gradually transforms into an annulus of radius $R$, which grows
with $\tilde A$. According to Eq. (\ref{match2d}), when $R$ increases,
we have $\bar\theta \sim R^{-1} \ll 1$ and $\bar\eta \sim \bar\theta$,
so the potential $V_0 \sim A^2 / R$ starts to dominate. The analysis
similar to that for the three-dimensional AS shows that for $\tilde A
\gg 1$ the parameters of the AS are given by
\begin{eqnarray} \label{r2d}
R^* = {\tilde A^2 \over 6}, ~~~\bar\eta_s^* = 1.
\end{eqnarray}
These results are also supported by the numerical solution of the
equations for the sharp distributions. The dependences of
$\tilde\eta(0)$ and $\tilde\theta(0)$ on $\tilde{A}$ are presented in
Fig. \ref{ett2d}.
\begin{figure}
\centerline{\psfig{figure=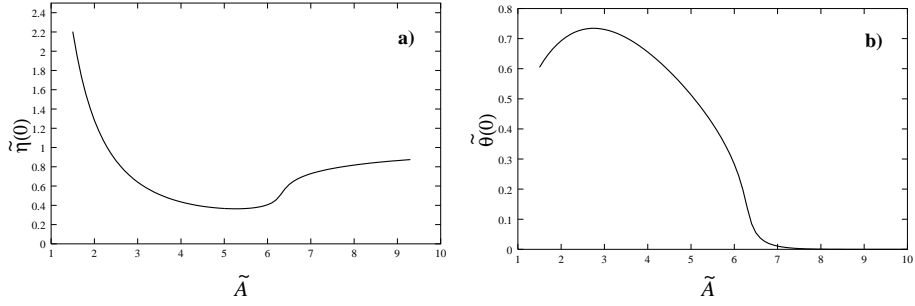,angle=-90,width=12cm}}
\begin{center}
\caption{The values of $\tilde\eta(0)$ (a) and $\tilde\theta(0)$ (b)
as functions of $\tilde{A}$ for the two-dimensional radially-symmetric
AS obtained from the numerical solution of Eqs. (\ref{sh3da}) and
(\ref{sh3db}).}
\label{ett2d}
\end{center}
\end{figure}
The solution of Eqs. (\ref{sh3da}) and (\ref{sh3db}) in the form of
the two-dimensional radially-symmetric static spike AS at a particular
value of $\tilde{A}$ is also presented in Fig. \ref{sh2df}.
\begin{figure}
\centerline{\psfig{figure=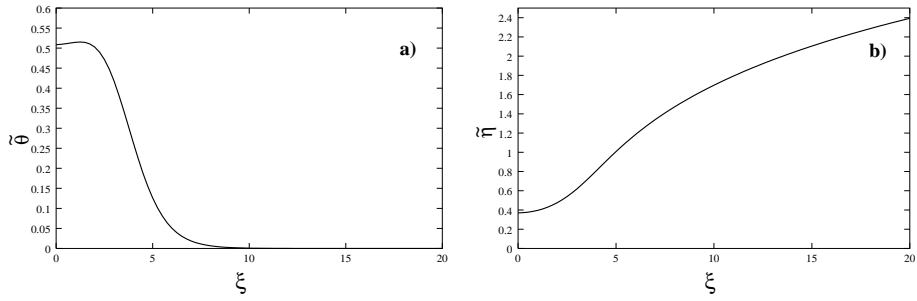,angle=-90,width=12cm}}
\begin{center}
\caption{Sharp distributions $\tilde\theta(\xi)$ (a) and
$\tilde\eta(\xi)$ (b) in a two-dimensional radially-symmetric static
AS, obtained from the numerical solution of Eqs. (\ref{sh3da}) and
(\ref{sh3db}) with $\tilde A = 5$. }
\label{sh2df}
\end{center}
\end{figure}
Of course, when $R \sim \ep^{-1}$, the approximations made in the
derivation of the equations for the sharp distributions are no longer
valid, so, as in the case of the three-dimensional radially-symmetric
AS, the two-dimensional radially-symmetric AS of radius $R^*$ will
transform into a quasi one-dimensional AS of infinite
radius. According to Eq. (\ref{r2d}), this will happen when $A > A_d
\sim \ep^{1/2} \ln \ep^{-1}$.

Finally, we note that the small parameter of the singular perturbation
expansion in the two-dimensional case turned out to be $1/ \ln
\ep^{-1}$, so one should expect it to give a good quantitative
agreement with the actual solutions only for extremely small values of
$\ep$. Nevertheless, the leading scaling given by Eq. (\ref{scale2d})
(up to the logarithmic terms) should be in good agreement even for not
very small $\ep$. Also, it is not difficult to modify the theory in
such a way that it uses $\ep$ as a small parameter in the
expansion. In that case the sharp distributions will contain a weak
logarithmic dependence on $\ep$. 

\section{Traveling spike autosolitons} \label{s:trav}

Up to now, we only considered the solutions of Eqs. (\ref{act0}) and
(\ref{inh0}) that correspond to the static ASs. At the same time, when
$\al = \tau_\theta / \tau_\eta \ll 1$ and $\ep$ is sufficiently large,
there exist solutions that propagate with a constant speed without
decay --- the traveling ASs \cite{KO89,KO90,KO94}, so now we are going
to look for these solutions. Throughout this section we assume that
$\al \ll 1$.

As we will show below, in the Gray-Scott model the traveling ASs are
realized for sufficiently small $\al$ and have the shape of narrow
spikes of high amplitude which strongly depends on $\al$. To analyze
the traveling spike ASs, it is convenient to measure length and time
in the units of $l$ and $\tau_\theta$, respectively. Then, the
equations describing the AS traveling with the constant speed $c$
along the $x$-axis will take the form
\begin{eqnarray}
{d^2 \theta \over d z^2} + c {d \theta \over d z} + A \theta^2 \eta -
\theta = 0, \label{acttrav} \\ \ep^{-2} {d^2 \eta \over d z^2} +
\al^{-1} c {d \eta \over d z} - \theta^2 \eta + 1 - \eta = 0,
\label{inhtrav}
\end{eqnarray}
where we introduced a self-similar variable $z = x - ct$. The solution
with $c > 0$ travels from left to right. The distributions of $\theta$
and $\eta$ should go to the homogeneous state $\theta_h = 0$ and
$\eta_h = 1$ [Eq. (\ref{hom})] for $z \rightarrow \pm \infty$.

\subsection{Non-diffusive inhibitor: $\ep \gg \al^{1/2}$} \label{trav:s1}

There are two qualitatively different types of traveling spike ASs in
the Gray-Scott model. First we consider the ultrafast traveling spike
AS, which is realized when the inhibitor does not diffuse, that is,
when $L = 0$ (or $\ep = \infty$). Such an AS was recently discovered
by us in a similar reaction-diffusion model (the Brusselator)
\cite{Mur95}. A remarkable property of this AS is that it has the
shape of a narrow spike whose velocity $c \sim A \al^{-1/2}$ is much
higher than the characteristic speed $l/\tau_\theta$ (which in these
units is of order 1) determined by the physical parameters of the
problem, and whose amplitude goes to infinity as $\al \rightarrow 0$.

\subsubsection{Case $\al^{1/2} \ll A \ll \al^{-1/2}$: ultrafast
traveling spike autosoliton} \label{trav:ss1} 

If we assume that $\theta \gg 1$, we can drop the last term from
Eq. (\ref{acttrav}) and neglect the last two terms in
Eq. (\ref{inhtrav}) (with the term involving the second derivative of
$\eta$ dropped in the limit of large $\ep$) in the front of the spike
where $\eta \sim \eta_h = 1$. If we then multiply the latter equation
by $A$ and add it to the former equation, we will get
\begin{eqnarray}
{d^2 \theta \over d z^2} + c {d \theta \over d z} + A \al^{-1} c {d
\eta \over d z} = 0.
\end{eqnarray}
This equation can be straightforwardly integrated. If we introduce the
variables
\begin{eqnarray} \label{scaletrav}
\tilde\theta = \al A^{-1} \theta, ~~~ \tilde c = \al^{1/2} A^{-1} c,
~~~ \xi = A \al^{-1/2} z,
\end{eqnarray}
we can write the solution for $\eta$ as
\begin{eqnarray} \label{supereta}
\eta = 1 - \tilde\theta - {1 \over \tilde c} {d \tilde\theta \over d
\xi},
\end{eqnarray}
where we took into account the boundary condition
$\tilde\theta(+\infty) = \tilde\theta_\xi(+\infty) = 0$,
$\eta(+\infty) = 1$.  Substituting this expression back to
Eq. (\ref{acttrav}) (with the last term dropped), we arrive at the
following equation
\begin{eqnarray} \label{supertheta}
{d^2 \tilde\theta \over d \xi^2} + {d \tilde\theta \over d \xi} \left(
\tilde c - {1 \over \tilde c} \theta^2 \right) + \tilde\theta^2 -
\tilde\theta^3 = 0.
\end{eqnarray}
One can see that in Eq. (\ref{supertheta}) all the $\al$- and
$A$-dependence is absent, so Eq. (\ref{scaletrav}) (with all the tilde
quantities of order 1) in fact determines the scaling of the main
parameters of the traveling spike AS for $\al \ll 1$. As was expected,
the AS will have the speed which diverges as $\al \rightarrow
0$. Also, note that the width of the front of the AS, which is of
order $\al^{1/2} A^{-1}$ goes to zero as $\al \rightarrow 0$. Thus,
the distributions of $\theta$ and $\eta$ in the front of the ultrafast
traveling spike AS will be given by the ``supersharp'' distributions
(in the sense that their characteristic length scale is much smaller
than 1) described by Eqs. (\ref{supereta}) and (\ref{supertheta}).

Let us take a closer look at Eq. (\ref{supertheta}). This equation has
the form of an equation of motion for a particle with the coordinate
$\tilde\theta$ and time $\xi$ in the potential $U = {\tilde\theta^3
\over 3} - {\tilde\theta^4 \over 4}$, with the nonlinear friction with
the coefficient $\tilde c - {1 \over \tilde c} \tilde\theta^2$. Since
the derivative of the friction coefficient is positive for all $\tilde
c$, the friction increases as $\tilde c$ grows, so there are no
special features associated with its nonlinearity. For $\tilde\theta >
0$ the potential $U$ has a maximum at $\tilde\theta = 1$ and a minimum
at an inflection point $\tilde\theta = 0$. The supersharp distribution
of $\theta$ will therefore be the heteroclinic trajectory going from
$\tilde\theta = 1$ to $\tilde\theta = 0$.

It is clear that if the friction is not strong enough, the particle
starting from $\tilde\theta = 1$ will miss the point $\tilde\theta =
0$ and go to minus infinity, so we must have $\tilde c \geq \tilde
c^*$, where $\tilde c^*$ is some positive constant of order 1. On the
other hand, it is clear that when $\tilde c > \tilde c^*$, the
particle will always get from $\tilde\theta = 1$ to $\tilde\theta =
0$, so in fact there is a continuous family of such solutions. Thus,
we have a multiplicity of the front solutions and, therefore, a
selection problem \cite{Cross}. To answer the question about the front
selection, we need to consider higher-order corrections to the
solution of Eq. (\ref{supertheta}) coming from Eqs. (\ref{acttrav})
and (\ref{inhtrav}). According to these equations, for small
$\tilde{\theta}$ the next order correction will amount to adding the
term $ - \al A^{-2} \tilde\theta$ to Eq. (\ref{supertheta}). In this
situation the potential $U$ will actually have a maximum at
$\tilde\theta = 0$ and a minimum at $\tilde\theta_{\mathrm min} \sim
\al A^{-2}$, so only the trajectory with the minimum velocity $\tilde
c$ will reach $\tilde\theta = 0$, whereas all other trajectories will
be stuck at $\tilde\theta = \tilde\theta_{\mathrm min}$.  Thus, we can
conclude that in the limit $\al \rightarrow 0$ the selected front
solution in our problem has the velocity $\tilde c = \tilde c^*$. The
numerical solution of Eq. (\ref{supertheta}) shows that the value of
$\tilde c^*$ is $\tilde c^* = 0.86$. The numerical simulations of
Eqs. (\ref{act0}) and (\ref{inh0}) confirm these conclusions. The main
parameters of the traveling spike AS, therefore, are
\begin{eqnarray} \label{ultrac}
\theta_{\mathrm max} = A \al^{-1}, ~~~~c = 0.86 \times A \al^{-1/2}.
\end{eqnarray} 
Note that the numerical solution of Eq. (\ref{supertheta}) in the form
of the supersharp front differs from $\tilde\theta_{\mathrm ssh} =
\frac{1}{2}[1 -\tanh (0.50 \xi)]$ by less than 1\%. Also note that the
results given by Eq. (\ref{ultrac}) precisely coincide with those
obtained by us for the Brusselator \cite{Mur95}. This is due to the
fact that the supersharp distributions in these two models are
described by the same equations.

In the back of the supersharp front the value of $\tilde\theta$ goes
exponentially to 1, and $\eta$ goes exponentially to 0 [see
Eq. (\ref{supereta})]. Note, however, that in writing the equations
describing the supersharp distributions we neglected the last two
terms in Eq. (\ref{inhtrav}). When the value of $\eta$ decreases, at
$\eta \sim \al^2 A^{-2}$ the term $\theta^2 \eta$ becomes of order 1,
and the equations for the supersharp distributions cease to be
valid. This will happen at a distance of order $\al^{1/2} A^{-1} \ln A
\al^{-1}$ behind the location of the supersharp front. We can
therefore call the region of this size right after the front where
$\eta$ exponentially decays to some value $\eta_{\mathrm min}$ the
secondary region of the supersharp distributions. Since the width of
this region is still much smaller than 1, we can assume that $\theta =
\theta_{\mathrm max}$ there. Then, the distribution of $\eta$ in the
secondary region of the supersharp distributions is given by
Eq. (\ref{inhtrav}) in which we should drop the last term, since $\eta
\ll 1$ there. We obtain
\begin{eqnarray}
\eta_{\mathrm ssh2} = \al^2 A^{-2} + C \e^{\xi / \tilde c},
\end{eqnarray}
where the constant $C$ should be determined by matching to the
asymptotics of the supersharp distribution of $\eta$ at $\xi
\rightarrow -\infty$ (this requires an explicit knowledge of the
solution in the supersharp region). As can be seen from this equation,
we have $\eta_{\mathrm min} = \al^2 A^{-2}$. 

As $z$ passes the secondary region of the supersharp distributions,
$\theta^2 \eta$ becomes of order 1, and therefore can be dropped from
Eq. (\ref{acttrav}). Then the activator and the inhibitor become
decoupled, so the characteristic length scale of the variation of
$\theta$ significantly increases. According to Eq. (\ref{acttrav}),
for $c \gg 1$ the characteristic length scale of the decay of $\theta$
behind the supersharp front is of order $c \sim A \al^{-1/2}$, which
is still much smaller than the length scale of the variation of $\eta$
behind the spike (the refractory region), which is of order $\al^{-1}
c$ (see below). This means that after the secondary region of the
supersharp distributions we should find the region of the sharp
distributions. According to Eq. (\ref{acttrav}) with the terms $d^2
\theta / dz^2$ and $A \theta^2 \eta$ dropped, the solution for
$\theta$ in this region will be
\begin{eqnarray} \label{shthetatrav}
\theta_{\mathrm sh} (z) = \al^{-1} A \e^{z/c},
\end{eqnarray}
where we chose the position of the supersharp front to be at $z = 0$
(with the accuracy of $\al$). This expression for $\theta_{\mathrm
sh}$ can be substituted back into Eq. (\ref{inhtrav}) to calculate
$\eta_{\mathrm sh}$. The analysis of this equation then shows that one
can neglect both $\al^{-1} c~ d \eta / d z$ and $-\eta$ in the region
of the sharp distributions, so $\eta_{\mathrm sh}$ and
$\theta_{\mathrm sh}$ are related locally. The resulting expression
for the sharp distribution of $\eta$ takes the following form
\begin{eqnarray} \label{shetatrav}
\eta_{\mathrm sh1} = \al^2 A^{-2} \e^{-2 z/c}.
\end{eqnarray}
As will be shown in the next paragraph, this equation is in fact valid
only in the part of the sharp distributions region, so we will call it
the primary sharp distribution of $\eta$. 

According to Eq. (\ref{shetatrav}), the value of $\eta$ exponentially
grows behind the region of the sharp distributions, so at some
distance of order $\al^{-1/2} A \ln \al^{-1} A^2$ one can no longer
neglect the term $\al^{-1} c ~ d \eta / dz$ in Eq. (\ref{inhtrav}). If
we take this derivative into account, we can solve
Eq. (\ref{inhtrav}), provided that $\theta$ is still given by
Eq. (\ref{shthetatrav}). The solution will have the following form
\begin{eqnarray} \label{sh2trav}
\eta_{\mathrm sh2}(z) = {\al \over 2} \Gamma \left( 0, \e^{{2 \over c}
(z - z_0)} \right) \e^{\e^{{2 \over c} (z - z_0)}}, ~~~z_0 = \frac{c}{2}
\ln 2 \al A^{-2},
\end{eqnarray}
where $\Gamma(a, x)$ is the incomplete gamma function. In writing the
last equation we matched this solution with the one from
Eq. (\ref{shetatrav}) at large $z - z_0$. We will call this
distribution of $\eta$ the secondary sharp distribution.

For yet more negative values of $z$ the distribution of $\eta$
approaches $\eta \simeq - \al c^{-1} z$ [see Eq. (\ref{sh2trav})], so
the characteristic length scale of the variation of $\eta$ becomes of
order $\al^{-1} c \gg c$. This means that we enter the refractory tail
of the AS where $\eta$ relaxes to $\eta_h$, that is, the region of the
smooth distributions. For these values of $z$ the distribution of
$\theta$ already relaxed to zero, so Eq. (\ref{inhtrav}) can be easily
solved. To do that we should recall that up to $c \ll \al^{-1} c$ the
region of the sharp distributions is located at $z = 0$, and to the
leading order in $\al$ we have $\eta(0) = 0$. This immediately gives
us the solution for $\eta$ in the region of the smooth distributions
\begin{eqnarray} \label{ultrasm}
\eta_{\mathrm sm} = 1 - \e^{\al z / c}.
\end{eqnarray}

The entire solution in the form of the ultrafast traveling spike AS is
presented in Fig. \ref{ultra}. 
\begin{figure}
\centerline{\psfig{figure=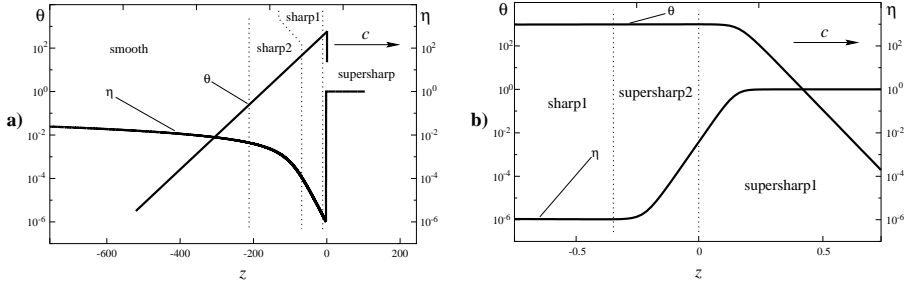,angle=-90,width=12cm}}
\begin{center}
\caption{Distributions of $\theta$ and $\eta$ in the ultrafast
traveling spike AS: (a) the back of the AS; (b) the front of the
AS. Results of the numerical solution of Eqs. (\ref{act}) and
(\ref{inh}) with $\ep = \infty$, $\al = 10^{-3}$, and $A = 1$. Length
is measured in the units of $l$. }
\label{ultra}
\end{center}
\end{figure}
This figure actually shows the result of the numerical simulations of
Eqs. (\ref{act}) and (\ref{inh}) with $\al = 10^{-3}$ and $A = 1$. One
can see an excellent agreement of this solution with the distributions
obtained above.

Thus, we introduced an asymptotic procedure for constructing the
solution in the form of the traveling spike AS in the Gray-Scott model
in the limit $\al \rightarrow 0$. This solution is considerably
different from the solutions in the form of the traveling ASs in
N-systems (see Sec. \ref{s:intro}). In N-systems the speed and the
distribution of $\theta$ in the AS front are determined only by the
equation for the activator with $\eta = \eta_h$ in the limit $\al
\rightarrow 0$, so the speed of the AS cannot exceed the value of
order 1 \cite{KO89,KO94}. The distribution of $\theta$ in such an AS
can be separated into two regions: the region of the sharp
distributions, which corresponds to the moving domain wall whose
characteristic size is of order 1, and the region of the smooth
distributions, where the distribution of $\theta$ is slaved by the
distribution of $\eta$ which varies on the length scale of $\al^{-1}$
\cite{KO94,Ross}. In other words, in the limit $\al \rightarrow 0$
there is only one boundary layer in the solution for $\theta$ (within
a single domain wall) and no singularities in the solution for
$\eta$. In contrast, in the Gray-Scott model the speed and the
amplitude of the ultrafast traveling spike AS become singular in the
limit $\al \rightarrow 0$. Moreover, there are {\em three} regions
with different behaviors for $\theta$ in the ultrafast traveling AS:
the region of the supersharp distributions, where $\theta$ varies on
the length scale of $\al^{1/2}$, the region of the sharp
distributions, in which the characteristic length scale of $\theta$ is
$\al^{-1/2}$, and the region of the smooth distributions, where
$\theta = 0$. The latter happens to be a specific property of the
Gray-Scott model, in more general models the distribution of $\theta$
is slaved by the distribution of $\eta$ in the smooth distributions
and thus has the characteristic length scale of its variation of order
$\al^{-1} c$ \cite{Mur95}. Moreover, the distribution of $\eta$ can be
separated into {\em five} regions where the asymptotic behavior of
$\eta$ is different. In other words, the solution in the form of the
ultrafast traveling spike AS contains four boundary layers in the
limit $\al \rightarrow 0$.

\subsubsection{Case $A \sim \al^{1/2}$: disappearance of solution} 
\label{trav:ss2} 

According to the procedure presented above, the main parameters of the
AS, such as the amplitude and the velocity are determined solely by
the supersharp distributions of $\theta$ and $\eta$. However,
according to Eq. (\ref{ultrac}), when $A$ becomes of order
$\al^{1/2}$, the velocity of the AS becomes of order 1, so the
separation of the distributions of $\theta$ and $\eta$ into the
supersharp and the sharp distributions in the spike becomes
invalid. For these values of $A$ the treatment of the spike region has
to be modified. Note that according to Eq. (\ref{ultrac}) we still
have $\theta_{\mathrm max} \sim \al^{-1/2} \gg 1$ for $A \sim
\al^{1/2}$. Let us introduce the following variables
\begin{eqnarray}
\tilde\theta = \al^{1/2} \theta, ~~~\tilde\eta = \eta, ~~~ \tilde A =
\al^{-1/2} A.
\end{eqnarray}
In these variables we can write Eqs. (\ref{acttrav}) and
(\ref{inhtrav}) as
\begin{eqnarray}
\tilde\theta_{zz} + c \tilde\theta_z + \tilde A \tilde\theta^2
\tilde\eta - \tilde\theta = 0, \label{travba} \\ c \tilde\eta_z -
\tilde\theta^2 \tilde\eta = 0, \label{travbb}
\end{eqnarray}
where we neglected the last two terms in Eq. (\ref{inhtrav}). These
equations with the boundary conditions $\tilde\theta(\pm \infty) = 0$
and $\tilde\eta(+\infty) = 1$ can be solved numerically. Figure
\ref{abt}(a) shows the solution of these equations for a particular
value of $\tilde A$. 
\begin{figure}
\centerline{\psfig{figure=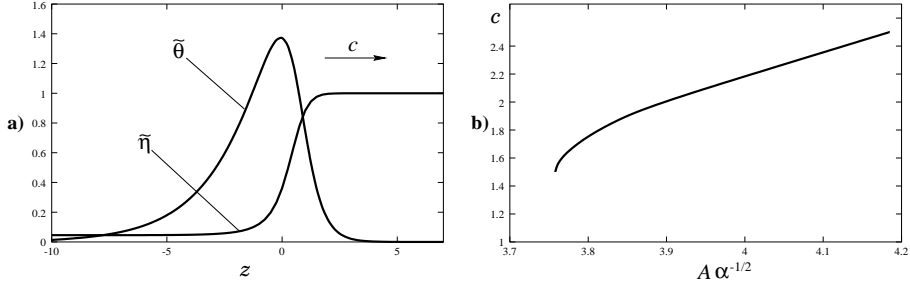,angle=-90,width=12cm}}
\begin{center}
\caption{ (a) Solution of Eqs. (\ref{travba}) and (\ref{travbb}) for
$\tilde A = 3.76$. (b) Dependence $c(\tilde A)$ obtained from
Eqs. (\ref{travba}) and (\ref{travbb}). }
\label{abt}
\end{center}
\end{figure}
One can see that the distribution of $\tilde\theta$ has the form of an
asymmetric spike, while the distribution of $\tilde\eta$ goes from
$\tilde\eta = 1$ at plus infinity to $\tilde\eta = \eta_{\mathrm min}$
at minus infinity. The numerical solution of Eqs. (\ref{travba}) and
(\ref{travbb}) shows that the traveling solution exists only for
$\tilde A > \tilde A_{bT} = 3.76$. The numerical solution also shows
that for $\tilde A > \tilde A_{bT}$ we have $\tilde\eta_{\mathrm min}
< 0.05$, which decreases with the increase of $\tilde A$, so with good
accuracy we can assume that $\eta_{\mathrm min} = 0$. Figure
\ref{abt}(b) shows the dependence $c(\tilde A)$ obtained from the
numerical solution of Eqs. (\ref{travba}) and (\ref{travbb}). Observe
that already for $\tilde A \gtrsim 1.4 \tilde A_{bT}$ the velocity of
the AS agrees with Eq. (\ref{ultrac}) with accuracy better than
$15\%$. Behind the spike region (in the region of the smooth
distributions) we will have $\theta = 0$ and $\eta = 1 - (1 -
\eta_{\mathrm min}) \e^{\al z/c}$. If one assumes that $\eta_{\mathrm
min} = 0$, one would arrive once again at Eq. (\ref{ultrasm}).

We would like to emphasize that even for $A \sim \al^{1/2} \ll 1$,
that is, for such a small deviation of the system from thermal
equilibrium, the amplitude of the AS is $\theta_{\mathrm max} \sim
\al^{-1/2} \gg 1$ and the velocity $c \sim 1$. In other words, the AS
remains a highly nonequilibrium object in the system only slightly
away from thermal equilibrium. 

\subsubsection{Case $A \sim \al^{-1/2}$: oscillatory tail}
\label{trav:ss3} 

In the other limiting case $A \gg 1$ the behavior of the secondary
sharp distributions in the back of the AS acquires special
features. This is due to the fact that for $A \gg 1$ the phase
trajectory $\theta(\eta)$ in the phase plain of $\theta$ and $\eta$
may pass close to the unstable fixed point [see Eq. (\ref{hom2})]
$\theta_{h3} \simeq A$, $\eta_{h2} \simeq A^{-2}$, so the behavior of
the distributions of $\theta$ and $\eta$ behind the spike becomes
oscillatory, so $\theta$ and $\eta$ will not be able to get back to
the homogeneous state $\theta = \theta_h$ and $\eta = \eta_h$ at $z =
-\infty$. To see that, let us introduce
\begin{eqnarray}
\tilde\theta = \al^{1/2} \theta, ~~~\tilde\eta = \al^{-1} \eta,
~~~\tilde A = \al^{1/2} A, ~~~\xi = {z \over c}.
\end{eqnarray}
Then, we can rewrite Eqs. (\ref{acttrav}) and (\ref{inhtrav}) behind
the spike as follows
\begin{eqnarray}
\tilde\theta_\xi + \tilde A \tilde\theta^2 \tilde\eta - \tilde\theta =
0, \label{travda} \\ \tilde\eta_\xi - \tilde\theta^2 \tilde\eta + 1 =
0, \label{travdb}
\end{eqnarray}
where we kept only the leading terms. In order for the solutions of
these equations to properly match with the primary sharp
distributions, we must have $\tilde\theta \sim \e^\xi$ and $\tilde\eta
\sim \e^{-2 \xi}$ as $\xi \rightarrow +\infty$ [see
Eqs. (\ref{shthetatrav}) and (\ref{shetatrav})]. The numerical
solution of Eqs. (\ref{travda}) and (\ref{travdb}) with these boundary
conditions shows that at $\tilde A > \tilde A_{dT} = 0.90$ the
distributions of $\tilde\theta$ and $\tilde\eta$ become oscillatory
behind the spike. Thus, we conclude that the ultrafast traveling spike
AS exists in a wide range $A_{bT} < A < A_{dT}$, where $A_{bT} = 3.76
\al^{1/2}$ and $A_{dT} = 0.90 \al^{-1/2}$. Notice that the oscillatory
behavior of the distributions of $\theta$ and $\eta$ behind the spike
of the AS is essentially related to the Hopf bifurcation of the
homogeneous state $\theta = \theta_{h3}$, $\eta = \eta_{h3}$ for $0.41
\al^{-1/2} < A < \al^{-1/2}$ (see Sec. \ref{s:mod}). On the other
hand, for $A > \al^{-1/2}$ this homogeneous state becomes stable, so
in that case the traveling spike AS transforms to a wave of switching
from one stable homogeneous state to the other.

\subsubsection{Justification of the condition $\ep \gg \al^{1/2}$ and
case $\al = \ep^2$} \label{trav:ss4}

Above we considered the case, in which the inhibitor does not
diffuse. Let us see how the diffusion of the inhibitor affects the
properties of the ultrafast traveling spike AS. Since the main
parameters of the AS are determined by the primary supersharp
distributions, the diffusion of the inhibitor should not significantly
affect these distributions. According to Eq. (\ref{inhtrav}), the term
$\ep^{-2} d^2 \eta / dz^2 \sim \ep^{-2} \al^{-1} A^2$ is small
compared to the leading terms which are of order $\al^{-2} A^2$ [see
Eq. (\ref{scaletrav})] if $\ep^2 \gg \al$ regardless of
$A$.\footnote{Note that these estimates remain valid even at $A \sim
A_{bT}$.} In terms of the physical parameters of the problem this
means that the ultrafast traveling AS will be described by the
solution obtained above as long as $D_\theta \gg D_\eta$, where
$D_\theta$ and $D_\eta$ are the diffusion coefficients of $\theta$ and
$\eta$, respectively. It is also clear that when $\al \sim \ep^2$, the
properties of the AS will not change qualitatively. An interesting
special case $\al = \ep^2$ which corresponds to the activator and the
inhibitor with equal diffusion coefficients can be treated in an
analogous way (see also \cite{MS}). The resulting equation for the
supersharp distributions will have the form of Eq. (\ref{supertheta}),
but without the nonlinear friction term. This equation can be solved
exactly, giving in this case the velocity $\tilde c = 1/ \sqrt{2}$
\cite{MS,BJ}, which is in fact close to the one obtained in the case
of the non-diffusing inhibitor. The explicit expression for the
supersharp distributions in this case are: $\tilde\theta_{\mathrm ssh}
= \frac{1}{2} \left(1 - \tanh {\xi \over 2 \sqrt{2}} \right)$,
$\eta_{\mathrm ssh1} = \frac{1}{2} \left(1 + \tanh {\xi \over 2
\sqrt{2}} \right)$, and $\eta_{\mathrm ssh2} = \al^2 A^{-2} + \e^{\xi
/ \sqrt{2}}$. The rest of the solution will be the same as in the case
$\ep = \infty$. All this allows us to conclude that the ultrafast
traveling spike AS in the Gray-Scott model exists when $\ep \gtrsim
\al^{1/2}$.
 
\subsection{Diffusive inhibitor: $\ep \ll \al^{1/2}$} \label{trav:s2}

Now let us analyze the second type of the traveling spike AS which is
realized when both $\al \ll 1$ and $\ep \ll 1$. It is obvious that in
this situation there exist a solution in the form of the spike AS
whose velocity is equal to zero (see Sec. \ref{stat:1d}). What we will
show below is that when $\al$ becomes small enough, in addition to the
static spike AS there are solutions in the form of the traveling spike
AS which propagates with constant velocity whose magnitude is $\gtrsim
l / \tau_\theta$.

Since $\ep \ll 1$, it is natural to separate the distributions of
$\theta$ and $\eta$ into the sharp and the smooth distributions. As in
the case of the one-dimensional static AS, in the spike region we
introduce the scaled variables from Eq. (\ref{scale1d}), with $\xi =
z$ in this case. In terms of these variables, Eqs. (\ref{acttrav}) and
(\ref{inhtrav}) become
\begin{eqnarray}
\tilde\theta_{zz} + c \tilde\theta_z + A \tilde\theta^2
\tilde\eta - \tilde\theta = 0, \label{thetashtrav} \\
\tilde\eta_{zz} - \tilde\theta^2 \tilde\eta = 0, \label{etashtrav}
\end{eqnarray}
where we assumed that $\ep \ll \al^{1/2}$ and $\theta^2 \eta \gg 1$ in
the spike and kept only the leading terms. Similarly to the
one-dimensional case, the scaling for the variables in the spike
region is given by Eq. (\ref{minmax1d}). According to
Eq. (\ref{etashtrav}), the asymptotic behavior of $\tilde\eta$ at
large $|z|$ is
\begin{eqnarray} \label{kappapm}
\tilde\eta \simeq \kappa_\pm z, ~~~z \rightarrow \pm \infty,
\end{eqnarray}
where $\kappa_\pm$ are some constants. Therefore, the distributions of
$\tilde\theta$ and $\tilde\eta$ in the spike will qualitatively have
the form shown in Fig. \ref{travqual}.
\begin{figure}
\centerline{\psfig{figure=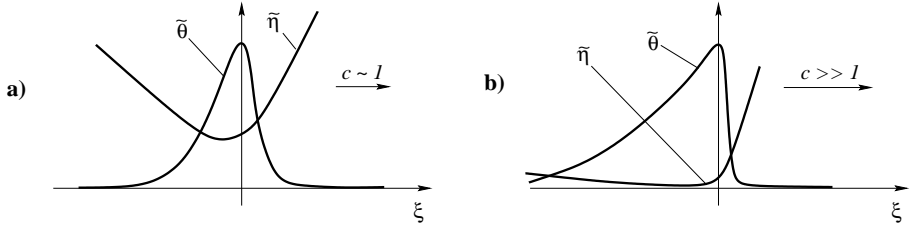,angle=-90,width=12cm}}
\begin{center}
\caption{ Qualitative form of the traveling spike AS at $\ep^2 \ll \al
\lesssim \ep$ for $c \sim 1$ (a) and $c \gg 1$ (b). }
\label{travqual}
\end{center}
\end{figure}

It is convenient to introduce the variables from
Eq. (\ref{bar1d}). Then, Eq. (\ref{etashtrav}) can be written as
\begin{eqnarray} \label{etashtravbar}
\bar\eta_{zz} = \bar\theta - c \bar\theta_z.
\end{eqnarray}
Because of the translational invariance, we have the freedom to choose
the position of the spike. We will do it in such a way that the
maximum of $\bar\theta$ is located at $z = 0$, that is, we have
$\bar\theta_z(0) = 0$. Also, according to Eq. (\ref{etashtravbar}), we
can add an arbitrary function $a + b z$ to its solution, so we may
replace $\bar\eta \rightarrow \bar\eta + \bar\eta_s + \kappa z$, where
$\bar\eta_s$ and $\kappa$ are arbitrary constants, and require that
$\bar\eta(0) = \bar\eta_z(0) = 0$, so the function $\bar\eta(z)$
is completely determined by $\bar\theta(z)$. In view of all this,
Eq. (\ref{thetashtrav}) becomes
\begin{eqnarray} \label{thetashtravbar}
\bar\theta_{zz} + c \bar\theta_z + A^2 \bar\theta^2 (\bar\eta_s
+ \kappa z + \bar\eta - \bar\theta) - \bar\theta = 0.
\end{eqnarray}

According to Eq. (\ref{kappapm}), we have $\bar\eta_z (\pm \infty) =
\kappa_\pm - \kappa$.  This implies that $\bar\eta_z(+\infty) -
\bar\eta_z(-\infty) = \kappa_+ - \kappa_-$. Integrating
Eq. (\ref{etashtravbar}) over $z$, we obtain an integral
representation of this condition
\begin{eqnarray} \label{matchtrav}
\int_{-\infty}^{+\infty} \bar\theta d z = \kappa_+ - \kappa_-.
\end{eqnarray} 
By a similar integration, the value of $\kappa$ is determined as
\begin{eqnarray} \label{kappa}
\kappa = \kappa_- + \int_{-\infty}^0 \bar\theta d z.
\end{eqnarray}

To study the solutions of Eqs. (\ref{etashtravbar}) and
(\ref{thetashtravbar}), we use the mechanical analogy. Equation
(\ref{thetashtravbar}) can be considered as an equation of motion for
a particle of unit mass with the coordinate $\bar\theta$ and time $z$
in the potential $U = - {\bar\theta^2 \over 2} + A^2 \bar\eta_s
{\bar\theta^3 \over 3} - A^2 {\bar\theta^4 \over 4}$ in the presence
of friction, with the friction coefficient $c$, and an external
time-dependent force $-A^2 \bar\theta^2 [\kappa z + \bar\eta(z)]$. For
$\bar\eta_s > 2 A^{-1}$ the potential has two maxima at $\bar\theta =
0$ and $\bar\theta = \bar\theta_m$, and a minimum in between. The
particle slides down from the maximum of the potential at $\bar\theta
= 0$ and after an excursion toward $\bar\theta = \bar\theta_{\mathrm
max} < \bar\theta_m$ at $z = 0$ returns to $\bar\theta = 0$. Notice
that since by definition $\bar\eta(0) = \bar\eta_z(0) = 0$, the value
of $\bar\eta$ should be rather small in the spike region, so one could
think of the second term in $-A^2 \bar\theta^2 [ \kappa z +
\bar\eta(z) ]$ as a small perturbation.

In the presence of friction the external time-dependent force acting
in Eq. (\ref{thetashtravbar}) must be such that it accelerates the
particle at some portion of the trajectory. According to
Eq. (\ref{etashtravbar}) and the fact that $\bar\theta_z <0$ for $z >
0$, we have $\bar\eta > 0$ for those values of $z$. Since the values
of $\kappa$ relevant to our analysis are positive (see below), in the
portion of the trajectory where $z > 0$ the external force does
accelerate the particle. All the kinetic energy that is gained by the
particle during this part of the motion must be dissipated by the
friction force, so that the particle arrives at $\bar\theta = 0$ with
zero velocity. This defines the precise value of the friction
coefficient $c$ as a function of $\bar\eta_s$ and $\kappa$. Recall
that in addition the distribution $\bar\theta(z)$ must satisfy the
integral condition in Eq. (\ref{matchtrav}), so in fact we are not
free to choose the value of $\bar\eta_s$. Thus, for given values of
$\kappa_\pm$ there may exist only a discrete set of the velocities
$c$.

Away from the spike $\theta$ is zero, and $\eta$ is described by the
smooth distributions. If we introduce the variable $\zeta = \ep z$, we
can write Eq. (\ref{inhtrav}) in the form
\begin{eqnarray}
\eta_{\zeta\zeta} + \beta^{-1} c \eta_\zeta + 1 - \eta = 0,
\end{eqnarray}
where $\beta = \al / \ep$. We should choose such a solution of this
equation that correctly matches with the sharp distribution. To do
that, we use the fact that to order $\ep$ the value of $\eta(0) = 0$,
so the smooth distribution of $\eta$ is
\begin{eqnarray} \label{smtrav}
\eta(\zeta) = 1 - \exp( - \kappa_\pm \zeta),
\end{eqnarray}
where 
\begin{eqnarray} \label{kappasm}
\kappa_\pm = {c \pm \sqrt{c^2 + 4 \beta^2} \over 2 \beta},
\end{eqnarray}
and one should take $\kappa_+$ if $\zeta > 0$, or $\kappa_-$
otherwise. Note that for $\beta \sim 1$ and $c \sim 1$ we have
$\kappa_\pm \sim 1$. From Eq. (\ref{smtrav}) one can see that when
$\zeta$ approaches the spike region, we have $\eta \simeq \kappa_\pm
\zeta = \ep \kappa_\pm z$. This means that we should use the values of
$\kappa_\pm$ given by Eq. (\ref{kappasm}) in solving the inner problem
given by Eqs. (\ref{etashtravbar}) -- (\ref{kappa}).

\subsubsection{Case $\ep A^2 \lesssim \al \ll \ep A$: bifurcation of
the static and the traveling autosolitons}
\label{trav:ss5} 

Equations (\ref{thetashtravbar}) -- (\ref{kappa}) are difficult to
deal with and in general can only be treated numerically. Such a
treatment was recently performed by Reynolds, Ponce-Dawson, and
Pearson, who also derived these equations in a different context
\cite{Reyn}. The problem can be significantly simplified in the case
$A \ll 1$. In this case there is a small factor of $A^2$ multiplying a
number of terms in Eq. (\ref{thetashtravbar}). It can be partially
compensated by choosing $\bar\eta_s \sim A^{-2} \gg 1$.  If we neglect
the other terms containing $A^2$ and put $c = 0$ in
Eq. (\ref{thetashtravbar}), we can solve this equation [together with
Eq. (\ref{matchtrav})] exactly. The result is
\begin{eqnarray} \label{th0}
\bar\theta_0(z) = {\kappa_+ - \kappa_- \over 4} \cosh^{-2} \left( {z
\over 2} \right), ~~~ \bar\eta_s = {6 \over A^2 (\kappa_+ -
\kappa_-)}.
\end{eqnarray}
Note that according to Eqs. (\ref{kappa}) and (\ref{th0}), we may put
$\kappa = (\kappa_+ + \kappa_-)/2 = c/(2 \beta)$.

The equation describing the small deviations $\delta \bar\theta =
\bar\theta - \bar\theta_0$ due to the order $A^2$ terms that were
dropped in the derivation of Eq. (\ref{th0}) can be obtained by the
linearization of Eq. (\ref{thetashtravbar}) around
$\bar\theta_0$. Assuming that $c \ll 1$ and retaining only the term
that gives the nontrivial contribution to $c$, we get
\begin{eqnarray} \label{trava}
\left[ -{d^2 \over d z^2} - 2 A^2 \bar\eta_s \bar\theta_0 + 1 \right]
\delta \bar\theta = A^2 \bar\theta_0^2 \kappa z + c {d \bar\theta_0
\over d z}.
\end{eqnarray}
The operator in the left-hand side of this equation has an eigenvalue
zero, corresponding to the translational mode $\delta \bar\theta = d
\bar\theta_0 / d z$. Therefore, in order for Eq. (\ref{trava}) to have
a solution, its right-hand side must be orthogonal to this
translational mode. The operator in Eq. (\ref{trava}) is self-adjoint,
so in order to get the solvability condition for this equation, we
should require that the integral of its right-hand side multiplied by
$d \bar\theta_0 / d z$ be equal to zero. With the use of
Eq. (\ref{th0}), this gives us the velocity $c$ as a function of
$\kappa_\pm$
\begin{eqnarray} \label{c}
c = \frac{A^2 (\kappa_+^2 - \kappa_-^2)}{6}.
\end{eqnarray}

To determine the velocities that are actually realized in the
traveling spike AS, we need to take into account the matching
conditions that are given by Eqs. (\ref{kappasm}). With the use of
these equations, Eq. (\ref{c}) becomes
\begin{eqnarray} \label{cmatch}
c = 2 \beta \sqrt{ {\beta^2 \over \beta_T^2} - 1 }, ~~~~\beta_T = {A^2
\over 3}.
\end{eqnarray}
Since in the derivation of this equation we assumed that $c \ll 1$, it
will be valid only for $\beta \ll A$. 

Two observations can be made from Eq. (\ref{cmatch}). First, at some
$\beta = \beta_T(A)$ [or at some $A = A_T(\beta)$] the velocity of the
AS becomes zero. Since this happens for $\beta \sim A^2$ and $\al \gg
\ep^2$, we have $\ep^{1/2} \ll A \ll 1$, so there is also a solution
with $c = 0$ (see Sec. \ref{stat:1d}). The presence of a point where
the velocity of the {\em traveling} solution goes to zero therefore
signifies a bifurcation of the static AS. Second, according to
Eq. (\ref{cmatch}) the velocity of the obtained solution is a {\em
decreasing} function of $A$ (or an increasing function of $\beta$). In
contrast, we would expect the velocity of the traveling spike AS to be
an {\em increasing} function of the excitation level $A$.

Let us consider an iterative map that takes $c$, substitutes it to
Eq. (\ref{kappasm}) with the fixed $\kappa_+ + \kappa_-$, calculates
$\kappa$ and uses Eq. (\ref{c}) to give the new value of $c$. Clearly,
those $c$ given by Eq. (\ref{cmatch}) (or $c = 0$) are the fixed
points of this map. However, an analysis of this map shows that an
arbitrarily small deviation of $c$ from that given by
Eq. (\ref{cmatch}) will lead to the unlimited growth of $c$ if the
deviation is positive, or to zero if the deviation is negative. In
other words, the fixed point given by Eq. (\ref{cmatch}) is
unstable. Also, one can easily see that for $A < A_T$ the fixed point
at $c = 0$ is stable for $A < A_T$ (or $\beta > \beta_T$) and unstable
otherwise. This means that the solution with non-zero velocity we
found above and the static solution at $A > A_T$ or $\beta < \beta_T$
should be unstable. Therefore, the stable traveling solutions should
have the velocity $c \gtrsim 1$ and may exist both when $A < A_T$ and
$A > A_T$. Also, when $A > A_T$, the solutions with $c = 0$ should be
unstable, so the static spike AS spontaneously transforms into the
traveling spike AS, whose speed $c \gtrsim 1$. These conclusions are
also supported by the numerical simulations of Eqs. (\ref{act0}) and
(\ref{inh0}).

\subsubsection{Case $\al \ll \ep A$: ultrafast traveling autosoliton}
\label{trav:ss6} 

Above we considered the situation in which $c \ll 1$. Let us now study
another possibility: $c \gg 1$. In this case the distribution of
$\bar\theta$ will become strongly asymmetric [see
Fig. \ref{travqual}(b)]. Indeed, according to
Eq. (\ref{thetashtravbar}), we will have $\bar\theta \sim \e^{-c z}$
at $z \rightarrow +\infty$ and $\bar\theta \sim \e^{z / c}$ at $z
\rightarrow -\infty$. In other words, we can once again separate the
distributions of $\bar\theta$ and $\bar\eta$ into the regions of the
supersharp distributions (in the front of the spike) and the sharp
distributions (in the back of the spike). In the region of the
supersharp distributions the supersharp front will have the width of
order $c^{-1} \ll 1$. Let us introduce $\xi = c z$. Then, we can write
Eq. (\ref{etashtravbar}) integrated over $\xi$ and
Eq. (\ref{thetashtravbar}) as
\begin{eqnarray}
\bar\theta_{\xi\xi} + \bar\theta_\xi + c^{-2} A^2 \bar\theta^2 (
\bar\eta_s + \bar\eta - \bar\theta) = 0, \label{fasta} \\ \bar\eta_\xi
= - \bar\theta + \bar\theta_{\mathrm max},
\label{fastb}
\end{eqnarray}
where we retained only the leading terms and moved the point where
$\bar\theta$ attains its maximum value to minus infinity [this amounts
to putting $\kappa = 0$ in Eq. (\ref{fasta}) and redefining the
boundary condition for $\bar\eta$ to be $\bar\eta(-\infty) =
\bar\eta_\xi(-\infty) = 0$].  One can see that by rescaling
$\bar\theta$ and $\bar\eta$ with $\bar\theta_{\mathrm max}$ all the
$A$- and $\bar\theta_{\mathrm max}$-dependence from Eqs. (\ref{fasta})
and (\ref{fastb}) can be absorbed into $c$. These equations have a
solution in the form of a supersharp front connecting $\bar\theta = 0$
ahead of the front with $\bar\theta = \bar\theta_{\mathrm max}$ behind
the front only when $\bar\eta_s = \bar\theta_{\mathrm max}$. In this
case we have $\bar\eta_\xi(+\infty) = \bar\theta_{\mathrm max}$ [see
Eq. (\ref{fastb})], what corresponds to $\kappa_- = 0$ and $\kappa_+ =
c \beta^{-1}$. According to Eq. (\ref{kappasm}), these values of
$\kappa_\pm$ can only be realized when $c \gg \beta$, with
$\bar\theta_{\mathrm max} = \beta^{-1}$. Note that the fact that
$\bar\eta_s = \bar\theta_{\mathrm max}$ behind the supersharp front
means that $\tilde\eta$ exponentially decays to zero at $\xi
\rightarrow -\infty$.

The numerical solution of Eqs. (\ref{fasta}) and (\ref{fastb}) shows that
the velocity of the supersharp front is
\begin{eqnarray} \label{c122}
c = 1.22 \times A \beta^{-1}. 
\end{eqnarray}
Since by assumption $c \gg 1$, we must have $\beta \ll A$. Recall that
in the derivation we also assumed that $c \gg \beta$. According to
Eq. (\ref{c122}), this leads to $\beta^2 \ll A$. Since, as we will
show in Sec. \ref{trav:ss7}, the solution in the form of the traveling
AS exists only when $\beta \lesssim 1$, this condition is always
satisfied when $\beta \ll A$. Note that the numerical solution of
Eqs. (\ref{fasta}) and (\ref{fastb}) in the form of the supersharp
front differs from $\bar\theta_{\mathrm ssh} = \frac{1}{2} [ 1 -
\tanh(0.44 \xi)]$ by less than 0.5\%.

In the region of the sharp distributions the characteristic length
scale of the variation of $\bar\theta$ is $c$, so we can neglect the
term $\bar\theta_{zz}$ in Eq. (\ref{thetashtravbar}). Recalling that
$\tilde\eta = 0$ in this region, we can write the solution of this
equation that represents the sharp distribution of $\theta$ behind the
supersharp front as $\bar\theta_{\mathrm sh} = \bar\theta_{\mathrm
max} \e^{z / c}$, where we assumed that the supersharp front is
located at $z = 0$.

When the value of $A$ is decreased, the velocity of the unstable
traveling solution grows and the velocity of the stable traveling
solution decreases until they reach the value of order 1 when the
approximations used in deriving the above equations become invalid.
According to Eq. (\ref{c122}), this will happen at $A \sim \beta$, so
at some $A = A_{bT} \sim \beta$ the solution in the form of the
traveling spike AS will disappear. Therefore, the velocity of the
traveling spike AS as a function of $A$ or $\beta$ should have the
form shown in Fig. \ref{fc}.
\begin{figure}
\centerline{\psfig{figure=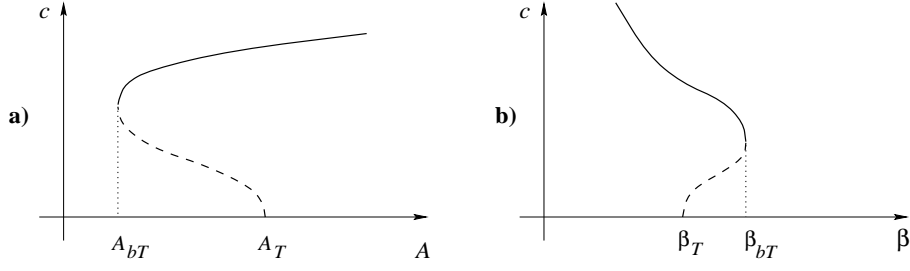,angle=-90,width=12cm}}
\begin{center}
\caption{ Qualitative form of the dependence $c(A)$ (a) and the
dependence $c(\beta)$ (b) for the traveling spike AS. } \label{fc}
\end{center}
\end{figure}

For $\beta \lesssim 1$ the traveling AS exists at $A \gtrsim
\beta$. On the other hand, for these values of $\beta$ the static
spike AS remains stable up to the values of $A \sim \beta^{1/2}$ [see
Eq. (\ref{cmatch}) and the discussion below]. Therefore, for $\beta
\lesssim A \lesssim \beta^{1/2}$ the static spike AS will coexist with
traveling.

\subsubsection{Case $A \sim \ep^{-1}$, $\al \lesssim \ep$: oscillatory
tail}
\label{trav:ss7} 

When the speed of the traveling spike AS increases, the behavior of
the distributions of $\theta$ and $\eta$ in the back of the AS changes
in the way similar to the case of the ultrafast traveling spike AS
(Sec. \ref{trav:ss3}). For large enough values of $A$ the sharp
distributions in the back of the spike become oscillatory. To see that
let us introduce the variables
\begin{eqnarray}
\tilde\theta = \ep^{1/2} \theta, ~~~ \tilde\eta = \ep^{-1} \eta,
~~~\tilde A = \ep^{1/2} A, ~~~\tilde c = \ep^{1/2} c, ~~~\xi = {z
\over c},
\end{eqnarray}
where $c$ is given by Eq. (\ref{c122}). Then, keeping only the
leading terms, we can write Eqs. (\ref{acttrav}) and (\ref{inhtrav})
in the region of the sharp distributions as follows
\begin{eqnarray}
\tilde\theta_\xi + \tilde A \tilde\theta^2 \tilde\eta - \tilde\theta =
0, \label{taila} \\ {1 \over \tilde c^2} \tilde\eta_{\xi\xi} + {1
\over \beta} \tilde\eta_\xi - \tilde\theta^2 \tilde\eta - 1 =
0. \label{tailb}
\end{eqnarray}
To match the solutions of these equations with the supersharp
distributions we must have $\eta \rightarrow 0$ for $\xi \rightarrow
+\infty$, so $\tilde\theta \sim \e^\xi$ and $\tilde\eta \sim \e^{-2
\xi}$ for large $\xi$. Similarly, in order for the distributions of
$\tilde\theta$ and $\tilde\eta$ to properly match with the smooth
distributions, for $\xi \rightarrow -\infty$ we must have
$\tilde\theta(-\infty) = 0$ and $\tilde\eta_\xi = -\beta$ [see
Eq. (\ref{kappasm}) with $c \gg 1$]. Equations (\ref{taila}) and
(\ref{tailb}) with these boundary conditions can then be solved
numerically. As a result, we find the values of $\beta = \beta_d$ as a
function of $\tilde A$ at which the behavior of the distributions of
$\tilde\theta$ and $\tilde\eta$ in the back of the spike changes to
oscillatory. The dependence $\beta_d(\tilde A)$ is plotted in
Fig. \ref{ftail}.
\begin{figure}
\centerline{\psfig{figure=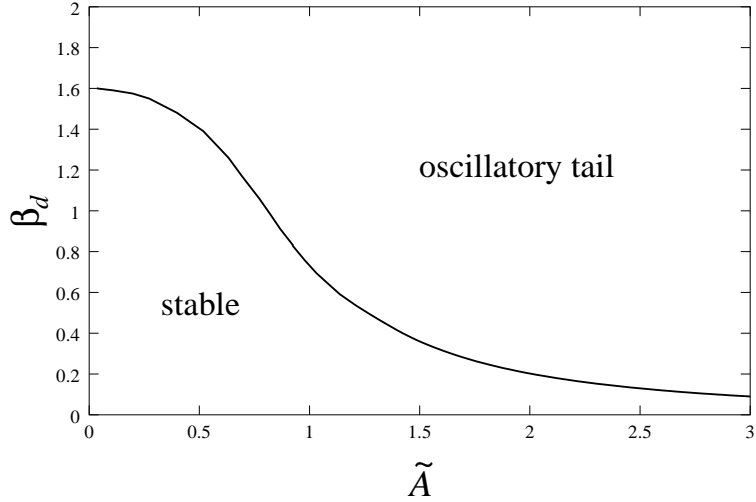,width=10cm}}
\begin{center}
\caption{ Dependence $\beta_d(\tilde A)$ for the traveling spike
AS. } \label{ftail}
\end{center}
\end{figure}
From this figure one can also see that the solution in the form of the
traveling spike AS exists only at $\beta < \beta_c \simeq 1.6$. The
analysis of Eqs. (\ref{taila}) and (\ref{tailb}) also shows that for
$\tilde A \gg \beta^{3/2}$ the second derivative of $\tilde\eta$ in
Eq. (\ref{tailb}) can be dropped, so the problem reduces to that given
by Eqs. (\ref{travda}) and (\ref{travdb}) with $\tilde A$ replaced by
$\tilde A \beta^{1/2}$. Therefore, for large enough values of $\tilde
A$ we should have $\beta_d \simeq 0.81 \tilde A^{-2}$. The numerical
simulations of Eqs. (\ref{act}) and (\ref{inh}) show that the onset of
the oscillatory behavior in the back of the spike in the traveling
spike AS results in the splitting of the AS as it moves (see
Sec. \ref{s:pf1d}).

\subsubsection{Case $\al \sim \ep$, $A \sim 1$: qualitative
considerations} \label{trav:ss8} 

So far, we concentrated on the situations in which either $c \ll 1$ or
$c \gg 1$. We found that only the solutions with $c \gg 1$ should be
stable and can be realized only if $\beta \lesssim 1$ and $\beta \ll
A$ (Sec. \ref{trav:ss6}). As was shown in Sec. \ref{trav:ss7}, in the
case $A \gg 1$ the solution in the form of the traveling spike AS will
exist only if $\beta < \beta_c \sim 1$. It is clear that qualitatively
these conclusions can also be made when both $A \sim 1$ and $\beta
\sim 1$, what corresponds to $c \sim 1$. In this case the AS will not
be able to propagate at $A < A_{bT} \sim 1$ and will transform into
the static AS. On the other hand, for $A > A_{dT} \sim 1$ the
traveling spike AS will start splitting. When the value of $\beta$ is
increased, the value of $A_{bT}$ will grow and the value of $A_{dT}$
will decrease, so at some $\beta = \beta_c$ the solution in the form
of the traveling spike AS will disappear. This conclusion is supported
by the numerical simulations of Eqs. (\ref{act}) and (\ref{inh}), with
the value of $\beta_c$ found to be very close to the one obtained in
the preceding paragraph (Sec. \ref{s:pf1d}).

The reasons for stopping of the traveling spike AS at $A < A_{bT}$ or
splitting at $A > A_{dT}$ are the following. Consider
Eqs. (\ref{etashtravbar}) and (\ref{thetashtravbar}), using the
mechanical analogy. Let us see what happens as the value of $A$
decreases when $c \sim 1$. In this case the external force $-A^2
\bar\theta (\kappa z + \bar\eta)$ accelerating the particle on the way
from $\bar\theta_{\mathrm max}$ to 0 becomes smaller [see
Eq. (\ref{thetashtravbar})], while the friction force remains the
same. When the value of $A$ becomes small enough, the dissipation will
exceed the acceleration, so the particle will not be able to reach
$\bar\theta = 0$ and the solution traveling with constant velocity
will disappear. On the other hand, as long as $A_b < A < A_d$, where
$A_b$ is the point where the solution in the form of the static
one-dimensional AS disappears and $A_d$ the point where the static
one-dimensional AS starts splitting (see Sec. \ref{stat:ss2}), the
static solution will exist, so the traveling AS will be able to stop
and transform into static when the value of $A$ is decreased.

According to the definition of $\bar\theta$, when $z$ is close to
zero, the time-dependent external force is dominated by $-A^2
\bar\theta^2 \kappa z$, which is positive in some interval $z_0 < z <
0$. This means that the particle is accelerated by this force before
it stops at $\bar\theta = \bar\theta_{\mathrm max}$. If the value of
$A$ is big enough and the friction coefficient $c \sim 1$, the forces
from the potential $U$ and the friction may not be enough to stop the
particle before it reaches the maximum of the potential at $\bar\theta
= \bar\theta_m$. Then, if the particle moves past $\bar\theta_m$, it
will keep on moving toward plus infinity and will never be able to
return to $\bar\theta = 0$. This means that the solution in the form
of the traveling spike AS must disappear at some $A = A_{dT} \sim
1$. Notice that the same argument can be applied to the static AS at
$A = A_d$, so this bifurcation point should indeed correspond to the
onset of splitting.

\section{Stability of static spike autosolitons} \label{s:stab}

In Sec. \ref{s:stat} we showed that in the Gray-Scott model the
solutions in the form of the static spike ASs exist in certain ranges
of the parameter $A$ in arbitrary dimensions. According to the general
qualitative theory of ASs \cite{KO89,KO90,KO94}, these static ASs are
stable in a wide range of the system's parameters. However, when these
parameters are varied, the ASs may become unstable with respect to
various kinds of instabilities \cite{KO89,KO90,KO94}. So, now we are
going to study the stability of these solutions against different
types of perturbations in the full system given by Eqs. (\ref{act})
and (\ref{inh}), with $\ep \ll 1$.

\subsection{Static one-dimensional autosoliton} \label{stab:1d}

We start by analyzing the one-dimensional static spike AS in higher
dimensions. The equations describing small deviations $\delta \theta =
\theta - \theta_0$ and $\delta \eta = \eta - \eta_0$ of the activator
and the inhibitor, respectively, from the distributions $\theta_0(x)$
and $\eta_0(x)$ in the form of the static one-dimensional AS are
obtained by linearizing Eqs. (\ref{act}) and (\ref{inh}) (we chose the
axes so that $\theta_0$ and $\eta_0$ depend only on $x$). Let us take
\begin{eqnarray} 
\delta \theta = \delta \theta_{k\omega} (x) \e^{\i \omega t - \i k y},
~~~\delta \eta = \delta \eta_{k\omega} (x) \e^{\i \omega t - \i k y},
\end{eqnarray} 
where $\omega$ is the complex frequency and $k$ is the wave vector
that characterizes the transverse perturbations of the AS. Then, after
some algebra, Eqs. (\ref{act}) and (\ref{inh}) linearized around
$\theta_0$ and $\eta_0$ can be written as
\begin{eqnarray}
\left[ - {d^2 \over d x^2} + 1 + \i \omega + k^2 - 2 A \theta_0 \eta_0
\right] \delta \theta_{k\omega} & = & A \theta_0^2 \delta
\eta_{k\omega},
\label{actlin} \\ \left[ - \ep^{-2} {d^2 \over d x^2} + 1 + \i \al^{-1}
\omega + \ep^{-2} k^2 \right] \delta \eta_{k\omega} & = & \nonumber \\
&& \hspace{-2cm} - A^{-1} \left[ - {d^2 \over d x^2} + 1 + \i \omega +
k^2\right] \delta \theta_{k\omega},
\label{inhlin} 
\end{eqnarray}
where we measured length and time in the units of $l$ and
$\tau_\theta$, respectively.

Equation (\ref{inhlin}) can be solved by means of the Green's
function
\begin{eqnarray}
\delta \eta_{k\omega} = - {\ep^2 \over A} \delta \theta_{k\omega} &&
\nonumber \\ && \hspace{-1cm} - {\ep \left(1 + \i \omega - \i \ep^2
\al^{-1} \omega \right) \over 2 A \sqrt{ 1 + \i \al^{-1} \omega +
\ep^{-2} k^2} } \int_{-\infty}^{+\infty} \e^{ - \ep \sqrt{1 + \i
\al^{-1} \omega + \ep^{-2} k^2} |x - x'|} \delta \theta_{k\omega}(x')
dx', \nonumber \\
\end{eqnarray}
where we neglected a term of order $\ep^2$. This expression for
$\delta \eta_{k\omega}$ can be substituted back into
Eq. (\ref{actlin}) to get a single operator equation in terms of
$\delta \theta_{k\omega}$ alone. Note that the operator in the
left-hand side of Eq. (\ref{actlin}) is a Schr\"odinger type operator
with the potential in the form of the well of the depth and the size
of order 1. Indeed, according to the results of Sec. \ref{stat:1d}, the
characteristic length scale of the variation of $\theta_0$ is of order
1, and we have $\theta_0 \sim A \ep^{-1}$ and $\eta_0 \sim A^{-2} \ep$
in the spike, so $A \theta_0 \eta_0 \sim 1$ and exponentially decays
away from the spike. Also, the function $\delta \eta_{k\omega}$ in the
right-hand side of Eq. (\ref{inhlin}) is multiplied by the function
$\theta_0^2$ which also exponentially decays away from the spike, so
in fact we are only interested in the behavior of $\delta
\theta_{k\omega}$ in this region. Therefore, in solving the operator
equation mentioned above one should only know the sharp distributions
of $\theta$ and $\eta$ in the spike. If we take these distributions
written in terms of the variables from Eq. (\ref{bar1d}), we can
reduce Eqs. (\ref{actlin}) and (\ref{inhlin}) to
\begin{eqnarray} 
&& \left[ - {d^2 \over d x^2} + 1 + \i \omega + k^2 - 2 A^2 \bar\eta_s
\bar\theta_0 - 2 A^2 \bar\theta_0 \bar\eta_0 + 3 A^2 \bar\theta_0^2
\right] \delta \theta_{k\omega} = \nonumber \\ && \hspace{1cm} -
{\ep^{-1} A^2 \bar\theta_0^2 (1 + \i \omega - \i \ep^2 \al^{-1} \omega
) \over 2 \sqrt{1 + \i \al^{-1} \omega + \ep^{-2} k^2 } }
\int_{-\infty}^{+\infty} \e^{ - \ep \sqrt{1 + \i \al^{-1} \omega +
\ep^{-2} k^2} |x - x'|} \delta \theta_{k\omega}(x') dx'. \nonumber \\
\label{disp1d}
\end{eqnarray}
This is the basic equation for studying the stability of the static
one-dimensional AS in higher-dimensional systems, which is
asymptotically valid for $\ep \ll 1$. One has to solve this equation
as an eigenvalue problem, find the modes $\delta \theta_n$ and the
values of $\omega = \omega_n (k)$ corresponding to them. The
instability of the AS will occur when the real part of $\gamma = - \i
\omega$ is negative.\footnote{${\mathrm Re}~\gamma$ is the damping
decrement (decay rate) of a fluctuation. We will use $\omega$ and
$\gamma$ interchangeably throughout this section.} To analyze the
stability of the static spike AS in one dimension, one should simply
put $k = 0$ in Eq. (\ref{disp1d}). Note that since the right-hand side
of Eq. (\ref{disp1d}) is an exponentially decaying function of $x$, a
mode $\delta\theta_n$ can be unstable only if it is localized since
otherwise we would have $\gamma \geq 1 + k^2 > 0$. In other words, the
unstable modes should be in the {\em discrete} spectrum of the
solutions of Eq. (\ref{disp1d}).

\subsubsection{Dangerous modes} \label{stab:ss1}

To solve Eq. (\ref{disp1d}) in general is a formidable task. Since we
are mostly interested in the instabilities, we can simplify the
problem by identifying the most {\em dangerous} modes and try to solve
Eq. (\ref{disp1d}) for these modes. In order to get an idea as to what
these dangerous modes might be, we first consider Eq. (\ref{disp1d})
with $k = 0$ and $\omega = 0$. In this case Eq. (\ref{disp1d}) can be
written asymptotically as an eigenvalue problem
\begin{eqnarray} \label{disp1d0}
\left[ - {d^2 \over d x^2} + 1 - 2 A^2 \bar\eta_s \bar\theta_0 - 2 A^2
\bar\theta_0 \bar\eta_0 + 3 A^2 \bar\theta_0^2 \right] \delta \theta_n
&& \nonumber \\ && \hspace{-3cm} + {\ep^{-1} A^2 \bar\theta_0^2 \over
2} \int_{-\infty}^{+\infty} \delta \theta_n(x') dx' = \lambda_n
\delta\theta_n,
\end{eqnarray}
where the unstable modes $\delta\theta_n$ will correspond to
$\lambda_n = 0$.  This equation determines the instabilities of the
spike AS in one dimension for sufficiently large values of $\al$. Note
that these modes (in the PDE sense) should be in correspondence with
the bifurcation points of the solution $\theta_0$ and $\eta_0$ (in the
ODE sense). Also note that equations of the type of
Eq. (\ref{disp1d0}) were analyzed by Kerner and Osipov in the context
of the stability of the spike AS in systems of small size
\cite{KO89,KO90,KO94,KO78}.

When $A_b \lesssim A \ll 1$, one can neglect the last two terms inside
the brackets in Eq. (\ref{disp1d0}). Since the solution $\bar\theta_0$
and $\bar\eta_s$ in this region is known [see Eqs. (\ref{th1dsh0}) and
(\ref{pm1d})], the potential in the Schr\"odinger operator becomes
$V(x) = 1 - 3 \cosh^{-2} (x/2)$.  The eigenfunctions of this operator
are well-known \cite{Landau}. The spectrum of this operator contains
three discrete eigenvalues
\begin{eqnarray}
\lambda_0^{(0)} = & -\frac{5}{4}, & ~~~\delta \theta_0^{(0)} =
\sqrt{15 \over 32} \cosh^{-3} \frac{x}{2}, \nonumber \\
\lambda_1^{(0)} = & 0, & ~~~\delta \theta_1^{(0)} = \sqrt{15 \over 8}
\tanh \frac{x}{2} \cosh^{-2} \frac{x}{2}, \label{schr0} \\
\lambda_2^{(0)} = & \frac{3}{4}, & ~~~\delta \theta_2^{(0)} = \sqrt{3
\over 32} \left(5 \tanh^2 \frac{x}{2} - 1 \right) \cosh^{-1}
\frac{x}{2}, \nonumber
\end{eqnarray}
and a continuous spectrum of the eigenvalues $\lambda^{(0)}_k = 1 +
k^2 \geq 1$, characterized by a wave vector $k$. The shape of the
functions from Eq. (\ref{schr0}) is shown in Fig. \ref{fluct}.
\begin{figure}
\centerline{\psfig{figure=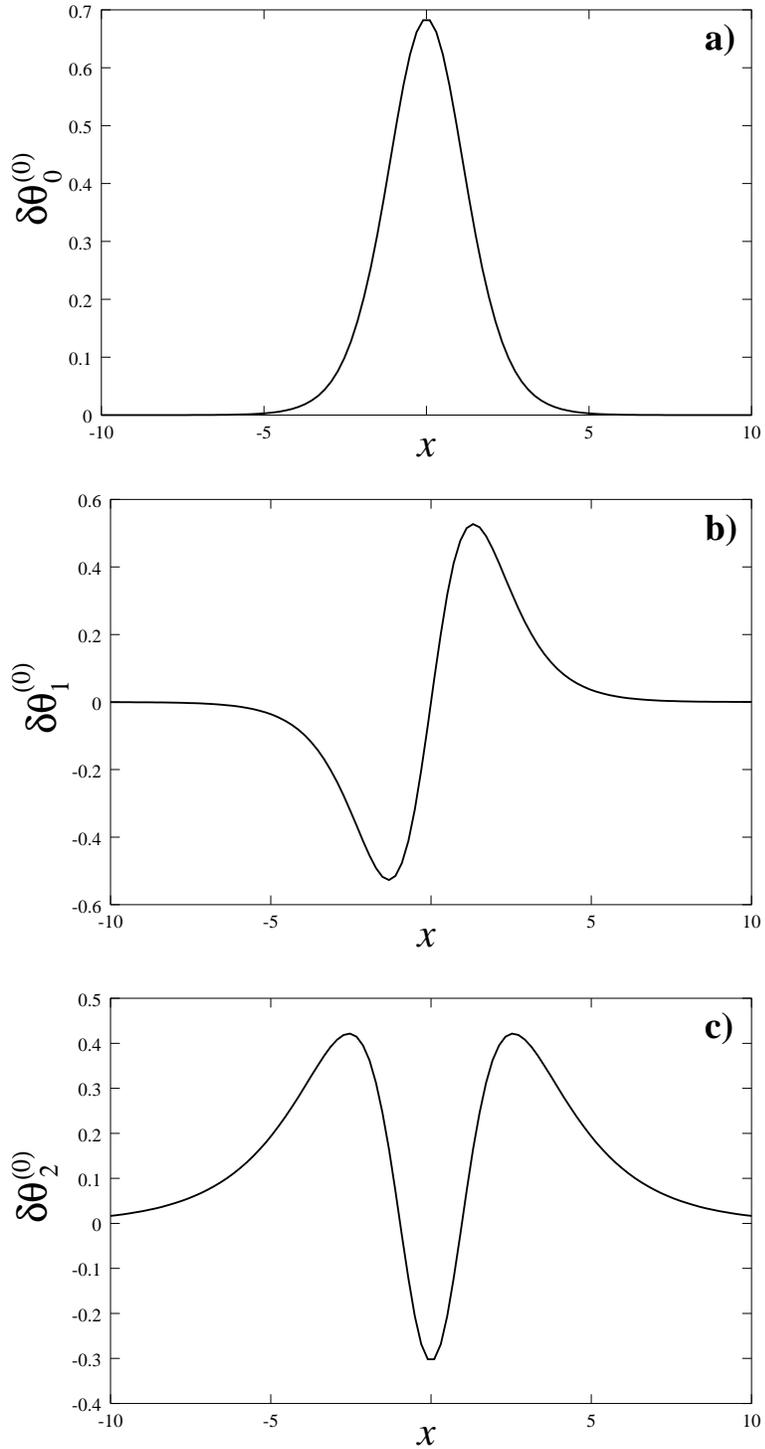,width=10cm}}
\begin{center}
\caption{Fluctuations $\delta\theta_0^{(0)}$ (a),
$\delta\theta_1^{(0)}$ (b), and $\delta\theta_2^{(0)}$ (c) from
Eq. (\ref{schr0}). } \label{fluct}
\end{center}
\end{figure}
The integral operator in Eq. (\ref{disp1d0}) has a discrete eigenvalue
$\lambda_s = - \frac{A^2}{A_b^2} \left[ 1 + \sqrt{1 - {A^2 \over
A_b^2} }~ \right]^2$ with the eigenfunction $\delta \theta_s = {3
\over 8} \cosh^{-4} (x/2) \propto \bar\theta_0^2$ with the adjoint
$\delta \theta_s^* = 1$, and a degenerate continuous spectrum of the
eigenfunctions satisfying $\int_{-\infty}^{+\infty} \delta \theta(x)
dx = 0$ (and for the adjoint $\int_{-\infty}^{+\infty}
\bar\theta_0^2(x) \delta \theta^*(x) dx = 0$) which all correspond to
$\lambda = 0$.

If the value of $A$ were much smaller than $A_b$ (ignoring for the
moment the fact that $\bar\theta_0$ is defined only for $A \geq A_b$
and thinking about it as $A$-independent), the integral operator in
Eq. (\ref{disp1d0}) would be small, so the solution of
Eq. (\ref{disp1d0}) would be determined by the the spectrum of the
Schr\"odinger operator [Eq. (\ref{schr0})]. When the value of $A$ is
increased, at $A \sim A_b$ the integral operator starts mixing these
eigenmodes. From the form of the eigenfunctions $\delta\theta^{(0)}_n$
and $\delta\theta_s$ it is clear that most of the mixing will occur
between $\delta \theta_0^{(0)}$ and $\delta \theta_s$. It is also
clear that because of the above mentioned properties of this integral
operator Eq. (\ref{disp1d0}) [as well as Eq. (\ref{disp1d})] should
generally have a discrete spectrum containing three different
eigenvalues with the eigenfunctions that look like
$\delta\theta_0^{(0)}$, $\delta\theta_1^{(0)}$, and
$\delta\theta_2^{(0)}$, respectively. Therefore, we will denote these
functions as $\delta \theta_0$, $\delta \theta_1$, and $\delta
\theta_2$, respectively. The fluctuation $\delta \theta_0$ will result
in the variation in the AS amplitude, the fluctuation $\delta
\theta_1$ in the shift of the AS along the $x$-axis, and $\delta
\theta_2$ in the widening of the AS.

It is not difficult to see that Eq. (\ref{disp1d0}) with the above
mentioned simplifications is satisfied for $\delta \theta_0 =
\cosh^{-2} (x/2)$ and $\lambda_0 = 0$ at $A = A_b$. This is in
agreement with the fact that the solution in the form of the static
spike AS disappears at $A = A_b$. Also, this equation has a solution
with $\lambda_1 = 0$ and $\delta \theta_1 = \delta
\theta_1^{(0)}$. This is in fact the consequence of the translational
symmetry of the problem \cite{KO89,KO90,KO94}. In general,
Eq. (\ref{disp1d0}) is always identically satisfied with $\lambda_1 =
0$ for $\delta \theta_1 = d \bar\theta_0 / dx$, which for $A \sim A_b$
is up to a coefficient given by $\delta \theta_1^{(0)}$. Note that
this fact alone does not correspond to the actual instability, for
which we should have ${\mathrm Re}~\gamma < 0$, so a careful analysis
of the full Eq. (\ref{disp1d}) is needed to determine whether the
function $\delta \theta_1$ is in fact a dangerous mode. Also note that
by symmetry $\delta \theta_1$ is completely decoupled from $\delta
\theta_0$ and $\delta \theta_2$.

When $A_b \ll A \ll 1$, we will have $|\lambda_s| \simeq 4 A^2/A_b^2
\gg 1$, so in this case it is the Schr\"odinger operator that becomes
a perturbation to the integral operator in
Eq. (\ref{disp1d0}). Therefore, the mode $\delta \theta_0 \cong \delta
\theta_s$ with $\lambda_0 \simeq \lambda_s$ can no longer give an
instability. On the other hand, we still have the mode $\delta
\theta_2$ that for $A \gg A_b$ should be orthogonal to $\delta
\theta_s$. Since $\delta \theta_2$ is not strongly coupled with
$\delta \theta_s$, it is natural to expect that the value of
$\lambda_2$ should be positive for $A_b \ll A \ll 1$ (recall that for
$A = 0$ in Eq. (\ref{disp1d0}) we would have it equal to
$\lambda_2^{(0)} = 3/4 > 0$), so this mode should not give an
instability for these values of $A$ either.

In contrast, when $A$ becomes of order 1, one should also consider the
last two terms in the integral operator in Eq. (\ref{disp1d0}). These
terms will give a negative contribution to $\lambda_2$ (see
Sec. \ref{stab:ss2}), so at some $A \sim 1$ it will become zero,
signifying an instability of the static spike AS in one dimension with
respect to the $\delta \theta_2$ mode. Comparing this with the results
of Sec. \ref{stat:1d}, we conclude that this will happen precisely at
$A = A_d = 1.35$ at which the solution in the form of the static
one-dimensional spike AS disappears. Thus, we would expect that the
modes $\delta \theta_{0,1,2}$ are the only dangerous modes in the
whole range of existence of the static spike AS.

\subsubsection{Case $\al \gg 1$ and $k = 0$: stability of the autosoliton
in one dimension} \label{stab:ss2}

Let us now go back to the analysis of Eq. (\ref{disp1d}).  In the case
$\al \gg 1$ and $k = 0$ it can be asymptotically written as
\begin{eqnarray} 
\left[ - {d^2 \over d x^2} + 1 - \gamma - 2 A^2 \bar\eta_s
\bar\theta_0 - 2 A^2 \bar\theta_0 \bar\eta_0 + 3 A^2 \bar\theta_0^2
\right] \delta \theta = && \nonumber \\ && \hspace{-4cm} - {\ep^{-1}
A^2 \bar\theta_0^2 (1 - \gamma ) \over 2} \int_{-\infty}^{+\infty}
\delta \theta(x') dx'.
\label{disp1d:stab}
\end{eqnarray}
When $A \sim A_b$, Eq. (\ref{disp1d:stab}) can be further simplified
by noting that in this case its right-hand side is of order 1 and the
last two terms in the square bracket in its left-hand side can be
neglected. Using $\bar\theta_0$ and $\bar\eta_s$ from
Eqs. (\ref{th1dsh0}) and (\ref{pm1d}), we can write
Eq. (\ref{disp1d:stab}) in the case $A \sim A_b$ asymptotically as
\begin{eqnarray}
\left[ - {d^2 \over d x^2} + 1 - \gamma - 3 \cosh^{-2} \left({x \over
2} \right) \right] \delta \theta = && \nonumber \\ && \hspace{-6cm} -
{3 A^2 (1 - \gamma ) \over 8 A_b^2} \left[ 1 + \sqrt{1 - {A_b^2 \over
A^2}} ~\right]^2 \cosh^{-4} \left({x \over 2} \right)
\int_{-\infty}^{+\infty} \delta \theta(x') dx'.
\label{disp1d:stab0}
\end{eqnarray}
Naturally, Eq. (\ref{disp1d:stab}) in the case $A \ll 1$ should give
the same results as Eq. (\ref{disp1d:stab0}) in the case $A \gg A_b$.

To see that the AS is actually stable in one dimension for $A_b < A <
A_d$ in the case of sufficiently large $\al$ we solved
Eqs. (\ref{disp1d:stab}) and (\ref{disp1d:stab0}) numerically, using
the asymptotic distributions $\bar\theta_0$ and $\bar\eta_0$ obtained
numerically in Sec. \ref{stat:ss2}. We solve these equations by
discretizing the operators and diagonalizing the obtained matrices
(taking the limit $\ep \rightarrow 0$ in the case $A \sim 1$). This
gives $\gamma_n$ as functions of $A$. The numerical solution indeed
shows that for $A_b < A < A_d$ the damping decrement ${\mathrm Re}~
\gamma_n$ is positive for all $\delta\theta_n$ and that it goes to
zero for $\delta\theta_0$ as $A \rightarrow A_b$ or for
$\delta\theta_2$ as $A \rightarrow A_d$ signifying the instabilities
of the AS at these points. The shape of the fluctuation
$\delta\theta_0$ suggests that the instability at $A = A_b$ will
result in the collapse of the AS, while the shape of $\delta\theta_2$
suggests that the instability at $A = A_d$ will result in the local
breakdown in the AS center and the onset of splitting.

The damping decrements ${\mathrm Re}~ \gamma_{0,2}(A)$ obtained from
the numerical solution of Eqs.(\ref{disp1d:stab}) and
(\ref{disp1d:stab0}) are shown in Fig. \ref{gam1d}.
\begin{figure}
\centerline{\psfig{figure=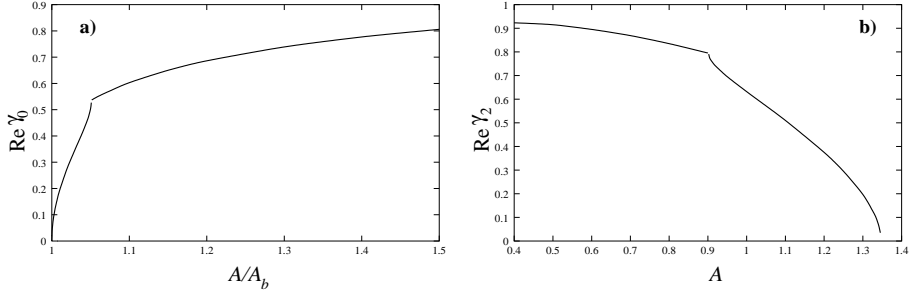,width=12cm,angle=-90}}
\begin{center}
\caption{Dependence ${\mathrm Re}~ \gamma_{0,2}(A)$ for the static
one-dimensional AS for $A \sim A_b$ (a) obtained from the numerical
solution of Eq. (\ref{disp1d:stab0}), and $A \sim A_d$ (b) obtained
from Eq. (\ref{disp1d:stab}). } \label{gam1d}
\end{center}
\end{figure}
Notice that it turns out that only at $A_b < A < 1.06 A_b$ or at $0.90
< A < A_d$ we have ${\mathrm Re}~\omega_{0,2} = 0$, so that there are
two distinct modes $\delta\theta_{0,2}$ with different values of
$\gamma$. For all other values of $A$ we have ${\mathrm Re}~\omega
\not= 0$ and there are two eigenfunctions: $\delta\theta_2$ and its
complex-conjugate, which correspond to the two complex frequencies
$\omega$ and $-\omega^*$. For $A_b \ll A \ll A_d$ we have ${\mathrm
Re}~ \gamma \simeq 1$ and ${\mathrm Re}~\omega \simeq 0.15$. 

Observe that Eq. (\ref{disp1d:stab0}) can be analyzed rigorously. This
analysis turns out to be rather involved, so we present it in appendix
A. It agrees with the conclusions of this section concerning the case
$A \ll 1$.

\subsubsection{Case $\al \lesssim 1$ and $k = 0$: instability with
respect to the pulsations} \label{stab:ss3}

General qualitative theory of ASs suggests that for small enough
values of $\al$ the static spike AS may become unstable with respect
to the fluctuations with ${\mathrm Re}~\omega \not= 0$, resulting in
the onset of the AS pulsations \cite{KO89,KO90,KO94,Os93,Sev}. The
analysis of Eq. (\ref{disp1d}) shows that in order for it to have a
solution with ${\mathrm Im}~\omega = 0$ and ${\mathrm Re}~\omega =
\omega_0 \not= 0$ we must have $\ep^2 \lesssim \al \lesssim 1$ and
$\omega_0 \sim 1$. Let us introduce $\kappa^2 = \ep^2 / \alpha
\lesssim 1$. Then, dropping 1 in the square roots in the right-hand
side of Eq. (\ref{disp1d}), and putting $k = 0$, we get asymptotically
\begin{eqnarray} \label{om1d}
\left[ -{d^2 \over dx^2} + 1 + \i \omega_0 - 2 A^2 \bar\eta_s
\bar\theta_0 - 2 A^2 \bar\theta_0 \bar\eta_0 + 3 A^2 \bar\theta_0^2
\right] \delta \theta = && \nonumber \\ && \hspace{-8cm} - \frac{A^2
(1 + \i \omega_0 - \i \kappa^2 \omega_0) \bar\theta_0^2}{2 \kappa
\sqrt{\i \omega_0}} \int_{-\infty}^{+\infty} \e^{-\kappa \sqrt{\i
\omega_0} |x - x'|} \delta \theta(x') dx'.
\end{eqnarray}
In order for the exponential in the right-hand side of this equation
to decay at large distances, we must choose the analytic branch of the
square root that has a positive real part for all values of
$\omega_0$, what is achieved by making a branch cut along the positive
imaginary axis. Obviously, the solutions of Eq. (\ref{om1d}) will come
in pairs, for each solution with $\omega$ there will be another
solution with $-\omega^*$. Note that because of the choice of the
branch cut this equation cannot have solutions with ${\mathrm Im~}
\omega > 0$ for which ${\mathrm Re~} \omega = 0$. Therefore, an
instability of the AS described by Eq. (\ref{om1d}) must necessarily
be a Hopf bifurcation (the exception here is the $\delta \theta_1$
mode).
\footnote{Strictly speaking, the branch cut should begin at $\omega =
i \al$ [see Eq. (\ref{disp1d}) with $k = 0$], which for sufficiently
small $\al$ can be considered to be at zero in the analysis of most of
the dangerous modes. This, however, cannot be done for the mode
$\delta\theta_1$ where the bifurcation is saddle-node (see the
discussion below).}

To find the instabilities of the static spike AS with respect to the
pulsations, we solved Eq. (\ref{om1d}) numerically with $\omega =
\omega_0$, with $\omega_0$ real. This numerical solution of
Eq. (\ref{om1d}) shows that the static spike AS indeed becomes
unstable with respect to the $\delta \theta_0$ mode with ${\mathrm
Re}~\omega = \omega_0(A)$ at $\al < \al_\omega(A) \sim \ep^2$ when $A
\sim 1$. The plots of $\al_\omega(A)$ and $\omega_0(A)$ are shown in
Figs. \ref{omega1d}(a) and \ref{omega1d}(b), respectively.
\begin{figure}
\centerline{\psfig{figure=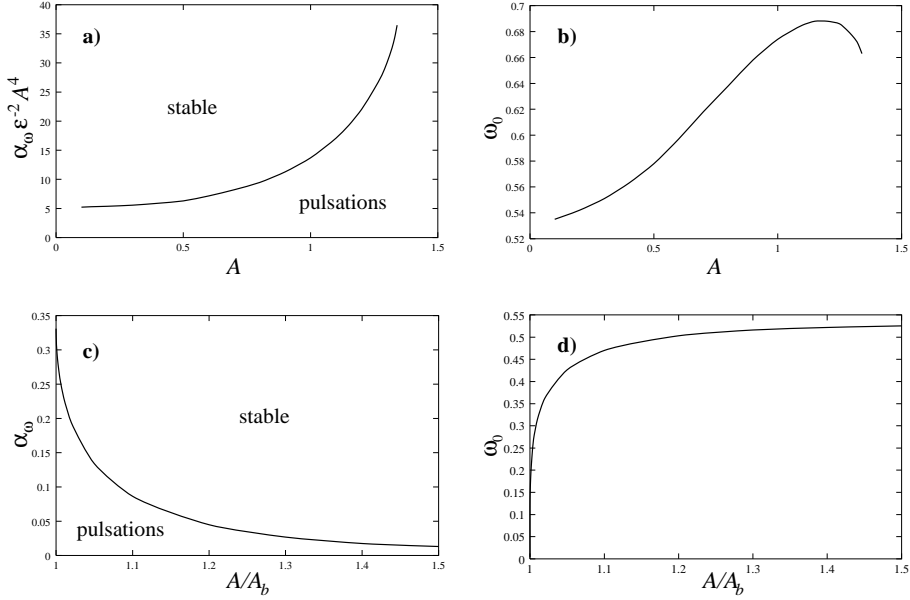,width=12cm,angle=-90}}
\begin{center}
\caption{Dependences $\al_\omega(A)$ (a) and $\omega_0(A)$ (b) for the
static one-dimensional AS obtained from the numerical solution of
Eq. (\ref{om1d}); the dependences $\al_\omega(A)$ (c) and $\omega_0$
(d) obtained from the numerical solution of Eq. (\ref{om1d00}) at $A
\sim A_b$. } \label{omega1d}
\end{center}
\end{figure}
Alternatively, for a given value of $\al \sim \ep^2$ the static spike
AS becomes unstable with respect to the pulsations when $A < A_\omega
\lesssim 1$. Note that the onset of the instability of the static
spike AS with respect to the pulsations in the Gierer-Meinhardt model
was studied by Osipov by the multifunctional variational method
\cite{Os93}. 

Equation (\ref{om1d}) can be simplified in the case $A_b \ll A \ll 1$
and $\ep^2 \ll \al \ll 1$. In this case one can neglect the last two
terms in the left-hand side of Eq. (\ref{om1d}), the exponential and
the term $\i \kappa^2 \omega_0$ in its right-hand side, and use the
distribution of $\bar\theta_0$ given by Eq. (\ref{sh1d00}). As a
result, Eq. (\ref{om1d}) can be written in this case as
\begin{eqnarray} \label{om1d0}
\left[ -{d^2 \over dx^2} + 1 + \i \omega_0 - 3 \cosh^{-2} \left( { x
\over 2} \right) \right] \delta \theta = && \nonumber \\ &&
\hspace{-4cm} - \frac{A^2 (1 + \i \omega_0 ) \cosh^{-4} (x/2)}{8
\kappa \sqrt{\i \omega_0}} \int_{-\infty}^{+\infty} \delta
\theta(x') dx'.
\end{eqnarray}
One can see that the dependence on the system's parameters enters this
equation only via the combination $A^2/\kappa = \al^{1/2} A^2 /
\ep$. Solving Eq. (\ref{om1d0}) numerically, we obtain that the AS
destabilizes at
\begin{eqnarray} \label{aom1d}
\al_\omega = 5.15 \times \ep^2 A^{-4}, ~~~\omega_0 = 0.534,~~~{\mathrm
for}~A_b \ll A \ll 1.
\end{eqnarray}
Note that Eq. (\ref{om1d0}) can be rigorously analyzed by the method
similar to the one used for Eq. (\ref{disp1d:stab0}). This analysis is
presented in appendix B. It yields the same conclusions as above and
Eq. (\ref{aom1d}).

The result of Eq. (\ref{aom1d}) is in agreement with the one obtained
in \cite{kaper} by a direct solution of Eq. (\ref{om1d0}). Notice that
this equation gives a good approximation for $\al_\omega$
[Fig. \ref{omega1d}(a)] only for $A \lesssim 0.5$. Recalling that $A
\gg A_b = \sqrt{12 \ep}$ [Eq. (\ref{Ab1d})], we see that
Eq. (\ref{aom1d}) can give a good approximation only for $\ep \lesssim
0.01$ in a limited range of $A$.

According to Eq. (\ref{aom1d}), as $A$ decreases the value of
$\al_\omega$ increases, and when $A$ reaches the value of order $A_b$
we must have $\al_\omega \sim 1$, so for these values of the
parameters one can no longer neglect 1 compared to $\al^{-1} \omega$
in the square roots in Eq. (\ref{disp1d}). On the other hand, in this
case one can still use the same approximations as in deriving
Eqs. (\ref{om1d0}), except we should use Eqs. (\ref{th1dsh0}) and
(\ref{pm1d}) instead of Eq. (\ref{sh1d00}). As a result, for $\al \sim
1$ and $A \sim A_b$ we get asymptotically
\begin{eqnarray} \label{om1d00}
\left[ -{d^2 \over dx^2} + 1 + \i \omega_0 - 3 \cosh^{-2} \left( { x
\over 2} \right) \right] \delta \theta = && \nonumber \\ &&
\hspace{-6cm} - {3 A^2 (1 + \i \omega_0) \over 8 A_b^2 \sqrt{1 + \i
\al^{-1} \omega_0}} \left[ 1 + \sqrt{1 - {A_b^2 \over A^2}} ~\right]^2
\cosh^{-4} \left( { x \over 2} \right) \int_{-\infty}^{+\infty} \delta
\theta(x') dx'.  \nonumber \\
\end{eqnarray}

The results of the numerical solution of this equation for
$\al_\omega$ and $\omega_0$ are presented in Figs. \ref{omega1d}(c)
and (d), respectively. From these figures one can see that for $\al >
\al_0 = 0.33$ the considered instability of the static spike AS cannot
be realized for any value of $A$. This result is in agreement with the
general qualitative theory of ASs \cite{KO89,KO90,KO94}. The numerical
solution of Eq. (\ref{om1d}) also shows that all other instabilities
with ${\mathrm Re~} \omega \not= 0$ occur at significantly lower
values of $\al$, when the AS is already unstable with respect to the
pulsations.

\subsubsection{Case $\al \gg 1$ and $k \not= 0$: instability with
respect to the corrugation} \label{stab:ss4}

Let us now see what happens in higher dimensions when $A \sim 1$ and
$\al$ is sufficiently large [so that the terms proportional to
$\al^{-1}$ in Eq. (\ref{disp1d}) can be dropped] as the value of $k$
is varied. Since the mode $\delta \theta_1$ generally requires a
careful consideration, we will first look only at the modes $\delta
\theta_{0,2}$. If $k \lesssim \ep$, the operator in the right-hand
side of Eq. (\ref{disp1d}) contains a large factor of order $\ep^{-1}$
and the exponential in Eq. (\ref{disp1d}) can be neglected. Therefore,
for these values of $k$ we should either have $\gamma_k \simeq 1$,
what corresponds to the modes at the bottom of the continuous spectrum
(see Sec. \ref{stab:ss1}), or the modes $\delta\theta_k$ should be
orthogonal to $\delta\theta_s$. In the latter case the $k$-dependence
in the right-hand side of Eq. (\ref{disp1d}) becomes inessential, so
for these $k$ the values of $\gamma_k$ will be close to $\gamma_n$
obtained for $k = 0$. Therefore, according to the results of
Sec. \ref{stab:ss2}, neither of the modes $\delta\theta_k$ should be
unstable for $k \lesssim \ep$ for $A_b < A < A_d$.

The situation changes when $\ep \ll k \lesssim 1$. Then, one can
neglect 1 compared to $\ep^{-2} k^2$ in the square roots in
Eq. (\ref{disp1d}) and write it asymptotically as
\begin{eqnarray} 
\left[ -{d^2 \over dx^2} + 1 - \gamma_k + k^2 - 2 A^2 \bar\eta_s
\bar\theta_0 - 2 A^2 \bar\theta_0 \bar\eta_0 + 3 A^2 \bar\theta_0^2
\right] \delta \theta_k = && \nonumber \\ && \hspace{-6cm} - \frac{A^2
\bar\theta_0^2 (1 - \gamma_k) }{2 k} \int_{-\infty}^{+\infty} \e^{-k |x
- x'|} \delta \theta_k(x') dx'. \label{k1d}
\end{eqnarray}
Note that for $k \sim 1$ the coefficient multiplying the right-hand
side of Eq. (\ref{k1d}) becomes of order 1. Here we should expect a
corrugation instability with respect to the mode $\delta \theta_0$
with some $k = k_0 \sim 1$ \cite{KO89,KO90,KO94,Sev}. Indeed, as the
value of $k$ increases, the magnitude of the right-hand side of
Eq. (\ref{disp1d}) decreases as $1/k$, while the contribution to the
left-hand side of Eq. (\ref{disp1d}) increases as $k^2$. Therefore,
for some $k = k_0$ the contribution of both these two terms to
Eq. (\ref{disp1d}) will be minimal, so that we can get an instability:
${\mathrm Re}~\gamma_{k_0} < 0$.

To show that there is indeed an instability at $k \sim 1$, we solved
Eq. (\ref{k1d}) numerically. Figure \ref{gammak} shows the solutions
for ${\mathrm Re}~\gamma_k$ obtained for a particular value of $A =
1.2$.
\begin{figure}
\centerline{\psfig{figure=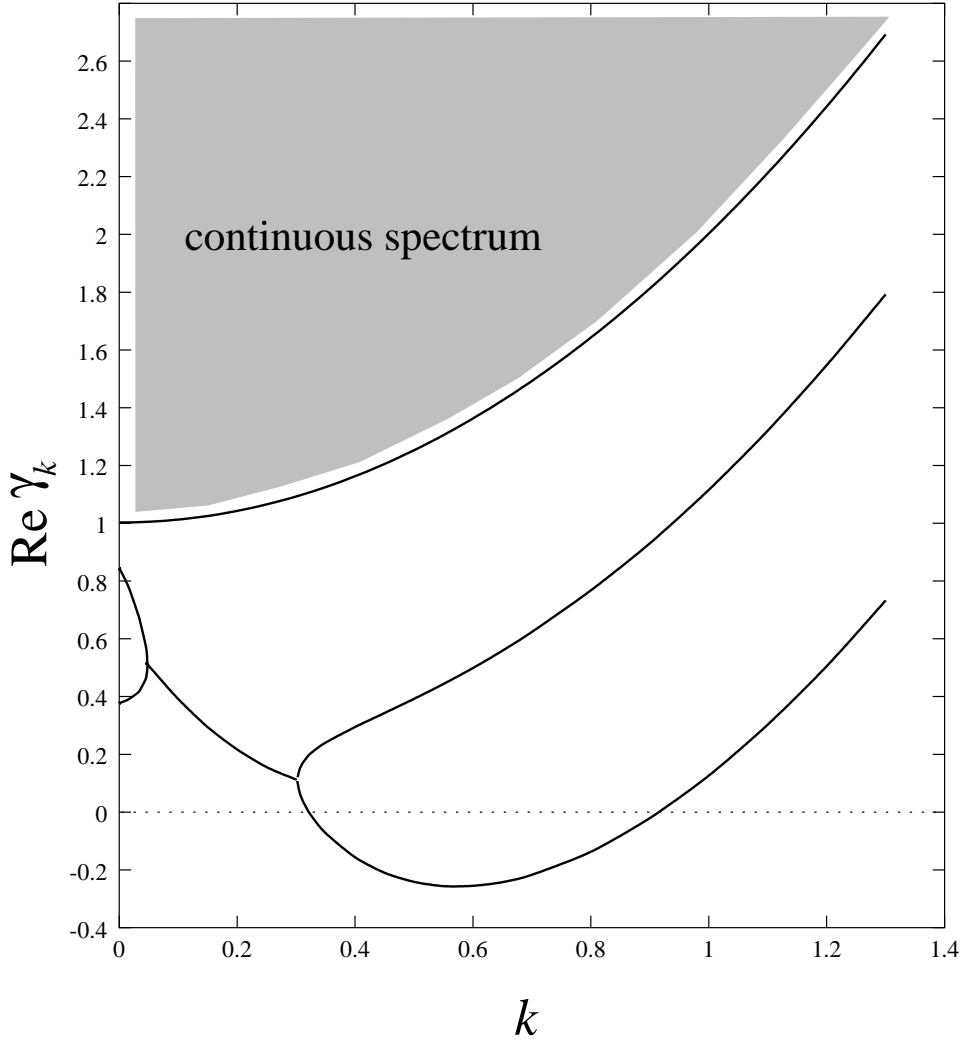}}
\begin{center}
\caption{Damping decrement ${\mathrm Re}~ \gamma_k$ for the static
one-dimensional AS obtained from the numerical solution of
Eq. (\ref{k1d}) for different modes at $A = 1.2$. } \label{gammak}
\end{center}
\end{figure}

For $k > 0.29$ Eq. (\ref{k1d}) has two localized solutions and a
continuous spectrum of $\gamma_k$ for $A = 1.2$, all having
${\mathrm Re}~\omega(k) = 0$. The curve at the bottom of the figure
corresponds to $\delta\theta_0$, the curve in the middle to
$\delta\theta_2$, and the curve on the top is the bottom of the
continuous spectrum. From Fig. \ref{gammak} one can see that the AS is
unstable with respect to $\delta\theta_0$ (the corrugation
instability) for a range of wave vectors $k \sim 1$. The analysis of
$\gamma_k$ for different values of $A$ shows that the static
one-dimensional spike AS in higher dimensions is unstable with respect
to corrugation for all values of $A$ at which it exists.

When $0.05 < k < 0.29$, the complex frequency $\omega(k)$ for the
modes $\delta\theta_{0,2}$ acquires a real part for $A = 1.2$. The
corresponding eigenfunctions for these modes, as well as $\gamma_k$,
are complex-conjugate. For yet smaller values of $k$ the real part of
$\omega$ vanishes once again, so we have two distinct solutions. The
latter is related to the presence of two distinct solutions in the
case of the one-dimensional AS studied in Sec. \ref{stab:ss2}, which
is obtained from Eq. (\ref{k1d}) in the limit $k \rightarrow 0$. When
the value of $A$ is decreased below $A = 0.90$, these two distinct
solutions disappear and the solution with a nonzero real part of
$\omega(k)$ goes all the way to $k = 0$.

Note that in the case $A_b \lesssim A \ll 1$ it can be easily shown
that the AS is unstable with respect to the corrugation
instability. Indeed, for $A \ll 1$ and $k \gg A^2$ the operator in
right-hand side of Eq. (\ref{k1d}) can be considered as a
perturbation. Therefore, in the zeroth approximation we will have
$\delta\theta_0 = \delta\theta_0^{(0)}$, where $\delta\theta_0^{(0)}$
is given by Eq. (\ref{schr0}). Then, in the first order of the
perturbation theory we will have
\begin{eqnarray}
\gamma_k = \left( -{5 \over 4} + {75 \pi^2 A^2 \over 2048 k} +
k^2\right) \left( 1 + {75 \pi^2 A^2 \over 2048 k} \right)^{-1},
\end{eqnarray}
which is negative for $k \lesssim 1$. Notice that according to this
equation the fastest growing mode has $k \simeq 0.60$, what coincides
with very good accuracy with the results of the numerical solution of
Eq. (\ref{k1d}) (see Fig. \ref{gammak}). Similarly, for $A_b \ll A \ll
1$ and $\al \ll \al_\omega$ it is easy to show that the AS will be
unstable. Here we can also treat the right-hand side of
Eq. (\ref{om1d0}) as a perturbation, so once again $\delta\theta_0 =
\delta\theta_0^{(0)}$. The solution of Eq. (\ref{om1d0}) in the first
order of the perturbation theory with $\al / \ep^2 \ll A^4$ and the
exponential set to 1 then gives us $\gamma \simeq - \frac{5}{4} + {27
\pi^2 \sqrt{5} \over 1024} A^2 \al^{1/2} \ep^{-1} < 0$. Notice that by
equating this expression to zero, one gets the value of $\al_\omega$
which differs from the one in Eq. (\ref{aom1d}) by only 10\%.

\subsubsection{Case $\al \gg 1$ and $k \not= 0$: instability with
respect to wriggling} \label{stab:ss5}

Let us now turn to the mode $\delta \theta_1$. Since, because of the
translational invariance Eq. (\ref{disp1d}) is identically satisfied
for $k = 0$ and $\gamma = 0$ with $\delta \theta_1 = d \bar\theta_0 /
dx$, it is of special interest to study its solutions for $|\gamma|
\ll 1$ and $k \ll 1$. The small values of $k$ and $\gamma$ introduce a
weak perturbation to the operators in Eq. (\ref{disp1d}) with $k = 0$
and $\gamma = 0$. Therefore, in the leading order of the perturbation
theory we must take $\delta \theta_1 = d \bar\theta_0 / dx$, and in
order to get $\gamma_k$ we must multiply Eq. (\ref{disp1d}) by the
adjoint function $\delta \theta_1^*$ and integrate over $x$. As a
result, in the first order in $\gamma_k$ and $k^2$ we obtain
\begin{eqnarray} 
&& - \gamma_k + k^2 + \gamma_k \frac{A^2 (1 - \ep^2 \al^{-1})}{2}
\int_{-\infty}^{+\infty} \int_{-\infty}^{+\infty} \bar\theta_0^2 (x)
\delta \theta_1^*(x) |x - x'| \delta \theta_1(x') dx dx' \nonumber \\
&& \quad = {\ep A^2 \over 8 \al} ( \gamma_k - \al \ep^{-2} k^2 )
\int_{-\infty}^{+\infty} \int_{-\infty}^{+\infty} \bar\theta_0^2(x)
\delta \theta_1^*(x) (x - x')^2 \delta \theta_1(x') dx dx',
\label{t1d} 
\end{eqnarray}
where we expanded the exponential in Eq. (\ref{disp1d}) up to the
second order in $\ep$ and used the normalization
$\int_{-\infty}^{+\infty} \delta \theta_1^* \delta \theta_1 dx =
1$. Recalling that $\delta \theta_1 = d \bar\theta_0 / dx$, we can
calculate the integral in the right-hand side of
Eq. (\ref{t1d}). Using the symmetry properties of $\delta \theta_1$,
we find this integral to be $2 \int_{-\infty}^{+\infty} x
\bar\theta_0^2 \delta \theta_1^* dx \int_{-\infty}^{+\infty}
\bar\theta_0 dx < 0$, since for $x < 0$ we have $\delta \theta_1^* >
0$ and vice versa. By a similar integration, one can show that the
integral in the left-hand side of Eq. (\ref{t1d}) is also
negative. Then, it is easy to see that when $\al \gtrsim \ep$
(otherwise the AS would be unstable in one-dimension, see
Sec. \ref{stab:ss6}), for small values of $k$ we will have $\gamma_k
\sim - \ep^{-1} A^2 k^2 < 0$, so the one-dimensional static spike AS
is in fact always unstable with respect to the $\delta \theta_1$ mode
with small $k$ for $A \gg A_b$. These fluctuations should lead to
wriggling of the AS. Moreover, it is possible to show that the static
spike AS which is stable in one dimension is in fact unstable with
respect to wriggling for all values of $A > A_b$. Indeed, for $A \sim
A_b$ the operator in the right-hand side of Eq. (\ref{disp1d}), which
is the only non self-adjoint operator there, can be considered as a
perturbation, so we can assume that $\delta \theta_1 = \delta
\theta_1^* = \delta \theta_1^{(0)}$ and take $\bar\theta_0$ to be
given by Eq. (\ref{th1dsh0}). Substituting this into Eq. (\ref{t1d}),
we obtain that for small values of $k$ the damping decrement
$\gamma_k$ of the fluctuations leading to wriggling is given by
\begin{eqnarray} 
\gamma_k \simeq k^2 \left[ 1 - {A^2 \over A_b^2} \left(1 + \sqrt{1 -
{A_b^2 \over A^2}} ~\right)^2 ~\right].
\end{eqnarray}
From this equation one can see that $\gamma_k < 0$, signifying an
instability, for $A > A_b$ and sufficiently small $k$. Thus, in
summary, the one-dimensional static spike AS is always unstable in
higher dimensions, so the instabilities that are realized for
sufficiently small $\al$ are meaningful only for the one-dimensional
system.

\subsubsection{Case $\al \ll 1$ and $k = 0$: instability with respect
to the onset of the traveling motion} \label{stab:ss6}

In addition to the instability of the AS with respect to wriggling,
another instability may be realized when $\al \ll 1$ \cite{Sev}. As
was already mentioned above, for $k = 0$ Eq. (\ref{t1d}) has a simple
zero solution for any $\al$ due to the translational
invariance. However, at some special value of $\al = \al_T$ the
solution $\gamma = 0$ may become doubly degenerate. This will happen
when the coefficient in front of $\gamma$ in Eq. (\ref{t1d})
vanishes. If this is the case, in addition to the trivial solution we
will have another non-trivial zero solution, for which the value of
$\gamma$ should {\em change sign} as the value of $\al$ passes through
$\al_T$. This signifies an instability that leads to the onset of the
traveling motion. It is not difficult to see that this instability
will occur when $\al < \al_T \sim \ep$ for $A \lesssim 1$. In
particular, when $A \ll 1$, we can put $\delta \theta_1^* = \delta
\theta_1 = \delta \theta_1^{(0)}$, since again the only non
self-adjoint operator is the operator in the right-hand side of
Eq. (\ref{disp1d}) and is small. After a little algebra, we obtain
that the instability will occur at
\begin{eqnarray} \label{at1d}
\al_T = \left\{
\begin{array}{lc}
{\ep^2 A^2 \over A_b^2} \left[ 1 + \sqrt{1 - {A_b^2 \over A^2} }
~\right]^2, & ~~~A \sim A_b, \\ \frac{1}{3} A^2 \ep, & ~~~A_b \ll A
\ll 1.
\end{array}
\right.
\end{eqnarray}
Note that the last formula is in agreement with the results of
Sec. \ref{trav:ss5}.  To analyze the dependence $\al_T(A)$ for $A \sim
1$ we solved for the conjugate function $\delta \theta_1^*$
numerically and then substituted it to Eq. (\ref{t1d}). The resulting
dependence $\al_T(A)$ is presented in Fig. \ref{travinst}.
\begin{figure}
\centerline{\psfig{figure=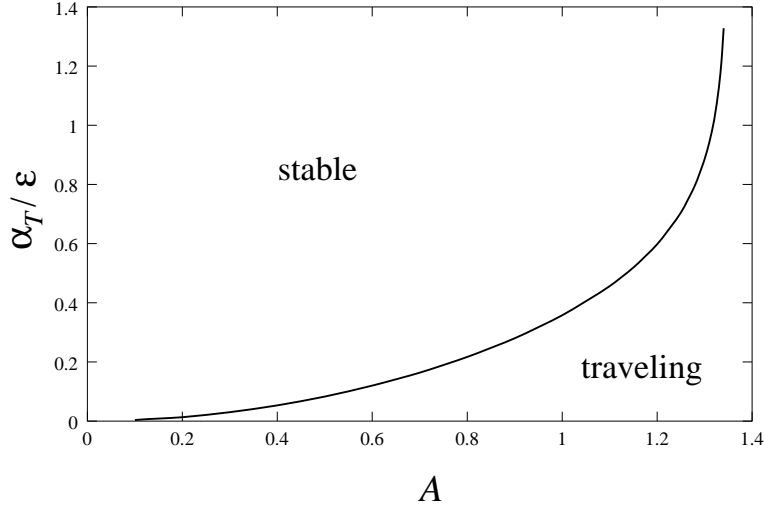,width=10cm}}
\begin{center}
\caption{Dependence $\al_T(A)$ for the static one-dimensional AS
obtained from the numerical solution of Eq. (\ref{t1d}). }
\label{travinst}
\end{center}
\end{figure}
Observe that for $A < 1$ the value of $\al_T$ is given by
Eq. (\ref{at1d}) with an accuracy better than 10\%.

\subsubsection{Comparison of the pulsation and the traveling
instabilities} \label{stab:ss7}

According to Eqs. (\ref{aom1d}) and (\ref{at1d}), for $A > 1.58
\ep^{1/6}$ we have $\al_T > \al_\omega$, so if one starts with the
static spike AS at $A \sim 1$ and sufficiently large value of $\al$
and then gradually decreases $\al$, the AS will destabilize with
respect to the fluctuation $\delta\theta_1$ and will transform into
traveling. If, on the other hand, we have $A < 1.58 \ep^{1/6}$ at the
start, the AS will destabilize with respect to the pulsations and will
collapse after a few oscillations of its amplitude (see
Sec. \ref{s:pf1d}). Also, according to Eqs. (\ref{aom1d}) and
(\ref{at1d}), for $\al < \al_c = 0.83 \ep^{4/3}$ the one-dimensional
static spike AS will be unstable regardless of the value of $A$, so we
conclude that this AS can be excited only when $\al > \al_c$.

\subsection{Stability of the three-dimensional radially-symmetric
static spike autosoliton} \label{stab:3d}

So far we were studying the one-dimensional static spike AS in higher
dimensions. Let us now turn to the radially-symmetric spike ASs. It
turns out that up to the logarithmic terms, all the results on the
stability of the two- and the three-dimensional radially-symmetric
static spike ASs have the same dependence on $\ep$, so we will
concentrate mostly on the three-dimensional AS. For the
three-dimensional radially-symmetric static spike AS the small
deviations $\delta \theta = \theta - \theta_0$ and $\delta \eta = \eta
- \eta_0$ in the spherical coordinates can be taken as
\begin{eqnarray}
\delta \theta = \e^{\i \omega t} Y_{nm} (\vartheta, \varphi) \delta
\theta_{n\omega}(r), ~~~ \delta \eta = \e^{\i \omega t} Y_{nm}
(\vartheta, \varphi) \delta \eta_{n\omega}(r),
\end{eqnarray}
where $Y_{nm}$ are the spherical harmonics. Then, Eqs. (\ref{act}) and
(\ref{inh}) linearized around $\theta_0$ and $\eta_0$ become
\begin{eqnarray}
\left[ -{d^2 \over dr^2} - {2 \over r} {d \over dr} + {n (n + 1) \over
r^2} + 1 + \i \omega - 2 \tilde{A} \tilde\theta_0 \tilde\eta_0 \right]
\delta \theta_{n\omega} & = & \ep^{-1} \tilde{A} \tilde\theta_0^2
\delta \eta_{n\omega}, \label{lin3da} \\ \left[ - {d^2 \over dr^2} -
{2 \over r} {d \over dr} + {n (n + 1) \over r^2} + \ep^2 + \i \ep^2
\al^{-1} \omega \right] \delta \eta_{n\omega} && \nonumber \\ &&
\hspace{-7cm} = - \ep \tilde{A}^{-1} \left[ - {d^2 \over dr^2} - {2
\over r} {d \over dr} + {n (n + 1) \over r^2} + 1 + \i \omega \right]
\delta \theta_{n\omega}, \label{lin3db}
\end{eqnarray}
where, once again, length and time are measured in the units of $l$
and $\tau_\theta$, respectively, and we used the scaled variables
given by Eq. (\ref{scale3d}). Equation (\ref{lin3db}) can be solved by
means of the Green's function, which in the case of the operator in
the left-hand side of Eq. (\ref{lin3db}) multiplied by $r^2$ is
(similar Green's function was used in \cite{Mur96a,Mur96b})
\begin{eqnarray} \label{G3d}
G_{n\omega} (r, r') = \left\{
\begin{array}{ll}
{I_{n + 1/2} (\ep r \sqrt{1 + \i \al^{-1} \omega} ) K_{n + 1/2} (\ep r'
\sqrt{1 + \i \al^{-1} \omega} ) \over \sqrt{r r'} }, & ~~~r \leq r', \\
{I_{n + 1/2} (\ep r' \sqrt{1 + \i \al^{-1} \omega} ) K_{n + 1/2} (\ep r
\sqrt{1 + \i \al^{-1} \omega} ) \over \sqrt{r r'} }, & ~~~r \geq r',
\end{array}
\right.
\end{eqnarray}
where $I_{n+1/2}(x)$ and $K_{n+1/2}(x)$ are the modified Bessel
functions. Solving Eq. (\ref{lin3db}) with the use of Eq. (\ref{G3d}),
we obtain
\begin{eqnarray}
\delta \eta_{n\omega} = - \ep \tilde{A}^{-1} \delta \theta_{n\omega} -
\ep \tilde{A}^{-1} (1 + \i \omega - \i \ep^2 \al^{-1} \omega)
\int_0^\infty G_{n\omega} (r, r') \delta \theta_{n\omega} (r') r'^2 d
r'. && \nonumber \\
\end{eqnarray}
Substituting this expression into Eq. (\ref{lin3da}), we obtain the
following equation
\begin{eqnarray}
\left[ -{d^2 \over dr^2} - {2 \over r} {d \over dr} + {n (n + 1) \over
r^2} + 1 + \i \omega - 2 \tilde{A} \tilde\theta_0 \tilde\eta_0 +
\tilde\theta_0^2 \right] \delta \theta_{n\omega} && \nonumber \\ &&
\hspace{-8cm} = - \tilde\theta_0^2 (1 + \i \omega - \i \ep^2 \al^{-1}
\omega) \int_0^\infty G_{n\omega} (r, r') \delta \theta_{n\omega} (r')
r'^2 d r'.
\label{disp3d}
\end{eqnarray}
This equation determines the complex frequencies of different
fluctuations as the functions of the control parameters and has to be
solved as an eigenvalue problem. The instability of the AS will occur
when ${\mathrm Re}~\gamma < 0$, with $\gamma = - \i \omega$.

\subsubsection{Case $\al \gg \ep^2$: instability with respect to the non
radially-symmetric fluctuations} \label{stab:ss8}

Let us first look at Eq. (\ref{disp3d}) at $\al \gg \ep^2$. In this
case the terms proportional to $\al^{-1}$ in the right-hand side of
Eq. (\ref{disp3d}) can be neglected, so the Green's function
$G_{n\omega}$ can be expanded in $\ep$. Then Eq. (\ref{disp3d})
becomes asymptotically
\begin{eqnarray}
\left[ -{d^2 \over dr^2} - {2 \over r} {d \over dr} + {n (n + 1)
\over r^2} + 1 - \gamma_n - 2 \tilde{A} \tilde\theta_0 \tilde\eta_0 +
\tilde\theta_0^2 \right] && \delta \theta_n \nonumber \\ &&
\hspace{-5cm} = - {\tilde\theta_0^2(r) (1 - \gamma_n) \over 2 n + 1}
\int_0^\infty g_n (r, r') \delta \theta_n (r') r' d
r',
\label{disp3d0}
\end{eqnarray}
where
\begin{eqnarray}
g_n(r, r') = \left\{
\begin{array}{ll}
(r / r')^{n}, & r \leq r', \\
(r'/r)^{n+1}, & r \geq r'.
\end{array}
\right.
\end{eqnarray}
The operator in the left-hand side of Eq. (\ref{disp3d0}) is a
Schr\"odinger type operator with the attractive potential $- 2 A
\tilde\theta_0 \tilde\eta_0$, repulsive potential $\tilde\theta_0^2$
and the centrifugal potential $n (n + 1)/r^2$.

As in the case of the one-dimensional AS, the modes $\delta\theta_n$
that can lead to instabilities are localized. From the translational
symmetry of the problem we know that there is a localized solution
$\delta\theta_1 = d \tilde\theta_0 / dr$ for $n = 1$ which corresponds
to $\gamma = 0$. It is clear that since the potential in the left-hand
side of Eq. (\ref{disp3d0}) is able to localize a state at $n = 1$ in
the presence of the centrifugal potential, it will also be able to
localize a state at $n = 0$ in its absence. According to
Sec. \ref{stat:3d}, we should in fact have an instability at the point
$\tilde A = \tilde A_b = 5.8$ where the solution in the form of the
static three-dimensional radially-symmetric AS disappears. Also, it is
natural to expect that the operator in the left-hand side of
Eq. (\ref{disp3d0}) will also be able to have a bound state for
sufficiently small $n$ and sufficiently large $\tilde{A}$, when the
role of the centrifugal potential decreases (see below). These will be
the dangerous modes that we need to consider.

It is not difficult to show that for $\tilde A \gg 1$ the AS will be
unstable with respect to the fluctuations with $n \not= 0$. Indeed,
according to the results of Sec. \ref{stat:3d}, for these values of
$\tilde A$ the AS has the form of an annulus of large radius $R \sim
\tilde A^2$ [see Eq. (\ref{fp3d})]. According to Eqs. (\ref{ann3d})
and (\ref{etaann3d}), the leading contribution to the potential in the
left-hand side of Eq. (\ref{disp3d0}) is $ V(r) \cong -3 \cosh^{-2}
\left( {r - R \over 2} \right)$. For $R \gg 1$ and $1 \ll n \ll R$ all
the other terms can be considered as a perturbation to the
Schr\"odinger operator with this potential, so in the first order of
the perturbation theory $\delta \theta_0 = \delta \theta_0^{(0)}(r -
R)$, where $\delta\theta_0^{(0)}$ is given by Eq. (\ref{schr0}). In
the first order of the perturbation theory we obtain
\begin{eqnarray} \label{gammal}
\gamma_n \simeq \left( - \frac{5}{4} + {n^2 \over R^2} + {225 \pi^2
\over 2048 n} \right) \left(1 + {225 \pi^2 \over 2048 n} \right)^{-1}.
\end{eqnarray}
This equation shows that for $R \gg 1$ there is a range of values of
$n$ for which $\gamma_n < 0$. The fact that the annulus must be
unstable with respect to the shape deformations is an obvious
consequence of its quasi one-dimensional character (see
Sec. \ref{stab:ss4}).

The presence of the localized solution $\delta \theta_n$ of
Eq. (\ref{disp3d0}) with $\gamma_n = 0$ for $n > 0$ signifies an
instability of the radially-symmetric static AS with respect to the
non radially-symmetric fluctuations resulting in the distortions of
the spike and leading to its splitting (see Sec. \ref{s:pf2d}). It is
clear that this instability will occur easier at smaller values of
$n$, so we have to solve Eq. (\ref{disp3d0}) for $n = 2$ (the case $n
= 1$ corresponds to the translation of the AS as a whole without any
deformation). The numerical solution of Eq. (\ref{disp3d0}) shows that
the AS indeed becomes unstable with respect to the fluctuation with $n
= 2$ when $\tilde{A} > \tilde{A}_{c2} = 8.4$. In general, the
numerical solution of Eq. (\ref{disp3d0}) shows that the static
radially-symmetric spike AS is stable only in the narrow range
$\tilde{A}_b < \tilde{A} < \tilde{A}_{c2}$. It is interesting to note
that Eq. (\ref{gammal}) with $n = 2$, together with Eq. (\ref{fp3d}),
gives the correct value of $A_{c2}$ with accuracy better than $5 \%$.

When the value of $\tilde A$ is increased beyond $A_{c2}$, the buildup
of the instability with $n = 2$ will result in the splitting of the
AS. After such a splitting event, the two newborn ASs will go apart
until they are separated by a sufficiently large distance, and then
will split again. Thus, the considered instability will result in
self-replication of ASs (see Sec. \ref{s:pf2d}). We would like to
emphasize that the character of the instability that leads to
self-replication in three-dimensional (and two-dimensional) systems is
significantly different from that in one dimension. In the former the
instability results in the buildup of the {\em shape} distortion that
eventually leads to splitting, while in the latter the instability
leads to the widening of the activator distribution profile and {\em
local breakdown} in the AS center (see Sec. \ref{s:pf1d}). Thus,
self-replication of the AS in one dimension is qualitatively different
from that in higher dimensions.

\subsubsection{Case $\al \lesssim \ep^2$: instability with respect to
the pulsations and the onset of the traveling motion} \label{stab:ss9}

According to the general qualitative theory of ASs, when $\al$ becomes
small, the AS may become unstable with respect to the pulsations with
${\mathrm Re}~\omega \not= 0$. The analysis of Eq. (\ref{disp3d}) with
$n = 0$ shows that this instability may be realized only when $\al
\sim \ep^2$ and ${\mathrm Re}~\omega = \omega_0 \sim 1$. In view of
this fact, we can drop 1 in the square roots in
Eq. (\ref{G3d}). Equation (\ref{disp3d}) can then be solved
numerically. The results of this numerical solution are presented in
Fig. \ref{omega3d}.
\begin{figure}
\centerline{\psfig{figure=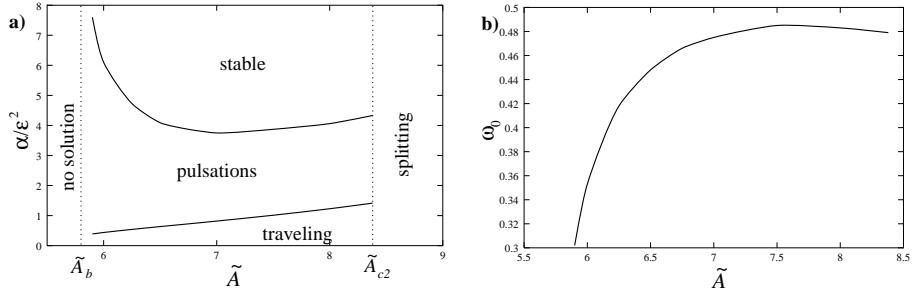,width=12cm,angle=-90}}
\begin{center}
\caption{ (a) Stability diagram for the three-dimensional
radially-symmetric AS. (b) Frequency $\omega_0$ at the threshold of
the pulsation instability. Results of the numerical solution of
Eq. (\ref{disp3d}). In (a) the upper solid line shows the values of
$\al_\omega / \ep^2$ corresponding to the onset of the pulsation
instability ($n = 0$) , the lower solid curve shows the values of
$\al_T / \ep^2$ corresponding to the instability leading to the onset
of the traveling motion ($n = 1$). The dotted lines correspond to the
instabilities of the static AS at $\tilde A = \tilde A_b$ and $\tilde
A = \tilde A_{c2}$. }
\label{omega3d}
\end{center}
\end{figure}
The upper curve in Fig. \ref{omega3d}(a) shows the critical values of
$\al_\omega$ for the onset of this instability for those values of
$\tilde A$ at which the AS is stable at $\al \gg \ep^2$. Figure
\ref{omega3d}(b) shows the frequency $\omega_0$ of the fluctuations at
the threshold of the instability. From Fig. \ref{omega3d}(a) one can
see that the static three-dimensional AS is unstable for all values of
$\tilde A$ if $\al < \al_c \simeq 3.7 \ep^2$.

As was already noted, in addition to the modes studied above, we
always have a dangerous mode $\delta \theta = d \tilde\theta / dr$
with $n = 1$. The analysis of Eq. (\ref{disp3d}) shows that in
addition to the trivial solution, at some $\al = \al_T \sim \ep^2$
this equation can have another solution with ${\mathrm Re}~\gamma =
0$. As in the case of the one-dimensional AS, this solution will
signify the instability which results in the AS starting to move as a
whole. To find this instability point, we need to know the behavior of
the Green's function $G_{1\omega}(r,r')$ at small values of $\i
\omega$. Expanding the Bessel functions in Eq. (\ref{G3d}) and finding
the adjoint function $\delta\theta_1^*$ numerically, we calculate the
coefficient in front of $\i \omega$ in the first order of the
perturbation theory, keeping only the leading terms in $\ep$. The
above mentioned instability will be realized when this coefficient
vanishes. The lower solid line in Fig. \ref{omega3d}(a) shows the
values of $\alpha_T/ \ep^2$ at which this happens. One can see that
this instability occurs when the AS is already unstable with respect
to the pulsations. Note that the numerical solution of
Eq. (\ref{disp3d}) shows that the instabilities with respect to the
fluctuations with $n \geq 2$ always happen for significantly lower
values of $\al$.

\subsection{Stability of the two-dimensional radially-symmetric static
spike autosoliton} \label{stab:2d}

Finally, we will briefly discuss the stability of the two-dimensional
static spike AS in the limit $\ep \rightarrow 0$. In the cylindrical
coordinates $r, \varphi$ the small deviations $\delta\theta = \theta -
\theta_0$ and $\delta\eta = \eta - \eta_0$ can be taken as
\begin{eqnarray}
\delta\theta = \delta\theta_{n\omega} (r) \e^{\i \omega t - \i n
\varphi}, ~~~\delta\eta = \delta\eta_{n\omega} (r) \e^{\i \omega t -
\i n \varphi},
\end{eqnarray}
where $n$ is an integer. 

The equation for $\delta\theta_{n\omega}$ is obtained by linearizing
Eqs. (\ref{act}) and (\ref{inh}) around $\theta_0$ and $\eta_0$ and
eliminating $\delta\eta_{n\omega}$ by inverting the equation for
$\delta\eta_{n\omega}$. As a result, using the scaled $\tilde\theta_0$
and $\tilde\eta_0$ given by Eq. (\ref{tilde2d}), we arrive at the
following equation for $\delta\theta_{n\omega}$
\begin{eqnarray} 
\left[ -{d^2 \over dr^2} - {1 \over r} {d \over dr} + {n^2 \over r^2}
+ 1 + \i \omega - 2 \tilde{A} \tilde\theta_0 \tilde\eta_0 +
\tilde\theta_0^2 \right] \delta \theta_{n\omega} && \nonumber \\ &&
\hspace{-7cm} = - \tilde\theta_0^2 (1 + \i \omega - \i \ep^2 \al^{-1}
\omega) \int_0^\infty G_{n\omega} (r, r') \delta \theta_{n\omega} (r')
r' d r',
\label{disp2d}
\end{eqnarray}
where $G_n(r, r')$ is given by
\begin{eqnarray}
G_n(r, r') = \left\{ 
\begin{array}{ll}
I_n(\ep r \sqrt{1 + \i \al^{-1} \omega} ) K_n(\ep r' \sqrt{1 + i
\al^{-1} \omega} ), & ~~~r < r', \\ I_n(\ep r' \sqrt{1 + \i \al^{-1}
\omega} ) K_n(\ep r \sqrt{1 + \i \al^{-1} \omega} ), & ~~~r > r',
\end{array}
\right.
\end{eqnarray}
where $I_n$ and $K_n$ are the modified Bessel functions. The analysis
of this equation can be performed in the way completely analogous to
that of Eq. (\ref{disp3d}). It is clear that the instabilities of the
two-dimensional static spike AS will be qualitatively the same as
those of the three-dimensional AS. We would like to emphasize that as
in the case of the three-dimensional AS, splitting and
self-replication of the static spike AS in the two-dimensional system
is related to the buildup of the fluctuation with $n = 2$ describing a
non-symmetric distortion of the AS. Because of the very slow
convergence of the asymptotic theory in two dimensions (recall that
the small parameter here is $1/\ln \ep^{-1}$) we will not present a
detailed study of Eq. (\ref{disp2d}).

\section{Pattern formation scenarios in one dimension} \label{s:pf1d}

In the following two Sections we present the results of the numerical
simulations of the Gray-Scott model for sufficiently small $\ep$ and $\al$
and compare them with the results of our asymptotic analysis.

A word needs to be said about the numerical methods we used in various
parts of the paper. Whenever it was not said otherwise, length and
time were measured in the units of $L$ and $\tau_\eta$,
respectively. To simulate the original reaction-diffusion equations we
used a simple explicit second-order scheme both in one and two
dimensions. We used neutral boundary conditions in all our
simulations. In order to resolve the details of the shape of the
spike, sufficiently small spatial discretization step was needed. It
was found that the step $\Delta x = 0.25 l$ gave the solutions with
accuracy of a few per cent. The stiffness of equations at small $\ep$
or $\al$ makes the simulations very time consuming, so two-dimensional
simulations were done on the massively parallel supercomputer
(SGI-Cray Origin 2000). A typical running time for the simulations
shown in the paper is about 1 hour on 16 processors.

In addition to the direct simulations of the original
reaction-diffusion equations, we used various methods in solving the
equations obtained from the asymptotic procedures. In solving for the
sharp distributions in the case of the static spike ASs we used
relaxation methods with second-order discretization. In solving the
nonlinear eigenvalue problems we used a combination of the relaxation
method and iterative method. In the case of the traveling spike ASs we
used shooting method to solve the obtained ODEs. In analyzing the
stability of the static spike ASs we used second-order discretization
of the linear operators in a sufficiently large finite domain and then
diagonalized the obtained matrices. The solution of the ODEs and the
diagonalization were performed using {\sf Mathematica 3.0}. In all the
cases the accuracy of the numerical solutions is better than a few per
cent.

\subsection{Properties of the static spike autosoliton}

In Sec. \ref{stat:1d} we found that the static spike ASs can form in
the Gray-Scott model when $\ep \ll 1$. In one dimension these AS are
stable when $\al > \al_0 = 0.33$ in the whole region of its existence
$A_b < A < A_d$. Our numerical simulations of Eqs. (\ref{act}) and
(\ref{inh}) show that the value of $A_b$ is given by Eq. (\ref{Ab1d})
with very good accuracy (less than 1\%) already for $\ep < 0.1$, while
the value of $A_d$ approaches its asymptotic value $A_d = 1.35$ for
$\ep \lesssim 0.01$ and increases somewhat for larger $\ep$ (for
example, for $\ep = 0.05$ we found $A_d = 1.48$). These results are
robust against decreasing $\ep$ from 0.01 to 0.001 and smaller. For
$A_b < A < 1$ the distributions of $\theta$ and $\eta$ are described
by Eqs. (\ref{th1dsh0}) and (\ref{esh1d0}) with accuracy better than
10\% for $\ep \lesssim 0.1$.

The numerical solution of Eqs. (\ref{act}) and (\ref{inh}) for $\ep =
0.05$ and $A = 1.4$ in the form of the one-dimensional static spike AS
is presented in Fig. \ref{as1d}.
\begin{figure}
\centerline{\psfig{figure=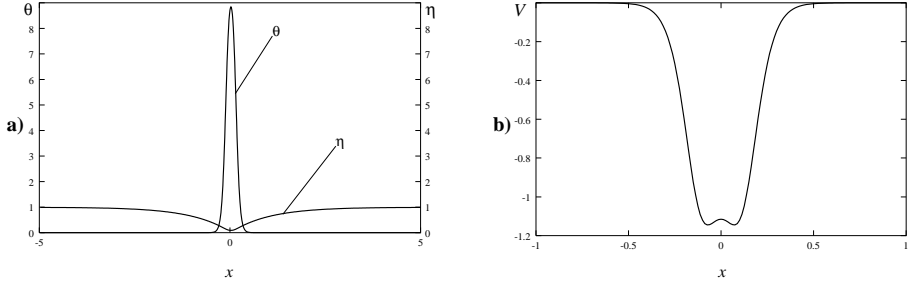,width=12cm,angle=-90}}
\begin{center}
\caption{(a) Distributions of $\theta$ and $\eta$ in a static spike
one-dimensional AS; (b) Potential $V$ from
Eq. (\ref{neigen1db}). Results of the numerical solution of
Eqs. (\ref{act}) and (\ref{inh}) with $\ep = 0.05$, $\al = 1$, and $A
= 1.4$. } \label{as1d}
\end{center}
\end{figure}
The shape of the distributions of $\theta$ and $\eta$ remains
qualitatively the same all the way up to $A_d = 1.48$, but when $A$
approaches $A_d$ the spike widens somewhat and its maximum becomes
flatter. One can see that the potential $V$ from the nonlinear
eigenvalue problem [see Eq. (\ref{neigen1db})] shown in
Fig. \ref{as1d}(b) indeed assumes a complicated shape predicted by the
asymptotic theory (Sec. \ref{s:stat}). According to the asymptotic
theory, the value of $\theta_{\mathrm max}$ should decrease with the
increase of $A$ for $A > 1.2$ (see Fig. \ref{ett1d}), and at $A > A_d$
the solution in the form of a single spike should disappear. This can
also be seen from the simulations of the one-dimensional Gray-Scott
model with $\ep = 0.05$ (Fig. \ref{ampl}).
\begin{figure}
\centerline{\psfig{figure=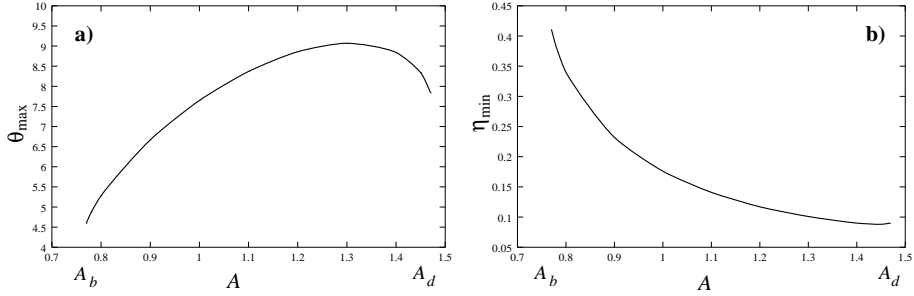,width=12cm,angle=-90}}
\begin{center}
\caption{Dependences $\theta_{\mathrm max}$ and $\eta_{\mathrm min}$
in the one-dimensional static spike AS obtained from the numerical
solution of Eqs. (\ref{act}) and (\ref{inh}) with $\ep = 0.05$. }
\label{ampl}
\end{center}
\end{figure}
Observe the similarity between Fig. \ref{ampl} and Fig. \ref{ett1d}
obtained from the asymptotic theory. Also observe that the values of
$\theta_{\mathrm max}$ for $A > 1$ are smaller for finite values of
$\ep$ than those predicted by the asymptotic theory. When the value of
$A$ becomes greater than $A_d$, a dip in the distribution of $\theta$
appears in the center of the spike followed by a rapid decrease of the
value of $\theta$ there (local breakdown) resulting in the AS
splitting into two spikes. These spikes go away from each other and
then split again, until the system becomes filled with a periodic
pattern of spikes. This process of self-replication of the static
spike ASs in the Gray-Scott model was studied in detail in
\cite{Reyn,nu}.

From the physical point of view, the existence of the static AS in
which for $\ep = l / L \ll 1$ the distribution of $\theta = X / X_0$
has the form of a spike [see Fig. \ref{as1d}(a)] is determined by the
diffusion processes and the special features of the autocatalytic
reaction in Eq. (\ref{reacta}) (see Sec. \ref{s:mod}). Due to this
reaction, the self-production of substance $X$ occurs accompanied by
the decrease of the amount of substance $Y$. The rate of this reaction
is $R \sim X^2 Y$, so seemingly the concentration of substance $Y$
should rapidly decrease. This, however, does not occur, and the high
rate of the reaction in the spike is maintained by the strong
diffusion influx of substance $Y$ from the neighboring regions of size
of order $L$. This ensures a finite concentration of substance $Y$ in
the spike which is necessary to maintain a high rate of the
self-production of substance $X$. The diffusion current of substance
$Y$ is roughly $Y_0 / L$, that is, it increases with $A \sim Y_0$ [see
Eq. (\ref{y0})], but the reaction rate $R$ grows as $X^2 Y$, so due to
the increase of $R$ with the increase of $X \sim A \sim Y_0$ the
concentration of $Y \sim A^{-2} \sim Y_0^{-2}$ [see Eqs. (\ref{pm1d})]
decreases faster in the spike as the value of $Y_0$ is
increased. Therefore, for large enough $Y_0$ the diffusion current
cannot maintain high reaction rate in the spike. This explains why for
large enough $A \sim Y_0$ the value of $\theta \sim X$ starts
decreasing with the increase of $A$ and at some $A = A_d$ the solution
in the form of the static spike AS disappears.

\subsection{Formation and collapse of the pulsating autosoliton}

In Sec. \ref{s:stab} we found that for $\al < 0.33$ the static spike
AS in the one-dimensional Gray-Scott model may become unstable with
respect to the pulsations, with the critical value of $\al$ rapidly
decreasing with the increase of $A$ [see Eq. (\ref{aom1d})]. Our
numerical simulations showed that in a sufficiently large system this
instability leads to the growth of self-oscillations of the AS
amplitude. In other words, this instability leads to the
transformation of the static AS into pulsating
\cite{KO89,KO90,KO94}. Figure \ref{puls} presents the results of the
simulation of such a process for $\ep = 0.05$, $\al = 0.024$, and $A =
1.2$.
\begin{figure}
\centerline{\psfig{figure=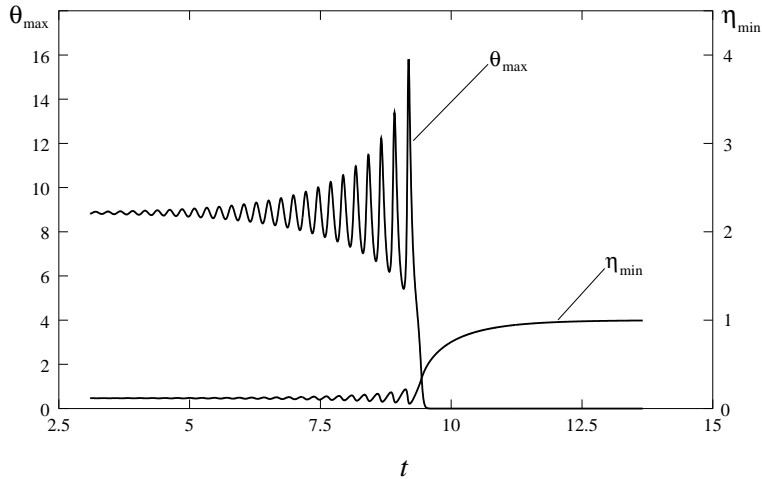,width=10cm}}
\begin{center}
\caption{ Dependences $\theta_{\mathrm max}(t)$ and
$\eta_{\mathrm min}(t)$ for the static spike AS at the onset of
the pulsations. Results of the numerical solution of Eqs. (\ref{act}) and
(\ref{inh}) for $\ep = 0.05$, $\al = 0.024$ and $A = 1.2$ }
\label{puls}
\end{center}
\end{figure}
The value of $\al_\omega = 0.0255$ found in the simulations for these
parameters agrees with the results of Sec. \ref{s:stab} with accuracy
better than 5\% (see Fig. \ref{omega1d}). Also, the critical frequency
agrees with the results of Sec. \ref{s:stab} within 1\%. Note that for
these values of $A$ the value of $\al_\omega$ given by
Eq. (\ref{aom1d}) (which is also predicted in \cite{kaper}) turns out
to be off by a factor of 4. According to the results of the
simulations, the pulsating AS seems to be unstable for all values of
the parameters. After a few oscillations of its amplitude, it
collapses into the homogeneous state (Fig. \ref{puls}).

\subsection{Properties of the traveling spike autosolitons}

The condition $\al \ll 1$ means that the inhibitor $Y$ is much slower
than the activator $X$. The appearance of pulsation instability is due
to the fact that because of its sluggishness the inhibitor cannot
control rapidly varying fluctuations of the activator. The lag in the
inhibitor reaction also determines the existence of the traveling
spike AS in the Gray-Scott model at $\al \ll 1$ and $\ep \gtrsim
\al^{1/2}$. The supply of the substance $Y$ into the spike of the
traveling AS which is necessary to maintain high reaction rate $R$
occurs by the AS running on the regions with high values of $Y$. The
situation here resembles combustion process. The numerical simulations
of Eqs. (\ref{act}) and (\ref{inh}) in one dimension confirm the
conclusions of Sec. \ref{s:trav} about the existence of the traveling
spike ASs. Figure \ref{trav}(a) shows the distributions of $\theta$
and $\eta$ in the form of an ultrafast traveling spike AS for $\ep =
\infty$, $\al = 0.05$, and $A = 2$.
\begin{figure}
\centerline{\psfig{figure=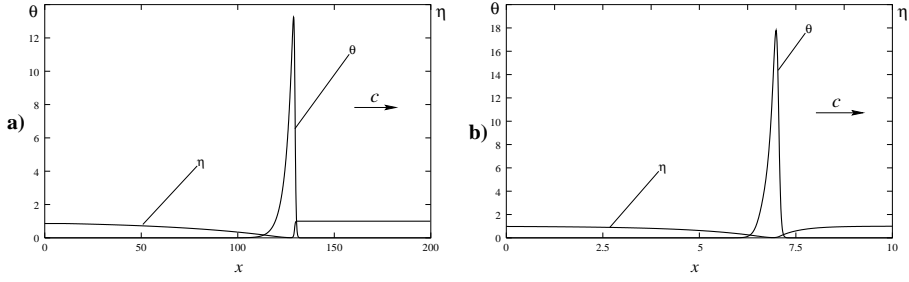,width=12cm,angle=-90}}
\begin{center}
\caption{ Two types of traveling spike AS: case $\ep \gg \al^{1/2}$
(a) and $\ep \sim \al$ (b). Results of the numerical solution of
Eqs. (\ref{act0}) and (\ref{inh0}). In (a) $L = 0$, $\al = 0.05$, $A =
2$, length is measured in the units of $l$. In (b) $\ep = 0.05$, $\al
= 0.05$, $A = 1.2$, length is measured in the units of $L$. }
\label{trav}
\end{center}
\end{figure}
The speed of this AS was found to be $c = 7.3$, which agrees within
5\% with that of Eq. (\ref{ultrac}). Also, the shape of the traveling
spike AS is in agreement with the predictions of Sec. \ref{trav:s1}.

When the value of $\ep$ becomes smaller, or, more precisely, the value
of $L$ becomes sufficiently large, the concentration of substance $Y$
in the front of the traveling spike decreases due to diffusion, so the
speed of the AS decreases. This explains the possibility of the
existence of two types of the traveling spike ASs studied in
Sec. \ref{s:trav}: the ultrafast traveling spike AS, which forms at
$\ep \gtrsim \al^{1/2}$, and the slower traveling spike AS, which
forms at $\ep^2 \lesssim \al \lesssim \ep$. The numerical simulations
confirm this statement. Figure \ref{trav}(b) shows the distributions
of $\theta$ and $\eta$ in the form of a slower traveling spike AS
obtained from the numerical solution of Eqs. (\ref{act}) and
(\ref{inh}) for $\ep = 0.05$, $\al = 0.05$ and $A = 2$. The speed of
this AS was found to be $c = 1.26$, much smaller than the speed of the
ultrafast traveling AS discussed in the preceding paragraph. The shape
of the slower traveling spike AS agrees with that found in the
asymptotic theory (see Sec. \ref{trav:s2}). Figure \ref{cnum} shows
the dependence of the AS speed $c$ on $A$ at different values of $\al$
obtained from the numerical solution of Eqs. (\ref{act}) and
(\ref{inh}) with $\ep = 0.05$.
\begin{figure}
\centerline{\psfig{figure=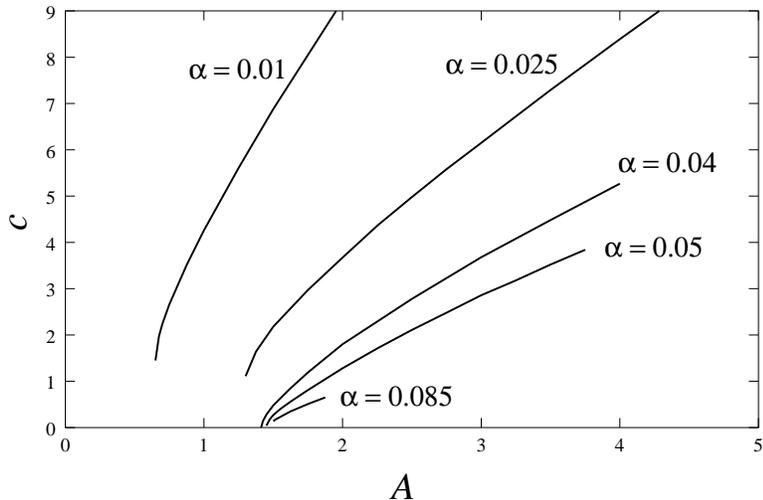,width=10cm}}
\begin{center}
\caption{ Dependences $c(A)$ for the traveling spike AS at different
values of $\al$ obtained from the numerical solution of
Eqs. (\ref{act}) and (\ref{inh}) with $\ep = 0.05$.}
\label{cnum}
\end{center}
\end{figure}
Note that the curves $c(A)$ in Fig. \ref{cnum} terminate at
sufficiently large values of $A$. This is due to the fact that when
the value of $A$ exceeds a certain critical value which depends on
$\al$, the traveling AS starts splitting as it moves. This is in
agreement with the predictions of Sec. \ref{trav:ss7} about the
disappearance of the solution in the form of the traveling AS due to
the onset of the oscillations in the back of the spike for
sufficiently large values of $A$.

Two ultrafast traveling ASs moving towards each other annihilate, what
follows from the physics of their existence. A much more diverse
situation is realized in the case $\ep^2 \lesssim \al \lesssim \ep$
when the slower traveling spike AS exist. Here the ASs moving towards
each other can annihilate before colliding or bounce off each other
and start traveling in the opposite direction as a result of the
interaction via the diffusion of the inhibitor (a diffusion
precursor). Also, as was shown in Sec. \ref{stab:ss7}, when $A > 1.58
\ep^{1/6}$, the static spike AS may spontaneously transform into
traveling when the value of $\al$ is decreased. This phenomenon was
studied by Osipov and Severtsev in \cite{Sev} in a simplified version
of the Gray-Scott model and is observed in our simulations of
Eqs. (\ref{act}) and (\ref{inh}) as well. Comparing Fig. \ref{cnum}
with Fig. \ref{ampl}, one can see that there exists a parameter region
where it is possible to excite both the static and the traveling ASs
simultaneously. 

According to Fig. \ref{cnum}, when the value of $A$ becomes
sufficiently close to $A_d$, the speed of the traveling spike AS may
go to zero for a range of $\al$. When $\ep = 0.05$, this happens when
$0.03 < \al < 0.06$. According to the simulations, at $\ep = 0.05$ and
$A = 1.34$ the bifurcation of the static and the traveling ASs changes
from subcritical to supercritical, so at higher values of $A$ their
coexistence is no longer possible.

In our numerical simulations we also found the phenomenon of
self-replication of the traveling spike ASs at large enough values of
$A$. This phenomenon is illustrated in Fig. \ref{travsplit} which
shows the density plots of the distributions of $\theta$ and $\eta$ as
functions of $x$ (horizontal axis) and $t$ (vertical axis) for $\ep =
0.05$, $A = 2$, and several values of $\al$.
\begin{figure}
\centerline{\psfig{figure=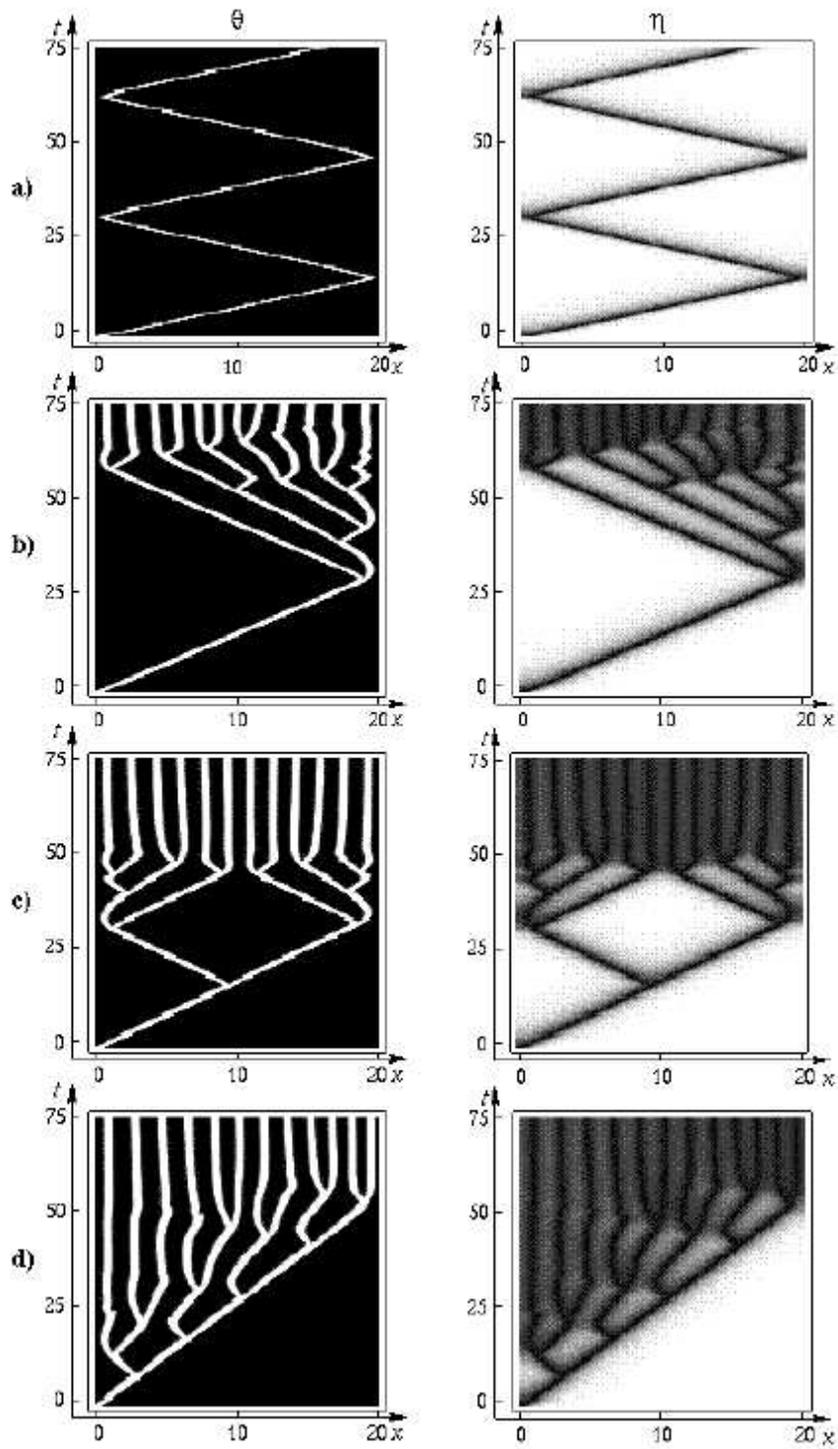,height=8in}}
\begin{center}
\caption{Self-replication of the traveling spike AS. Results of the
numerical solution of Eqs. (\ref{act}) and (\ref{inh}) with $\ep =
0.05$, $A = 2$; $\al = 0.05$ (a), $\al = 0.07$ (b), $\al = 0.075$ (c),
$\al = 0.1$ (d). Darker areas correspond to lower values of $\theta$
and $\eta$.}
\label{travsplit}
\end{center}
\end{figure}
One can see that for $\al = 0.05$ the traveling AS goes back and forth
between the system boundaries bouncing elastically off of them. When
the value of $\al$ is increased to $\al = 0.07$, a collision with the
boundary stimulates a splitting event, which is followed by
consecutive splitting of the newborn traveling spikes. Eventually, the
system becomes filled with a stationary periodic pattern of
spikes. Note that this pattern is stable in spite of the small value
of $\al$ and the large value of $A$ at which the solution in the form
of the static spike AS does not exist. When the value of $\al$ is
increased to $\al = 0.075$, the traveling AS starts splitting as it
moves, and when $\al = 0.1$, splitting occurs more frequently. Observe
that a small increase in the value of $\al$ leads to a significant
decrease of the distance between the consecutive splitting
events. Also, observe that despite splitting, the speed of the leading
traveling spike remains practically constant. This is in agreement
with the results of Sec. \ref{trav:ss7} which attribute the splitting
to the onset of the oscillatory behavior in the tail of the traveling
spike AS. The behavior of $\theta$ and $\eta$ in the back of the spike
only weakly affects the front of the spike, whose shape determines the
speed of the traveling spike AS for sufficiently small values of $\al$
and sufficiently large values of $A$ (see Sec. \ref{trav:ss6}). Notice
that self-replication of the traveling spike AS and wave reflection in
the Gray-Scott model was first observed numerically in \cite{pss}.

The results on the existence of the static and the traveling ASs, and
the stability of the static AS in one dimension obtained in the
previous sections, together with the results of the numerical
simulations, are summarized in Fig. \ref{fdiag}.
\begin{figure}
\centerline{\psfig{figure=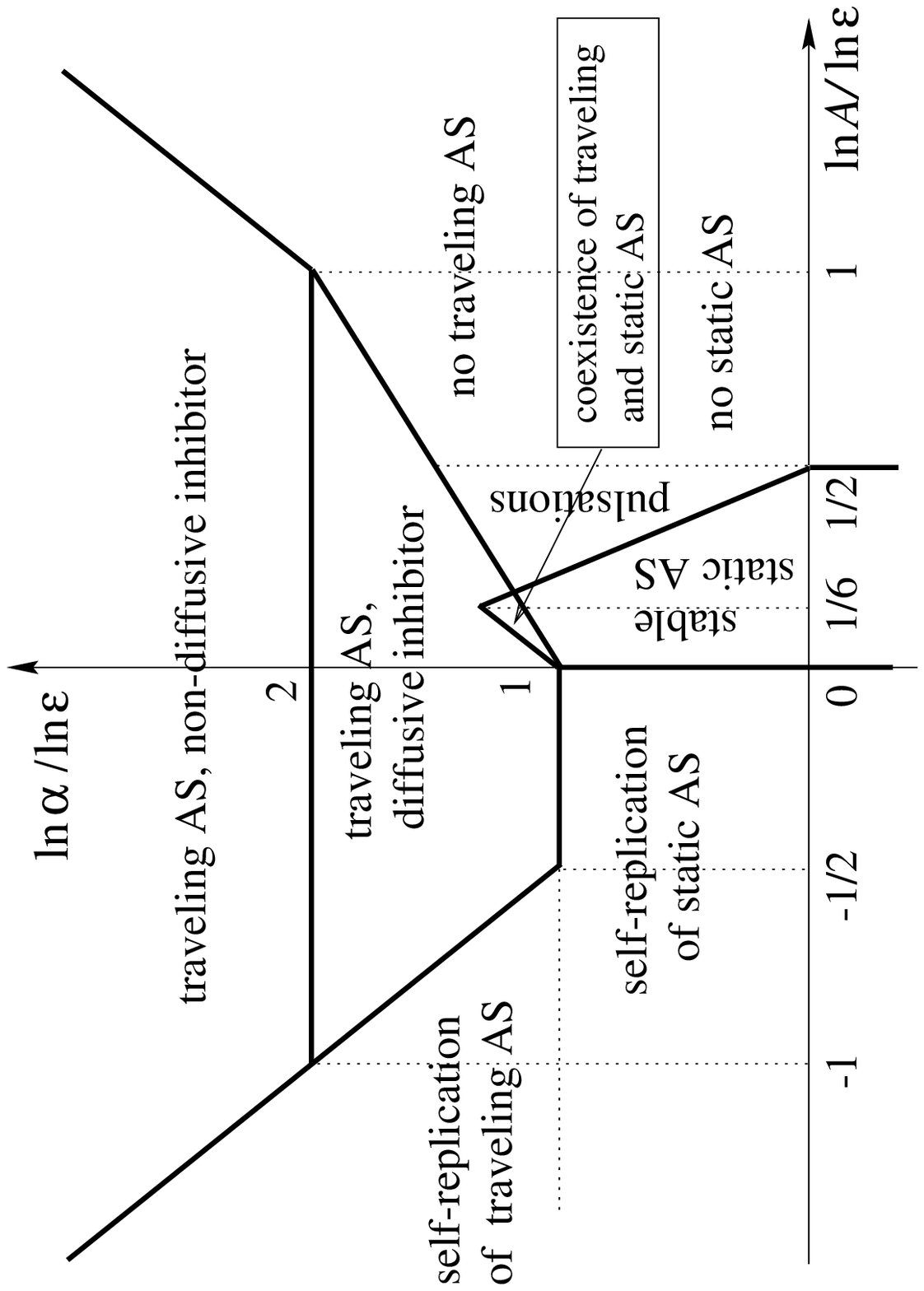,width=12cm,angle=-90}}
\caption{Diagram of the existence and the stability of different ASs
in one dimension in the limit $\ep \rightarrow 0$.}
\label{fdiag}
\end{figure}
This figure shows the domains of existence and the instability lines
for the ASs in the $\ln \al - \ln A$ plane in the limit $\ep
\rightarrow 0$. 

\section{Pattern formations scenarios in two dimensions}
\label{s:pf2d} 

Up to now, we studied the dynamics of the spike patterns in the
one-dimensional Gray-Scott model. Now we are going to study the
pattern formation scenarios in the two-dimensional Gray-Scott model.

\subsection{Granulation of the one-dimensional static spike
autosoliton} 

In Sec. \ref{stab:ss4} we showed that in two dimensions the static
spike AS in the form of a stripe is unstable in the whole region of
its existence $A_b < A < A_d$ with respect to the corrugation
instability, that is, the fluctuation $\delta \theta_0$ with $k \simeq
0.6 l^{-1}$, even for $\al > 0.33$. The growth of such a short-wave
fluctuation should lead to the granulation of the stripe into small
spots of size of order $l$. It is obvious that any wriggled stripe
will also granulate into small spots. Figure \ref{gran} shows such a
process obtained from the simulations with $\ep = 0.05$, $\al = 0.5$
and $A = 2$.
\begin{figure}
\centerline{\psfig{figure=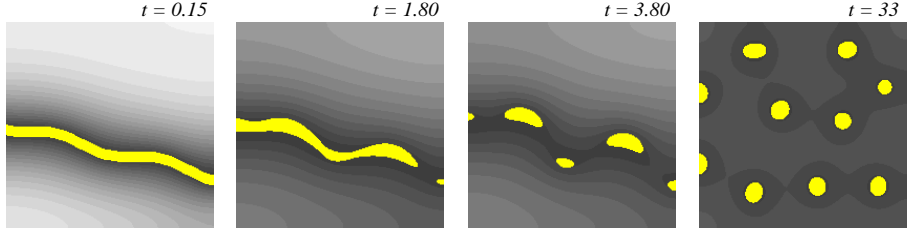,width=12cm,angle=-90}}
\begin{center}
\caption{Granulation of a stripe. Results of the numerical solution of
Eqs. (\ref{act}) and (\ref{inh}) with $\ep = 0.05$, $\al = 0.5$, and
$A = 2$. The system is $2.5 \times 2.5$. The shades of gray show the
distribution of $\eta$. The spots show the regions where $\theta > 10$.}
\label{gran}
\end{center}
\end{figure}
One can see that the stripe indeed granulates to spots of size of
order $l$ which then go away from each other until they become
uniformly distributed across the system. Self-replication of spots may
occur during this process (see below).

\subsection{Properities of the radially-symmetric static spike
autosoliton}

As was shown in Sec. \ref{stat:2d} and Sec. \ref{stab:2d}, static
radially-symmetric AS in two dimensions exists in a relatively narrow
range of the values of $A \sim \ep$. From our numerical simulations we
see that at $\ep = 0.05$ and sufficiently large values of $\al$ a
localized stimulus applied to the system at $t = 0$ evolves into a
stable static radially-symmetric spike AS at $0.38 < A < 0.65$. When
$\al$ becomes sufficiently small, this region becomes even narrower
since the AS becomes unstable with respect to the pulsations at
sufficiently small $A$. If this is the case, an initial stimulus may
produce a state which is close to a radially-symmetric AS, which after
several pulsations will collapse. The process here is similar to the
onset of the pulsations of the one-dimensional AS shown in
Fig. \ref{puls}. When the value of $\al$ becomes smaller than some
value $\al_c$, static radially-symmetric AS becomes unstable for all
values of $A$ and can no longer be excited.

\subsection{Self-replication of the static radially-symmetric
autosoliton} 

As was shown in Sec. \ref{stab:3d} and Sec. \ref{stab:2d}, when the
value of $A$ becomes greater than some critical value $A_{c2}$, the
radially-symmetric static spike AS looses stability with respect to
the radially non-symmetric fluctuations. The growth of such
fluctuations leads to splitting and self-replication of the AS. In our
simulations we found that for $\ep = 0.05$ and sufficiently large
$\al$ static radially-symmetric spike AS becomes unstable and
self-replicates at $A \geq 0.7$. Such a process for $\ep = 0.05$, $\al
= 0.5$, and $A = 2$ is shown in Fig. \ref{split}.
\begin{figure}
\centerline{\psfig{figure=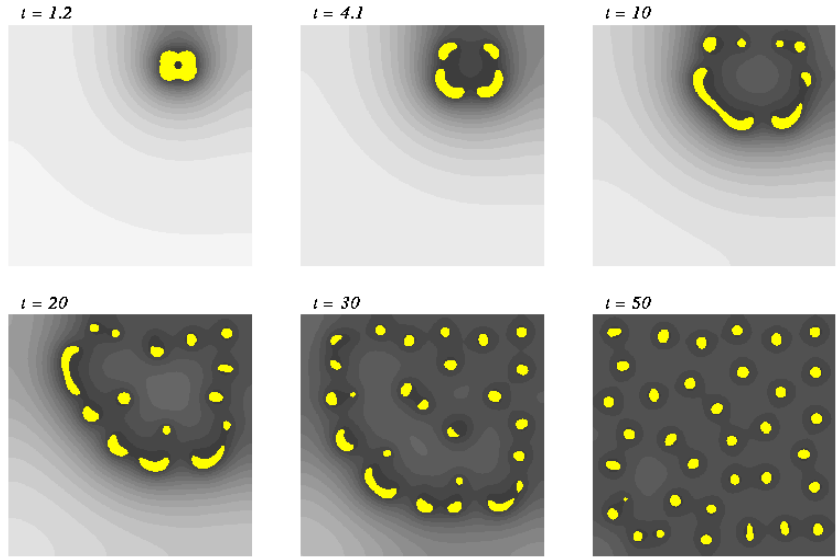,width=12cm,angle=-90}}
\begin{center}
\caption{ Self-replication of spots in two dimensions. Results of the
numerical solution of Eqs. (\ref{act}) and (\ref{inh}) with $\ep =
0.05$, $\al = 0.5$, and $A = 2$. The system is $5 \times 5$. The
shades of gray show the distribution of $\eta$. The spots show the
regions where $\theta > 10$.}
\label{split}
\end{center}
\end{figure}
From this figure one can see that the initial condition in the form of
a rectangle of size of a few $l$ splits into four (which is due to the
rectangular shape of the initial condition and the fact that the value
of $A$ is well above $A_{c2}$), and then the newborn spots go on
splitting until the system gets filled with an irregular arrangement
of spots, with the characteristic distance between the spots much less
than $L$. We would like to emphasize that the patterns observed in our
simulations are essentially different from the domain patterns that
form in N-systems \cite{Gol,Mur96a,Mur96b,Hag,thesis}. The
distributions of the activator in our simulations consist of the small
spots instead of the sharp interfaces, and in the spots they are close
to those in the radially-symmetric static spike AS. This is
illustrated in Fig. \ref{split3d} which shows the distribution of
$\theta$ at one of the moment of the simulation shown in
Fig. \ref{split}.
\begin{figure}
\centerline{\psfig{figure=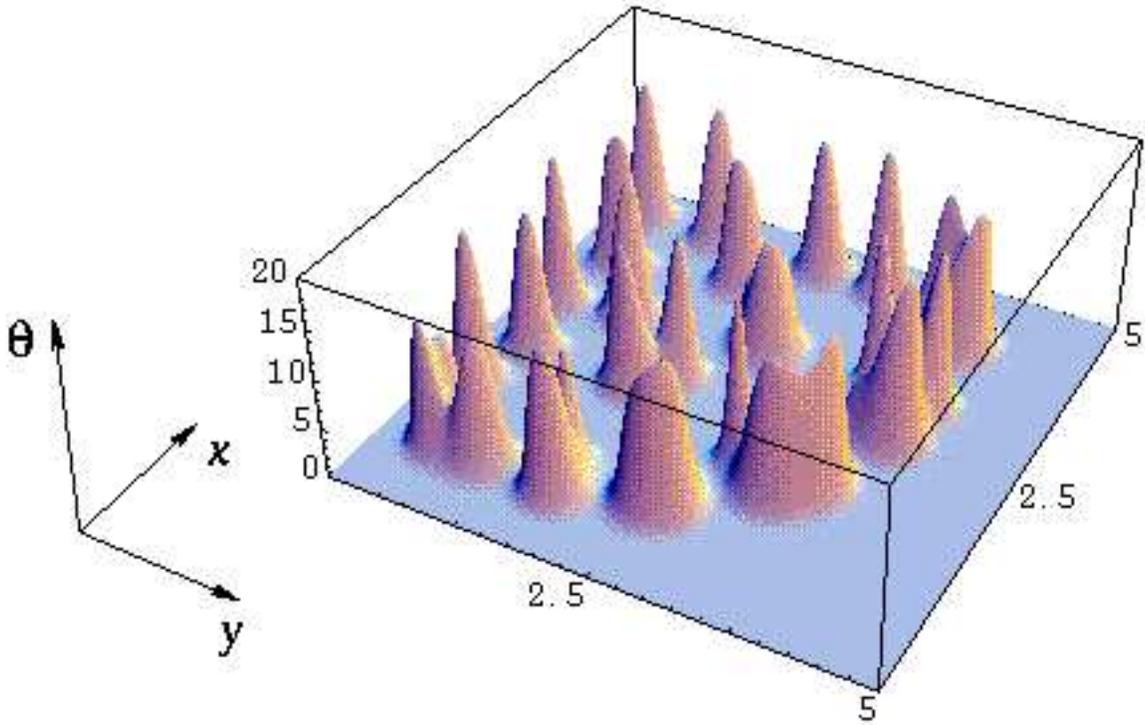}}
\begin{center}
\caption{Distribution of $\theta$ in the simulation of
Fig. \ref{split} at $t = 30$. }
\label{split3d}
\end{center}
\end{figure}

When the value of $\al$ becomes of order $\ep$, the dynamics of
splitting significantly changes. Figure \ref{wave} shows the evolution
of the system with $\ep = 0.05$, $\al = 0.1$, $A = 3$, and a localized
initial condition.
\begin{figure}
\centerline{\psfig{figure=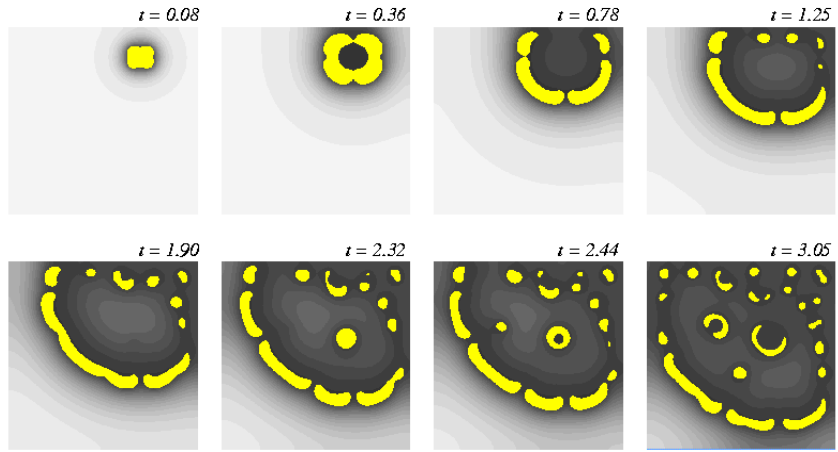,width=12cm,angle=-90}}
\begin{center}
\caption{Splitting as a result of the formation and the breakdown of a
quasi one-dimensional wave. Results of the numerical solution of
Eqs. (\ref{act}) and (\ref{inh}) with $\ep = 0.05$, $\al = 0.1$, and
$A = 3$. The system is $5 \times 5$. The shades of gray show the
distribution of $\eta$. The spots show the regions where $\theta >
10$. }
\label{wave}
\end{center}
\end{figure}
As was already mentioned in Sec. \ref{s:pf1d}, a decrease in $\al$
means greater sluggishness of the inhibitor, so the pieces that form
after splitting of an initial spot can go a greater distance apart and
become more elongated than in Fig. \ref{split} (where $\al \gg
\ep$). The state that forms here is close to a torn-up quasi
one-dimensional wave of width of order $l$. This is natural to expect
since, as we showed in Sec. \ref{trav:s2} and Sec. \ref{s:pf1d}, the
traveling spike ASs are realized in the system for these values of the
parameters. In this case the formation of new spots occurs as a result
of their pinching off of the tips of the quasi one-dimensional wave
pieces, that is, they sort of drip off from them. This process is
illustrated in Fig. \ref{wave3d} which shows the distribution of
$\theta$ at one moment of the simulation in Fig. \ref{wave}.
\begin{figure}
\centerline{\psfig{figure=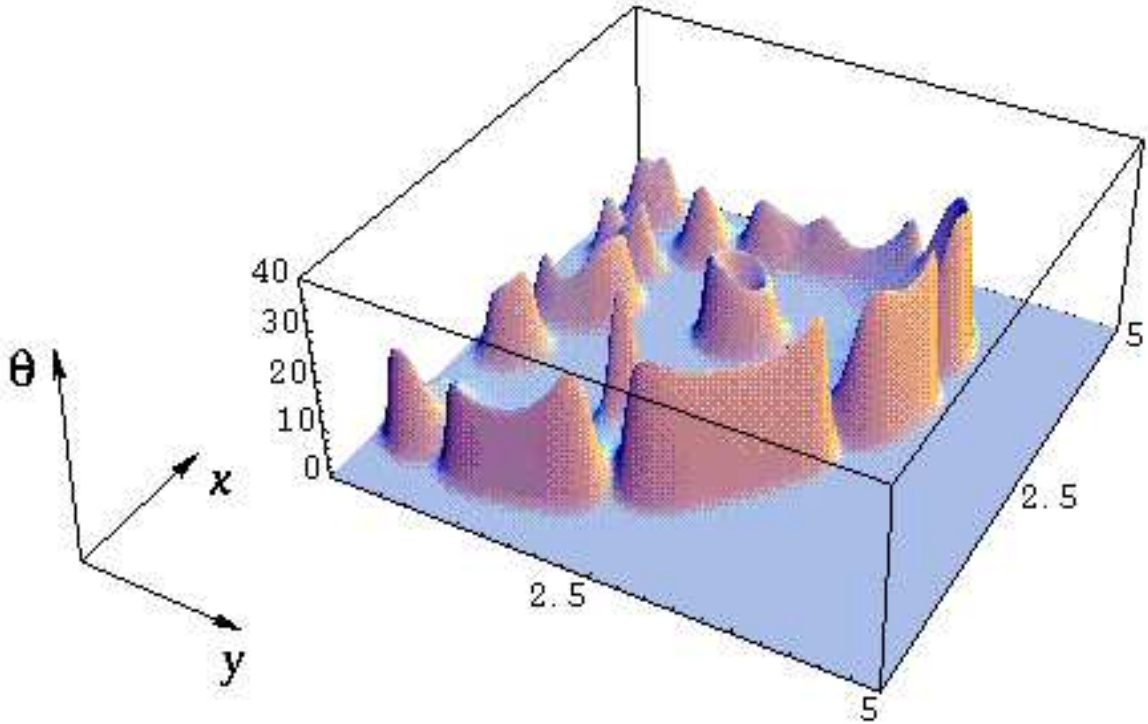}}
\begin{center}
\caption{Distribution of $\theta$ in the simulation of Fig. \ref{wave}
at $t = 2.44$.}
\label{wave3d}
\end{center}
\end{figure}
The spots that drip off from the wave pieces can further transform
into quasi one-dimensional waves which in turn break up. As a result,
the system becomes filled with a stable stationary pattern of spots,
just as in the case of large $\al$ (see the last in Fig. \ref{wave}).

\subsection{Spatio-temporal chaos}

For small enough values of $\al$ and $A$ we were able to observe
spatio-temporal chaos. Figure \ref{chaos} shows the development of a
chaotic pattern at $\ep = 0.1$, $\al = 0.04$, and $A = 1$. 
\begin{figure}
\centerline{\psfig{figure=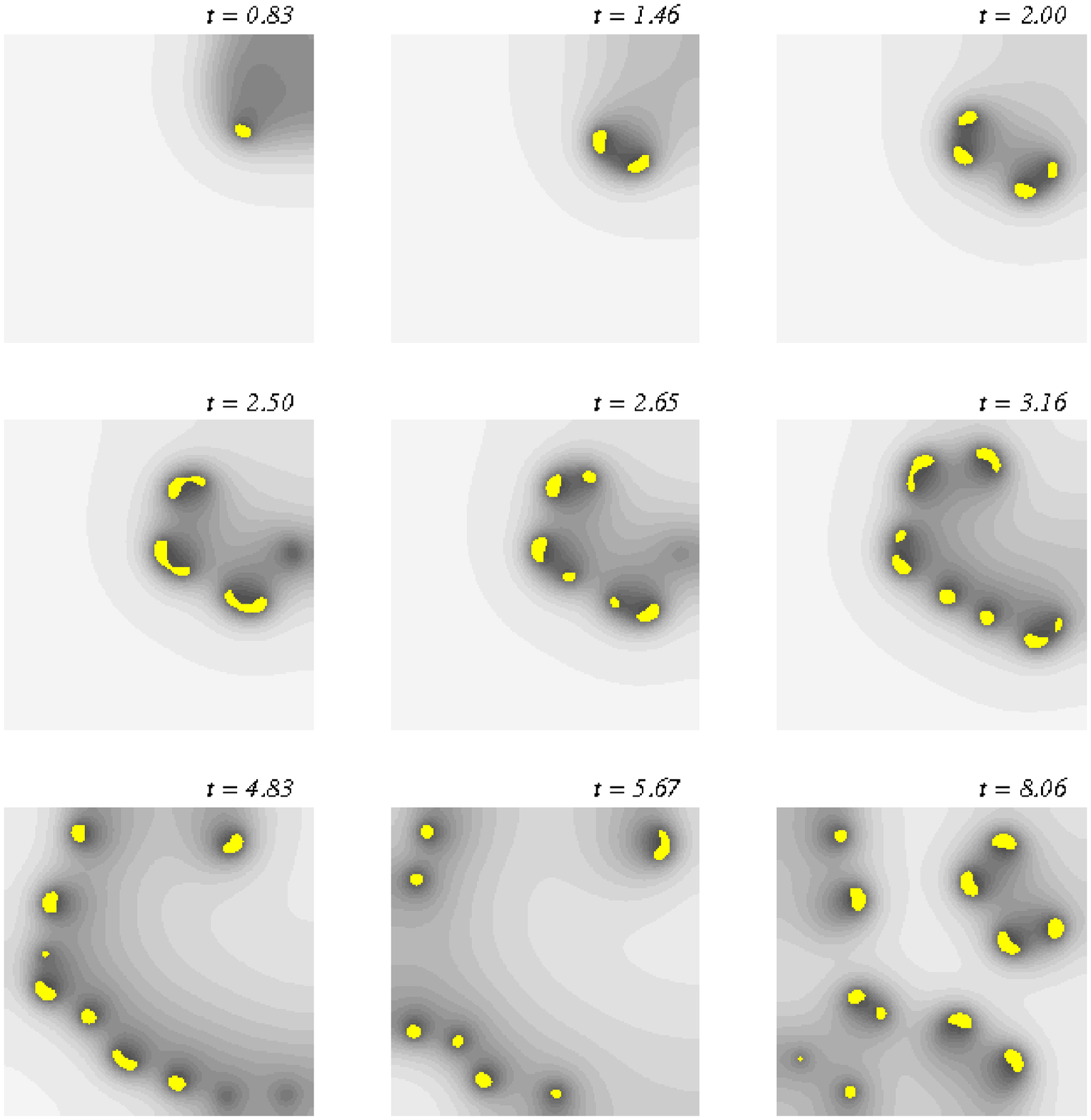,width=12cm}}
\begin{center}
\caption{Formation of spatio-temporal chaos. Results of the numerical
solution of Eqs. (\ref{act}) and (\ref{inh}) with $\ep = 0.1$, $\al =
0.04$, and $A = 1$. The system is $10 \times 10$. The shades of gray
show the distribution of $\eta$. The spots show the regions where
$\theta > 5$. }
\label{chaos}
\end{center}
\end{figure}
This pattern does not transform to a stationary pattern of spots even
for very long simulation times ($t > 100$). The stochastization of the
pattern is caused by random splitting of spots and the disappearance
of some of the spots due to their annihilation upon collision with the
bigger spots. We would expect that these effects will be most
pronounced at $\al \sim \ep^2$ and $A \sim A_{c2}$ when the static
radially-symmetric AS is close to the instabilities with respect to
the pulsations and the onset of the traveling motion
(Fig. \ref{omega3d}) and is unstable with respect to splitting. Notice
that such chaotic patterns are also observed in semiconductor
structures \cite{Semi}, combustion systems \cite{Gor}, and chemical
systems \cite{Lee}.

\subsection{Radially diverging waves}

For larger values of $A$ and sufficiently small $\al$ an initially
localized spot transforms into a circular stripe of growing radius
(Fig. \ref{ww}). 
\begin{figure}
\centerline{\psfig{figure=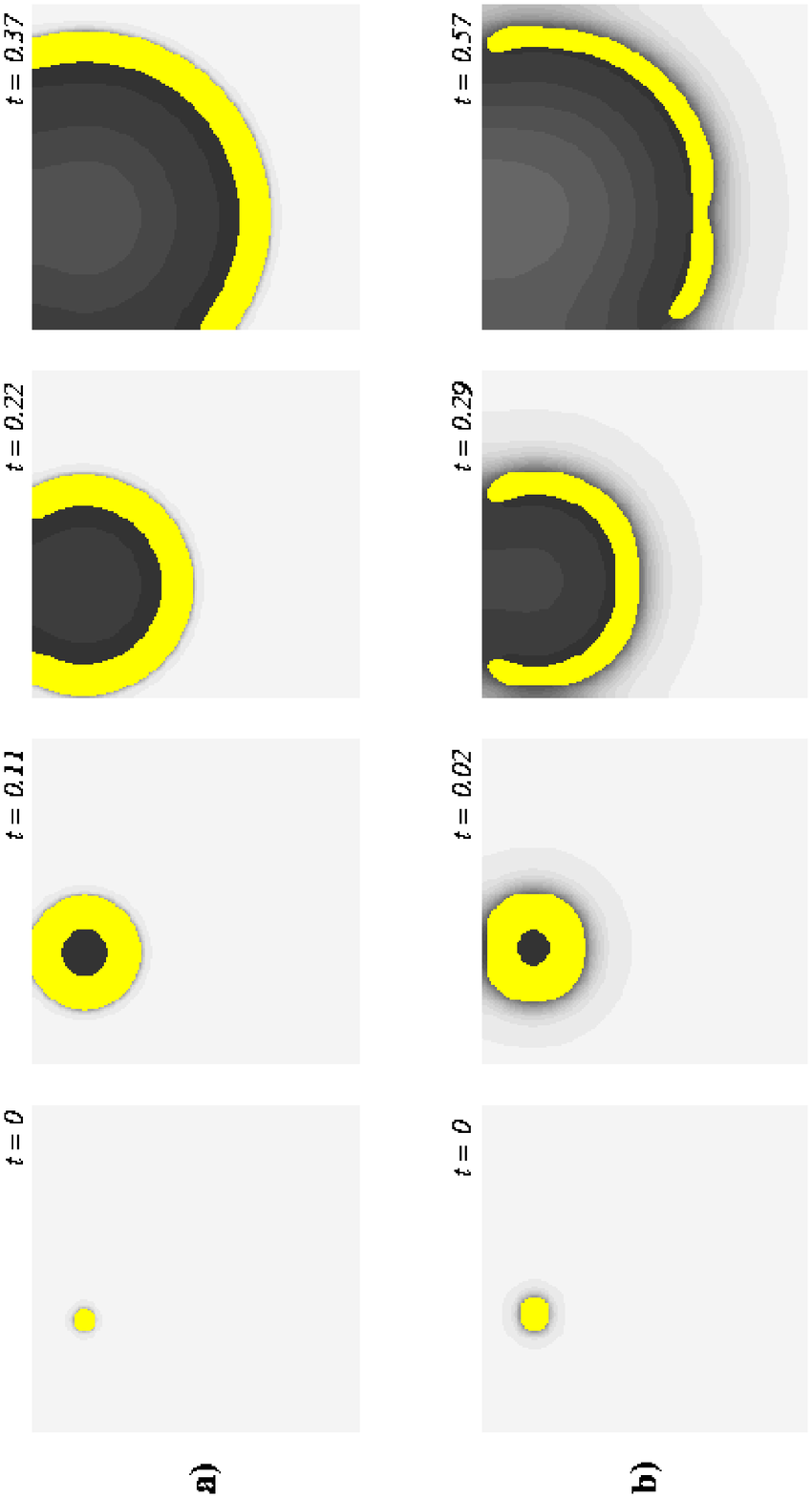,width=12cm,angle=-90}}
\begin{center}
\caption{Propagation of waves in the Gray-Scott model. Results of the
numerical solution of Eqs. (\ref{act}) and (\ref{inh}) with (a) $\ep =
0.2$, $\al = 0.05$, and $A = 1.5$; (b) $\ep = 0.1$, $\al = 0.05$, $A =
1.5$. In (a) the system is $10 \times 10$; in (b) the system is $5
\times 5$. The shades of gray show the distribution of $\eta$. The
spots show the regions where $\theta > 10$.}
\label{ww}
\end{center}
\end{figure}
When $\al$ is very small ($\al \lesssim \ep^2$), this wave does not
tear up and disappears at the system boundary
[Fig. \ref{ww}(a)]. When, on the other hand, we have $\al \sim \ep$,
the wave rebounds from the boundary and parts of it annihilate, so at
some moment it tears up [Fig. \ref{ww}(b)]. The tips of this wave get
repelled from the boundary, so the wave propagates across the system
until it annihilates upon collision with the boundary at its opposite
side.

\subsection{Spike spiral wave}

As was shown in Sec. \ref{trav:s1} and Sec. \ref{s:pf1d}, the
ultrafast traveling spike AS can form in one dimension when $\al \ll
1$ and $\ep = \infty$ (or $L = 0$). In our numerical simulations we
found that in two dimensions the solution in the form of the traveling
stripe is stable in a wide range of $A$, so it is natural to expect
that at the same parameters it is possible to excite a steadily
rotating spiral wave. The formation of such a wave at $\al = 0.1$ and
$A = 2$ is shown in Fig. \ref{spir}.
\begin{figure}
\centerline{\psfig{figure=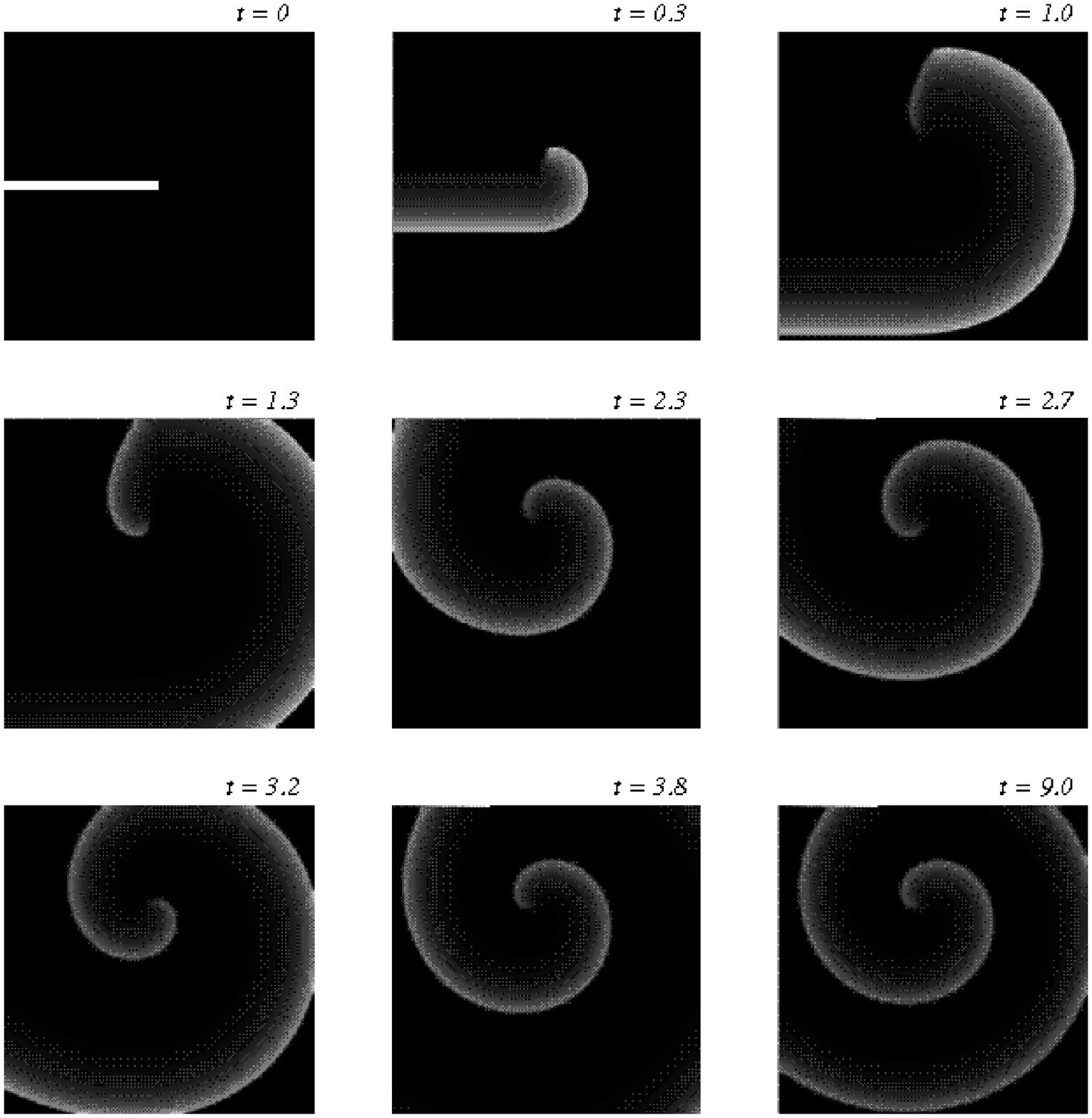,width=12cm}}
\begin{center}
\caption{Formation of a steadily rotating spiral wave. Distribution of
$\theta$ obtained from the numerical solution of Eqs. (\ref{act0}) and
(\ref{inh0}) with $L = 0$, $\al = 0.1$, and $A = 2$. Length and time
are measured in the units of $l$ and $\tau_\theta$, respectively. The
system is $100 \times 100$. }
\label{spir}
\end{center}
\end{figure}
We would like to emphasize that this wave is essentially different
from the spiral waves observed in N-systems in that in the
cross-section it has a form of a narrow spike and does not have a
front and a back separated by a large distance, as is the case in
N-systems (see, for example, \cite{Vas,Osc,Cross,Mik,Kapral}). Note
that the spiral wave shown in Fig. \ref{spir} is precisely the kind of
the wave that is observed in the experiments on Belousov-Zhabotinsky
reaction, and in the cardiac tissue, where such waves are thought to
cause abnormal heart rhythms (see, for example, \cite{Osc,Cross} and
references therein). Recently, we proposed an asymptotic theory of
these spiral wave in the Gray-Scott model \cite{spi}. 

\section{Discussion} \label{s:dis}

While the work on this paper was being done, a number of publications
on the Gray-Scott model appeared in the literature. The publications that
are most relevant to our analysis are those of \cite{Reyn,kaper}.

In \cite{kaper} Doelman {\em et al.} present an asymptotic study of
the static AS and periodic strata in the one-dimensional Gray-Scott
model. They perform a singular perturbation analysis of the localized
and spatially periodic stationary solutions in a limited region of the
parameter space. Their results are in agreement with ours in this
parameter region. However, the conclusions of the authors may be
somewhat confusing since they do not use the natural scaling of the
Gray-Scott model given by Eqs. (\ref{act}) and (\ref{inh}). Instead,
they introduce such dimensionless quantities that the original
Eqs. (\ref{X}) and (\ref{Y}) become
\begin{eqnarray} 
{\partial U \over \partial t} & = & {\partial^2 U \over \partial x^2}
- U V^2 + \delta^2 a (1 - U), \label{kapa} \\ {\partial V \over
\partial t} & = & \delta^2 {\partial^2 V \over \partial x^2} + U V^2 -
\delta^\beta b V, \label{kapb}
\end{eqnarray}
where $U$ is the inhibitor, $V$ is the activator, $\delta$ is the
small parameter, $a$ and $b$ are constants of order one, and $\beta$
($=\frac{2 \al}{3}$ from the first of \cite{kaper}) is a parameter
that lies in the interval from zero to one. The solutions are studied
in the limit $\delta \rightarrow 0$.

It is clear that the scaling introduced in Eqs. (\ref{kapa}) and
(\ref{kapb}) does not represent the true time and length scales of the
problem, which are represented by Eqs. (\ref{act}) and (\ref{inh}) of
our paper. It is easy to see that Eqs. (\ref{kapa}) and (\ref{kapb})
can be reduced to Eqs. (\ref{act}) and (\ref{inh}) with
\begin{eqnarray} \label{kapmo}
\ep = \delta^{4 - \beta \over 2} \sqrt{a \over b}, ~~~ \al = \delta^{2
- \beta} {a \over b}, ~~~ A = \delta^{1 - \beta} {\sqrt{a} \over b}.
\end{eqnarray}
Observe that the scaling in Eqs. (\ref{kapa}) and (\ref{kapb}) forces
certain relations between the true parameters $\ep$, $\al$, and $A$,
so in fact such a choice of scaling significantly restricts the
parameter space studied by Doelman {\em et al.}, leading them to
incorrect conclusions about non-existence of certain types of
solutions.

According to the results of the first paper in \cite{kaper}, the
static one-dimensional AS exists only when $\beta \in [0, 1)$ as
$\delta \rightarrow 0$. This statement is based on the fact that the
coefficient in front of the last term in the right-hand side of
Eq. (\ref{kapa}) scales as $\delta^2$. On the other hand, as we showed
in Sec. \ref{stat:1d}, this AS in fact exists in a wider range of the
parameters as long as $\ep \ll 1$, so in fact such a scaling is not a
necessary condition for the AS existence. This fact was noticed by the
authors in the more recent paper in \cite{kaper}.

The case $\beta = 0$ is equivalent to $A \sim A_b$ in our
notation. Our result that the solution exists only at $A > A_b =
\sqrt{12 \ep}$ is in agreement with the result of \cite{kaper} that
the solution exists at $a > 144 b^3$. Notice that the analogous method
of finding $A_b$ for the ASs in systems of small size was introduced
two decades ago by Kerner and Osipov \cite{KO78}. There they performed
an analysis of a very similar model --- the Brusselator. The results
obtained by them at $A \sim A_b$ for the Brusselator differ from those
presented in our paper only by numerical coefficients. A calculation
similar to that of Sec. \ref{stat:ss1} was performed by Dubitskii,
Kerner, and Osipov for the Gierer-Meinhardt model \cite{Dub} (see also
\cite{KO94}). Osipov and Severtsev also performed such an analysis for
a simplified version of the Gray-Scott model, which are also very
similar to the results of the first part of Sec. \ref{stat:1d}
\cite{Sev}. The advantage of the method of Doelman {\em et al.},
however, is that it rigorously shows the existence of the static spike
AS for $A_b \lesssim A \ll 1$ and $\ep \ll 1$.

Doelman {\em et al.} claim that the solution obtained at $\beta = 0$
is valid with accuracy to $\delta \sim \ep^{1/2}$. On the other hand,
the analysis of Sec. \ref{stat:1d} suggests that this solution is
actually valid with a much better accuracy $\ep$. In fact, $\ep$ is
the true small parameter of the problem, and the asymptotic procedure
for obtaining the solutions for $\ep \ll 1$ can be constructed {\em
regardless} of all other parameters of the problem with accuracy
better than $\ep^{1/2}$ (see the discussion at the end of
Sec. \ref{s:stat}). Note that this procedure for general V- and
$\Lambda$-systems was first introduced by Dubitskii, Kerner, and
Osipov in \cite{Dub} (see also \cite{KO94}).

For $\beta > 0$ Doelman {\em et al.} claim that the solution obtained
by them is accurate to order $\delta^{1 - \beta}$. They forget that
although the accuracy with which the asymptotic equations describe the
actual distribution of $V$ increases as $\beta$ decreases, the
accuracy of the matching condition, namely, the fact that the true
$U(0)$ is non-zero, decreases as $\beta$ decreases.\footnote{In the
first of \cite{kaper} this is equivalent to the approximation of
Eq. (3.6) by Eq. (3.7) of that paper. See also the remark after
Theorem 4.3 there.} It is not difficult to see that the latter
actually gives corrections of order $\delta^{3 \beta \over 2}$ which
become of order 1 as $\beta \rightarrow 0$, as should be
expected. This means that the best accuracy with which the asymptotic
procedure of Doelman {\em et al.}  is valid is $\delta^\frac{3}{5}$,
what in our notation corresponds to $\ep^{1/3}$. On the other hand, as
can be seen from Sec. \ref{stat:1d}, the asymptotic procedure
introduced by us always gives an accuracy better than $\ep^{1/2}$, and
in the parameter regions of interest the accuracy is of order
$\ep$. Furthermore, our asymptotic procedure remains valid even for $A
\sim 1$, which is equivalent to $\beta = 1$ of \cite{kaper}, where the
approach of the latter fails completely, thus being able to
quantitatively describe the local breakdown and the disappearance of
the solution at $A = A_d$. Notice that in the second of \cite{kaper}
the authors used a different approach to qualitatively analyze the
disappearance of solution at $A \gg 1$.

Doelman {\em et al.} make a statement that the traveling ASs do not
exist in the Gray-Scott model. Although this statement is true for the
scaling in Eqs. (\ref{kapa}) and (\ref{kapb}), it is generally
wrong. In fact, as we showed in Sec. \ref{s:trav}, there are {\em two}
qualitatively different traveling spike ASs that are realized in the
Gray-Scott model. The ultrafast traveling spike AS are realized when
$\al \lesssim \ep^2$ and $\al^{1/2} \lesssim A \lesssim \al^{-1/2}$,
and the slower traveling spike AS are realized when $\ep^2 \lesssim
\al \lesssim \ep$ and $\al \ep^{-1} \lesssim A \lesssim
\al^{-1/2}$. These parameter regimes lie outside of those studied in
\cite{kaper}. The more so, we find {\em coexistence} of the traveling
spike AS with static.

In the second of \cite{kaper} Doelman {\em et al.} study the stability
of the one-dimensional static spike AS. Their analysis is based on the
equations of the type of Eq. (\ref{disp1d:stab0}) and
(\ref{om1d0}). Equations of this type were studied by Kerner and
Osipov in the context of the stability of the spike AS in the systems
of small size \cite{KO89,KO90,KO94,KO78}. As we showed in the
appendices A and B, a straightforward extension of the method of
Kerner and Osipov allows rigorous analytic studies of these
equations. Doelman {\em et al.}  found in the scaling of
Eqs. (\ref{kapa}) and (\ref{kapb}) that the AS becomes unstable with
respect to the pulsations when $\beta = \frac{1}{2}$, what corresponds
to $A_\omega \sim \ep^{2/7}$. Their results agree with those obtained
by us in Sec. \ref{stab:ss3} in the case $A_b \ll A \ll A_d$ [see
Eq. (\ref{aom1d})]. However, as can be seen from Sec. \ref{stab:ss3}
and \ref{s:pf1d} of our paper, in practice this result has a very
limited applicability, so one should use the results presented in
Fig. \ref{omega1d} of our paper to find the values of $\al_\omega$ and
$\omega_0$.

Moreover, this is not the only possibility for the instability of the
static spike AS for the general choice of the parameters. What we
showed in Sec. \ref{stab:1d} is that, first, the instability with
respect to the pulsations is realized in a wider range of the system
parameters, and, second, there is another instability which leads to
the transformation of the static spike AS into traveling, which for $A
\gtrsim \ep^{1/6}$ precedes the pulsation instability. In general, we
performed a complete analysis of the stability of the static spike AS
in one dimension.

Reynolds, Ponce-Dawson, and Pearson derived asymptotic equation of
motion for the spike AS in the Gray-Scott model with $A \sim 1$, $\al
\sim 1$, and $\ep \ll 1$ (in our notation). The reason the AS moves is
its interaction with the boundary. However, for $A > A_d$ the
peculiarities of the internal dynamics result in the breakdown of the
asymptotic description which the authors attribute to splitting and
self-replication of the AS. The equations obtained by the authors for
the inner region in fact coincide with Eqs. (\ref{thetashtrav}) and
(\ref{etashtrav}) obtained by us in Sec. \ref{trav:s2} for the
traveling spike AS in the case $\ep^2 \ll \al \lesssim \ep$. They also
found the value of $A_d$ which agrees with the one obtained by
us. Thus, the peculiarities noticed in \cite{Reyn} should be important
in the case of the traveling spike AS in the system of infinite extent
and should explain the dynamics of splitting of the traveling AS for
$A \sim 1$ and $\al \sim \ep$.

Notice that both the authors of \cite{Reyn} and the authors of
\cite{kaper} studied self-replication and splitting of the static
spike ASs in one dimension and their results cannot be used to explain
self-replication in real chemical systems which are higher-dimensional
and in which the radially-symmetric static spike ASs (spots) rather
than the one-dimensional static spike ASs (pulses) form. Indeed, as we
showed in the Sec. \ref{stab:1d}, \ref{stab:3d}, \ref{s:pf1d} and
\ref{s:pf2d}, in one dimension splitting occurs as a result of the
local breakdown in the AS center, whereas in higher dimensions
splitting is the result of the buildup of radially non-symmetric
fluctuations and occurs at different values of the
parameters. Moreover, the local breakdown in the case of the
higher-dimensional radially-symmetric static spike AS cannot be
realized at all. Note that the same situation is realized in a
different class of reaction-diffusion systems studied by us
(N-systems) where splitting is also shown to be the consequence of the
buildup of radially non-symmetric fluctuations
\cite{Mur96a,Mur96b,m:pre96}. Thus, for $\ep \ll 1$ self-replication
and splitting in both N-systems and the Gray-Scott model are driven by
qualitatively the same mechanism.

Let us comment on the relationship between the numerical simulations
of the two-dimensional Gray-Scott model performed by Pearson in
\cite{Pear} and those of Sec. \ref{s:pf2d} performed by us. Pearson
uses a different non-dimensionalization of the Gray-Scott model, which
has the following correspondence with our parameters: $\ep^2 = {D_v F
\over D_u (F + k)}$, $\al = {F \over F + k}$, $A = {\sqrt{F} \over F +
k}$ (see \cite{Pear}). It is not difficult to see that for the
simulations of Pearson $\ep \simeq 0.45$, $\al \simeq 0.4$, and $A
\simeq 2$, so his choice of the parameters corresponds to $\ep \sim 1$
and $\al \sim 1$. This is different from our simulations for which
$\ep \ll 1$ and/or $\al \ll 1$. Note that the stationary patterns in
the Gray-Scott model with $\ep \sim 1$ should actually resemble those
forming in N-systems \cite{Mur96b}. Indeed, if one introduces the new
variables $\tilde\theta = {\theta \over A}$ and $\tilde\eta = \eta +
{\ep^2 \over A} \theta$, after simple algebra one can write
Eqs. (\ref{act}) and (\ref{inh}) as
\begin{eqnarray}
&& \al {\partial \tilde\theta \over \partial t} = \ep^2 \Delta
\tilde\theta + A^2 \tilde\theta^2 \tilde\eta - \ep^2 A^2
\tilde\theta^3 - \tilde\theta \label{changea} \\ && ~{\partial
\tilde\eta \over \partial t} = \Delta \tilde\eta + 1 - \tilde\eta - (1
- \ep^2) \tilde\theta - (\al - \ep^2) {\partial \tilde\theta \over
\partial t}. \label{changeb}
\end{eqnarray}
From Eqs. (\ref{changea}) and (\ref{changeb}) one can see that for
$\ep \sim 1$ the nullcline of the equation for $\tilde\theta$ is
actually N-like, and the coupling between $\tilde\theta$ and
$\tilde\eta$ becomes linear, so the stationary patterns in this case
should in fact look like those forming in N-systems
\cite{Mur96b}. This is the reason why Pearson observed the stationary
labyrinthine patterns, while in our simulations any stripe-like
pattern always granulates into spots. Also, the collective
oscillations of the space-filling patterns observed by Pearson are
similar to the collective oscillations of the domain patterns in
N-systems \cite{m1:pre97}. Pearson did not see the spiral waves
because in his simulations $\al \sim \ep^2$ and the spirals break up
as they form. We do not see any phase turbulence since in our
simulations the system is far away from the Hopf bifurcation of the
homogeneous state $\theta_{h3}$, $\eta_{h3}$. The rest of the patterns
observed in \cite{Pear} are similar to those observed by us.

Finally, we will mention recent studies by Hale, Peletier, and Troy
\cite{peletier}. They found the exact solutions in the form of the
solitary pulses and fronts and analyzed their stability in the
Gray-Scott model with $\epsilon = 1$ and $\alpha = \epsilon^2$. The
reason this can be done exactly for these values of $\ep$ and $\al$ is
because the equation for the activator and the inhibitor effectively
decouple from each other [see Eq. (\ref{changeb})], so the model
behaves as a scalar reaction-diffusion system
\cite{Cross,KO90,KO94}. It is then not surprising that the behavior of
the system remains the same if $\ep$ is close to, but not exactly
equal to 1 \cite{peletier}.

\section{Conclusion} \label{s:conc}

Let us now summarize the results of our analysis of the patterns in
the Gray-Scott model. As was emphasized in Sec. \ref{s:mod}, a unique
feature of the Gray-Scott model is the fact that in it the homogeneous
state $\theta_h = 0$, $\eta_h = 1$ is stable for all values of the
parameters. However, in such a stable homogeneous system it is
possible to excite various steady inhomogeneous states, including the
self-sustained solitary inhomogeneous states, autosolitons (ASs), by
applying a sufficiently strong external stimulus. The formation of
these inhomogeneous states is due to the self-production of substance
$X$ (which plays the role of the activator) controlled by the other
substance $Y$ (which plays the role of the inhibitor). The properties
of the patterns are determined by only three parameters: $\ep$, $\al$,
and $A$. The parameters $\ep = l / L$ and $\al = \tau_\theta /
\tau_\eta$ are the ratios of the characteristic length and time scales
of the activator and the inhibitor, respectively, and the control
parameter $A$ determines the degree of the deviation of the system
from equilibrium since it is proportional to the rate of supply of
substance $Y$, which plays the role of ``fuel'' for the reaction in
Eq. (\ref{reacta}). We emphasize that for the same values of the
system's parameters it is possible to excite {\em different} patterns
by choosing the form of the stimulus, which will be stable in certain
ranges of the parameters $\ep$, $\al$, and $A$. At the stability
margin the patterns spontaneously disappear or transform into the
patterns of different kind.

As follows from the general qualitative theory of the patterns in
reaction-diffusion systems, the necessary condition for the existence
of the persistent patterns of any kind is the smallness of the
parameters $\ep$ and/or $\al$ \cite{KO89,KO90,KO94}. In this paper we
took advantage of this fact and performed asymptotic analysis of the
simplest patterns (ASs) in the limit when either of these parameters
goes to zero. This analysis gave us the dependence of the main
parameters of the ASs on the system's parameters and their ranges of
existence and stability. Their asymptotic behavior is summarized in
Table \ref{tab}.
\begin{table}
\caption{Scaling of the main parameters of different types of ASs in
the Gray-Scott model. } \label{tab}
\begin{tabular}{c|c|c|c|c|c|c|c|c|c}
\hline\hline AS & $A_b$ & $A_d$ & $A_{c2}$ & $\al_\omega$ & $\al_T$ &
$\al_c$ & $\theta_{\mathrm max}$ & $\eta_{\mathrm min}$ & $c$ \\
\hline static 1-$d$ & $\ep^{1/2}$ & $\ep^0$ & & $\ep^2 A^{-4}$ & $\ep
A^2$ & $\ep^{4/3}$ & $\ep^{-1} A$ & $\ep A^{-2}$ & \\ static 2-$d$ &
$\ep (\ln \ep^{-1})^{1/2}$ & $\ep^{1/2} \ln \ep^{-1}$ & $\ep \ln
\ep^{-1}$ & $\ep^2$ & ${A^2 \over \ln \ep^{-1}}$ & $\ep^2$ & ${A
\ep^{-2} \over \ln \ep^{-1} }$ & $\ep^2 A^{-2} \ln \ep^{-1}$ & \\
static 3-$d$ & $\ep$ & $\ep^{1/2}$ & $\ep$ & $\ep^2$ & $\ep^2$ &
$\ep^2$ & $\ep^{-1}$ & $\ep^0$ & \\ traveling & $\al^{1/2}$ &
$\al^{-1/2}$ & & & & & $\al^{-1} A$ & & $\al^{-1/2} A$
\vspace{-4mm} \\ $( \al \lesssim \ep^2)$ & & & & & & & & \\ traveling
& $\al \ep^{-1}$ & $\al^{-1/2}$ & & & & & $\al^{-1} A$ & & $\ep \al^{-1} A$
\vspace{-4mm} \\ $( \ep^2 \lesssim \al \lesssim \ep)$ & & & & & & & &
\\ \hline\hline
\end{tabular}
\end{table}
The properties of the forming patterns strongly depend on the
parameters of the system. Depending on the values of $\ep$ and $\al$
one can distinguish three different cases.

The first case corresponds to $\ep \ll 1$ and $\al \gtrsim 1$. As was
expected from the general qualitative theory \cite{KO89,KO90,KO94},
for these values of $\ep$ and $\al$ one can excite only the static
spike ASs (Fig. \ref{as1d}). In one dimension these ASs are stable in
a wide range of $A$ from $A_b \sim \ep^{1/2}$ to $A_d \sim 1$
(Sec. \ref{stat:1d}). The characteristic size of the spike of the AS
is determined by the diffusion length $l$ of the activator and is
practically independent of $A$. When the value of $A$ is increased,
the amplitude of the static spike AS grows from $\theta_{\mathrm max}
\sim \ep^{-1/2}$ for $A \sim A_b$ to $\theta_{\mathrm max} \sim
\ep^{-1}$ for $A \sim 1$. According to the general qualitative theory
\cite{KO89,KO90,KO94}, at $A = A_b$ and $A = A_d$ we must have $d
\theta_{\mathrm max} / dA = \infty$. For this reason at $A = A_b$ the
static spike AS, having a large amplitude, abruptly disappears. When
$A$ approaches $A_d$, the AS widens somewhat, until at $A = A_d$ a
local breakdown occurs in its center leading to the consecutive
splitting of the static spike AS. As a result of this self-replication
effect a static pattern consisting of a periodic array of spikes forms
\cite{Reyn}.

When the value of $\al$ is decreased below $\al_0 \sim 1$, the range
of existence of the static spike AS narrows. When $A$ is decreased,
the AS looses its stability before reaching the point $A = A_b$
(Sec. \ref{s:stab}). The instability is realized with respect to the
pulsations leading to the AS collapse (Fig. \ref{puls}). On the other
hand, when $A$ is increased, the static spike AS may destabilize and
spontaneously transform into traveling before reaching the point $A =
A_d$ (see also \cite{Sev}).

In the two- and the three-dimensional Gray-Scott model with $\ep \ll
1$ and $\al \gtrsim 1$ one can excite the radially-symmetric static
spike ASs of size of order $l$ (Sec. \ref{stat:3d} and
Sec. \ref{stat:2d}). The range of the values of $A$ for which these
ASs exist is very narrow for $\ep \ll 1$ (see Table \ref{tab}). At the
point $A = A_b \sim \ep$ the static radially-symmetric spike AS,
having large amplitude $\theta_{\mathrm max} \sim \ep^{-1}$, abruptly
disappears. The range of $A$ at which the radially-symmetric static
spike AS exist becomes even narrower for $\al \sim \ep^2$ when the AS
becomes unstable with respect to the pulsations (Sec. \ref{stab:3d}
and Sec. \ref{stab:2d}). On the other hand, when the value of $A$ is
increased to $A = A_{c2} \sim \ep$, the static radially-symmetric
spike AS looses stability with respect to the radially non-symmetric
fluctuations. As a result of the development of such fluctuations the
AS splits into two, which then split in turn (self-replicate) until
the system gets filled with a multispot pattern (Fig. \ref{split}). We
would like to emphasize that for $\ep \ll 1$ a spot (a state close to
the radially-symmetric AS) is the dominant morphology, so that any
localized initial state such as a stripe or a square first granulates
into spots, and the evolution of the system is then governed by
self-replication of these spots (Figs. \ref{gran}, \ref{split}).  Let
us note that in N-systems one sees complex patterns in the form of
wriggling stripes, connected and disconnected labyrinthine patterns
\cite{Gol,Mur96a,Mur96b,Hag,thesis}, which are also observed in
chemical experiments \cite{Lee}. These patterns do not form from a
localized stimulus in the Gray-Scott model with $\ep \ll 1$. Note,
however, that when $\ep \sim 1$, the Gray-Scott model starts behaving
like an N-system (see the end of Sec. \ref{s:dis}), so for these
values of $\ep$ such patterns can in fact be excited \cite{Pear}.

In the second case we have $\al \ll 1$ and $\ep \gg \al^{1/2}$. In
this case one can excite different kinds of self-sustained waves
(autowaves) which have the form of the narrow spikes of size roughly
$l$ and the amplitude $\theta_{\mathrm max} \sim \al^{-1}$. In one
dimension the ultrafast traveling spike ASs are realized
[Fig. \ref{trav}(a)] whose speed $c \sim A \al^{-1/2} \times l /
\tau_\theta$, i. e., it is $A \al^{-1/2}$ times greater than the
typical speed of the traveling AS in N-systems
(Sec. \ref{trav:s1}). This ultrafast traveling spike AS can be excited
in a wide range of $A$ from $A_b \sim \al^{1/2} \ll 1$ to $A_d \sim
\al^{-1/2} \gg 1$. In two dimensions, besides the ultrafast traveling
spike AS, one can excite radially diverging waves (Fig. \ref{ww}) and
the steadily rotating spiral wave (Fig. \ref{spir}), which in the
cross-section looks like the ultrafast traveling spike AS. The spiral
wave observed by us is a new type of spiral waves, which is
essentially different from those forming in N-systems.

In the third case $\ep \lesssim \al \ll 1$ the behavior of the
patterns in the Gray-Scott model is most diverse. In one dimension,
besides the static spike AS one can excite the traveling spike AS
[Fig. \ref{trav}(b)] whose speed decreases with the decrease of
$\ep$. These ASs may elastically rebound upon collision with the
boundary or with each other (Fig. \ref{travsplit}). When $\al \gtrsim
\ep$, they start self-replicating. The greater the value of $\al$, the
smaller the distance between the consecutive splitting events in such
a self-replication process. We would like to emphasize that as a
result of this effect a {\em static} periodic pattern of spikes forms
(Fig. \ref{travsplit}). In two dimensions the traveling spike AS
undergoes a transversal breakup leading to the formation of the
stationary multispot patterns (Fig. \ref{wave}). Also, when the values
of $\al$ and $A$ are small enough, one can excite the turbulent
patterns (Fig. \ref{chaos}). The chaotic behavior of the latter is due
to the random creation of the new spots as a result of the
self-replication and the annihilation of some of the spots as they
collide with each other. This kind of turbulence is observed in
chemical experiments \cite{Lee} and is not unlike the one realized in
N-systems \cite{Mur96b}.

Above we considered the properties of the spike ASs, which are the
simplest patterns and are, therefore, the building blocks of the more
complex patterns forming in the Gray-Scott model. Their properties
allowed us to understand the pattern formation scenarios in the
Gray-Scott model subjected to a localized stimulus. We expect that the
asymptotic methods developed by us in this paper can be successfully
applied to other systems of this kind. Also, we found that in many
instances a localized stimulus results in the formation of the complex
patterns consisting of many strongly interacting AS-like states. The
effect of interaction of such states in the complex patterns is an
interesting open problem. We hope that the singular perturbation
techniques developed in this paper will be useful for studying these
interactions and their effects on the dynamics of the complex
space-filling patterns in the Gray-Scott model and other similar
models.

\ack

We would like to acknowledge the computational support from the Center
for Computational Science of Boston University.

\appendix

\section{Analysis of Eq. (\ref{disp1d:stab0})}

Equation (\ref{disp1d:stab0}) is of the kind studied by Kerner and
Osipov in the case of the ASs in systems of small size
\cite{KO89,KO90,KO94,KO78}. Here we perform a rigorous analysis of
Eq. (\ref{disp1d:stab0}) using their method.

Let us introduce the orthonormal basis set $\delta\theta_n$ of the
eigenfunctions of the Schr\"odinger operator in the left-hand side of
Eq. (\ref{disp1d:stab0}) \footnote{There should be no confusion
between the eigenfunctions $\delta\theta_n$ of this section, which
correspond to $\delta\theta^{(0)}_n$ of Sec. \ref{stab:1d} and
$\delta\theta_n$ of that section, which are the eigenfunctions of
Eq. (\ref{disp1d0}). }
\begin{eqnarray} \label{hth}
\left[ - {d^2 \over dx^2} + 1 - 3 \cosh^{-2} \left({x \over 2} \right)
\right] \delta\theta_n = \lambda_n \delta\theta_n.
\end{eqnarray}
For the simplicity of notation we omitted the superscript 0 in this
section. As was already discussed in Sec. \ref{stab:ss1}, this
operator has three discrete eigenvalues [Eq. (\ref{schr0})] and a
continuous spectrum for $\lambda_n > 1$.

Assuming for a moment that the problem is considered on a large but
finite domain, we can write the operators of Eq. (\ref{disp1d:stab0})
in this basis as
\begin{eqnarray} \label{bmn1}
B_{mn} = (\lambda_n - \gamma) \delta_{mn} + C (1 - \gamma) b^l_m
b^r_n,
\end{eqnarray}
where $\delta_{mn}$ is the Kronecker delta,
\begin{eqnarray}
b^l_n = \int_{-\infty}^{+\infty} \cosh^{-4} \left( {x \over 2} \right)
\delta\theta_n(x) dx, ~~~b^r_n = \int_{-\infty}^{+\infty}
\delta\theta_n(x) dx,
\end{eqnarray}
and 
\begin{eqnarray}
C = {3 A^2 \over 8 A_b^2} \left( 1 + \sqrt{1 - {A_b^2 \over A^2}}
\right)^2. 
\end{eqnarray}
Observe that $C$ is a monotonically increasing function of $A \geq
A_b$. 

In terms of $B_{mn}$ Eq. (\ref{disp1d:stab0}) becomes
\begin{eqnarray} \label{det}
\det B_{mn} = 0.
\end{eqnarray}
Note that since by symmetry $b^l_n$ and $b^r_n$ are identically zero
for odd functions $\delta\theta_n$, we immediately conclude that these
functions are the solutions of Eq. (\ref{disp1d:stab0}) with $\gamma_n
= \lambda_n$ corresponding to these functions.

It is not difficult to show that because of the special form of the
second matrix in Eq. (\ref{bmn1}) we have \cite{KO89,KO90,KO94,KO78}
\begin{eqnarray} \label{det1}
\det B_{mn} = \left[ 1 + C (1 - \gamma) \sum_n { a_n \over \lambda_n -
\gamma} \right] \prod_n (\lambda_n - \gamma),
\end{eqnarray}
where $a_n = b_n^l b_n^r$ and the summation is over the even states
only. Using Eq. (\ref{hth}), one can bring the expression for $a_n$ to
a symmetric form which is convenient for the further calculations
\begin{equation} \label{an}
a_n = {2 \lambda_n \over \lambda_n - 1} \left[
\int_{-\infty}^{+\infty} \cosh^{-2} \left( {x \over 2} \right)
\delta\theta_n(x) dx \right]^2.
\end{equation}
The values of $a_0$ and $a_2$ can be calculated explicitly with the
use of Eq. (\ref{schr0})
\begin{eqnarray} \label{a02}
a_0 = {75 \pi^2 \over 256}, ~~~a_2 = - {9 \pi^2 \over 256}.
\end{eqnarray}
The calculation of $a_k$ corresponding to the functions
$\delta\theta_k$ of the continuous spectrum (with the wave vector $k$
and $\lambda_k = 1 + k^2$) is rather involved. The functions
$\delta\theta_k$ can be written as linear combinations of the real and
the imaginary parts of
\begin{eqnarray}
u(y) = (1 - y^2)^{\i k} F \left( 2 \i k - 3, 2 \i k + 4, 2 \i k + 1,
{1 - y \over 2} \right),
\end{eqnarray}
where $y = \tanh (x/2)$ and $F(\alpha, \beta, \gamma, x)$ is the
hypergeometric function \cite{Landau}, to obtain the even functions
$\delta\theta_k$. The functions $\delta\theta_k$ should be normalized
in such a way that $\delta\theta_k(x) \rightarrow \cos(k x \pm
\delta)$ as $x \rightarrow \pm \infty$. Then, after calculating the
respective integrals, we arrive at
\begin{eqnarray} \label{ak}
a_k = {8 \pi^2 k^2 (k^2 + 1) \over (16 k^4 + 40 k^2 + 9) \sinh^2( \pi
k) } > 0.
\end{eqnarray}
Naturally, in the infinite domain one should replace the summation
over the continuous spectrum in Eq. (\ref{det1}) by integration:
$\sum_n \rightarrow \int_0^\infty {d k \over \pi}$.

To study the unstable solutions of Eq. (\ref{det}), we need to analyze
the zeros of the function
\begin{eqnarray}
D(\omega) = 1 + C(1 + \i \omega) \left( {a_0 \over \lambda_0 + \i
\omega} + {a_2 \over \lambda_2 + \i \omega} + \int_0^{\infty} {a_k dk
\over \pi (1 + k^2 + \i \omega)} \right)
\end{eqnarray}
in the lower half-plane of the complex frequency $\omega = \i
\gamma$. This can be done with the aid of the argument principle
\cite{KO89,KO90,KO94,KO78} which states that the number of zeros $N$
of the complex function $D(\omega)$ in this region of complex
frequency $\omega$ is equal to
\begin{eqnarray} \label{arg}
N = P + {1 \over 2 \pi} \Delta \arg D(\omega),
\end{eqnarray}
where $P$ is the number of poles there and $\Delta \arg D(\omega)$ is
the change of the argument of the function $D(\omega)$ as $\omega$
winds around this region of the complex frequency counterclockwise.

From the spectrum of the operator in Eq. (\ref{hth}) one can see that
only the pole at $\omega = \i \lambda_0$ lies in the lower-half plane
of the complex frequency, so we have $P = 1$
\cite{KO89,KO90,KO94,KO78}. Let us see how the function $D(\omega)$
with $\omega$ real varies as $\omega$ goes from $+\infty$ to
$-\infty$. Since $D(\omega)$ is symmetric with respect to the real
axis, one only needs to analyze the case of positive $\omega$. At
$\omega = \infty$ we have 
\begin{eqnarray}
D(\infty) = 1 + C \left( a_0 + a_2 + \int_0^\infty {a_k dk \over \pi}
\right) > 0,
\end{eqnarray}
where we used the explicit expressions for $a_{0,2,k}$ and
$\lambda_{0,2}$ to calculate the sign of $D(\infty)$ and evaluated the
integral in this equation to be $\int_0^\infty \pi^{-1} a_k dk \simeq
0.12$. On the other hand, at $\omega = 0$ we have
\begin{equation}
D(0) = 1 + C \left( {a_0 \over \lambda_0} + {a_2 \over \lambda_2} +
\int_0^{\infty} {a_k dk \over \pi (1 + k^2)} \right) = 1 - {8 \over 3}
C < 0
\end{equation}
for $A > A_b$. The latter expression can be obtained by recalling that
at $A = A_b$, for which $C = 3/8$, we have $D(0) = 0$ (see
Sec. \ref{stab:ss1}), and $C$ monotonically increases with $A$.

It is not difficult to show that the imaginary part of $D(\omega)$ is
negative for all $\omega > 0$:
\begin{eqnarray}
{ {\mathrm Im}~D(\omega) \over C \omega} = {a_0 (\lambda_0 - 1) \over
\lambda_0^2 + \omega^2} + {a_2 (\lambda_2 - 1) \over \lambda_2^2 +
\omega^2} + \int_0^{\infty} {k^2 a_k dk \over \pi [ (1 + k^2)^2 +
\omega^2]} \nonumber \\ < {a_0 (\lambda_0 - 1) \over \lambda_0^2 +
\omega^2} + {1 \over \lambda_2^2 + \omega^2} \left( a_2 (\lambda_2 -
1) + \int_0^{\infty} {k^2 a_k d k \over \pi} \right) < 0.
\end{eqnarray}
The last inequality is obtained by using the explicit expressions for
$a_{0,2,k}$, $\lambda_{0,2}$ and the evaluation of the last integral
$\int_0^\infty \pi^{-1} k^2 a_k dk \simeq 0.02$. Note that because of
the smallness of the contributions from the continuous spectrum one
can get very good approximations for the solutions of
Eq. (\ref{disp1d:stab0}) by restricting $\delta\theta_n$ to the
discrete spectrum only (see also \cite{Sev}).

From all this we conclude that the function $D(\omega)$ has the form
shown in Fig. \ref{cont1}(a),
\begin{figure}
\centerline{\psfig{figure=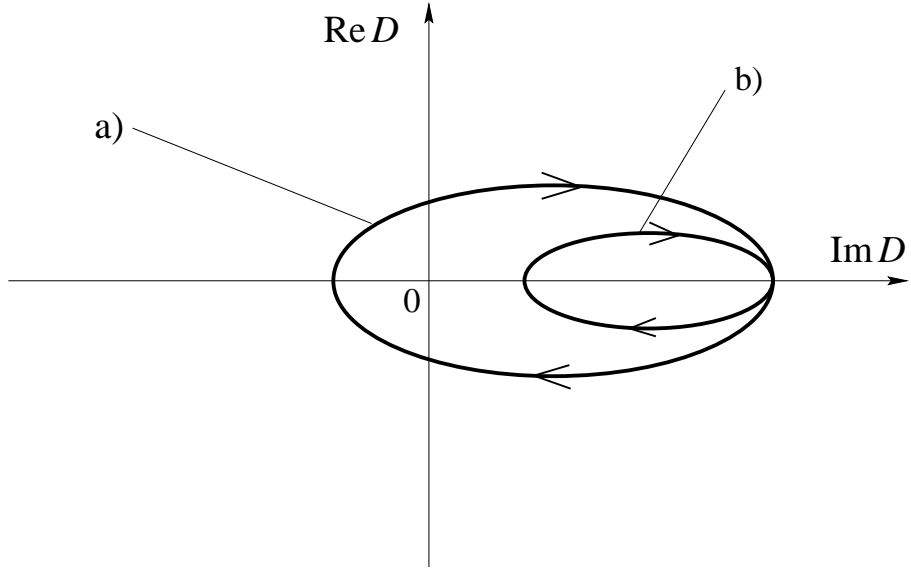,width=12cm,angle=-90}}
\caption{The behavior of the function $D(\omega)$ for the static AS
(a) and for the small amplitude state corresponding to the ``--'' sign
in Eq. (\ref{pm1d}). }
\label{cont1}
\end{figure}
so we have $\Delta \arg D(\omega) = - 2 \pi$ for the static spike
AS. This means that $N = 0$ and Eq. (\ref{disp1d:stab0}) does not have
solutions with ${\mathrm Re}~\gamma < 0$. Note that the same line of
arguments shows that the asymptotic stability problem for the
stationary solution with the smaller amplitude which corresponds to
the ``--'' sign in Eq. (\ref{pm1d}) always has a solution with
${\mathrm Re}~\gamma < 0$ since in that case $\Delta \arg D(\omega) =
0$ [see Fig. \ref{cont1}(b)], so this small-amplitude solution is
always unstable. These conclusions are in agreement with the general
qualitative theory of the ASs \cite{KO89,KO90,KO94}.

\section{Analysis of Eq. (\ref{om1d0})}

In this section we use the method of the previous section to analyze
the solutions of Eq. (\ref{om1d0}). This method was used by Kerner and
Osipov for studying the instabilities of the static ASs for small
values of $\alpha$ in the systems of small size \cite{KO89,KO90,KO94}.

After introducing the orthonormal basis set of the eigenfunctions of
Eq. (\ref{hth}), we get the following expression for $B_{mn}$ for
Eq. (\ref{om1d0}):
\begin{eqnarray} \label{bmn2}
B_{mn} = (\lambda_n + \i \omega) \delta_{mn} + C { 1 + \i \omega \over
\sqrt{\i \omega + \alpha}} b^l_m b^r_n,
\end{eqnarray}
where now
\begin{eqnarray}
C = {A^2 \al^{1/2} \over 8 \ep},
\end{eqnarray}
the rest is the same as in Eq. (\ref{bmn1}), and $\alpha \rightarrow +
0$ [cf. Eq. (\ref{disp1d}), $\alpha$ will determine the proper winding
direction, see below]. As in the previous section, we may write
\begin{eqnarray}
\det B_{mn} = \left[ 1 + C {1 + \i \omega \over \sqrt{\i \omega +
\alpha}} \sum_n {a_n \over \lambda_n + \i \omega} \right] \prod_n
(\lambda_n - \gamma),
\end{eqnarray}
where $a_n$ are given by Eq. (\ref{an}). To analyze the solutions of
Eq. (\ref{det}) with this $B_{mn}$, we will study the zeros of the
function
\begin{eqnarray} \label{dom}
D(\omega) = 1 + C {1 + \i \omega \over \sqrt{\i \omega + \alpha}}
\left( {a_0 \over \lambda_0 + \i \omega} + {a_2 \over \lambda_2 + \i
\omega} + \int_0^{\infty} {a_k dk \over \pi (1 + k^2 + \i
\omega)} \right),
\end{eqnarray}
where $a_k$ are given by Eq. (\ref{ak}), in the lower half-plane of
the complex frequency $\omega$ by using the argument principle
[Eq. (\ref{arg}), in which, as before, $P = 1$]. Of course, as in the
previous section, $D(\omega)$ should be symmetric with respect to the
real axis.

For $\omega > 0$ the real and the imaginary parts of $D(\omega)$ can
be written as
\begin{eqnarray} \label{omre}
{\mathrm Re}~{\sqrt{2 \omega} \over C} D(\omega) = {\sqrt{2 \omega}
\over C} + (1 + \omega) \left( {a_0 \lambda_0 \over \lambda_0^2 +
\omega^2} + {a_2 \lambda_2 \over \lambda_2^2 + \omega^2} +
\int_0^\infty {(1 + k^2) a_k dk \over \pi [ (1 + k^2)^2 + \omega^2 ]}
\right) \nonumber \\ + \omega (1 - \omega) \left( - {a_0 \over
\lambda_0^2 + \omega^2} - {a_2 \over \lambda_2^2 + \omega^2} -
\int_0^\infty {a_k dk \over \pi [ (1 + k^2)^2 + \omega^2 ]} \right),
\nonumber \\
\end{eqnarray}
and
\begin{eqnarray} \label{omim}
{\mathrm Im}~{\sqrt{2 \omega} \over C} D(\omega) = \omega (1 + \omega)
\left( - {a_0 \over \lambda_0^2 + \omega^2} - {a_2 \over \lambda_2^2 +
\omega^2} - \int_0^\infty {a_k dk \over \pi [ (1 + k^2)^2 + \omega^2
]} \right) \nonumber \\ + (\omega - 1) \left( {a_0 \lambda_0 \over
\lambda_0^2 + \omega^2} + {a_2 \lambda_2 \over \lambda_2^2 + \omega^2}
+ \int_0^\infty {(1 + k^2) a_k dk \over \pi [ (1 + k^2)^2 + \omega^2
]} \right). \nonumber \\
\end{eqnarray}
Using the explicit expressions for $a_{0,2,k}$ and $\lambda_{0,2}$
from the previous section, it is not difficult to show that the
expressions in the brackets above are negative for all values of
$\omega$. The analysis of Eq. (\ref{omim}) then shows that ${\mathrm
Im}~D(\omega)$ should change sign once when $0 < \omega < \infty$. Let
us denote the value of $\omega$ at which this happens as
$\omega_0$. Note that according to Eq. (\ref{omim}), we must have
$\omega_0 < 1$.

From the definition of $D(\omega)$ one can see that
\begin{eqnarray} \label{oo}
D(\omega) \rightarrow 1 + C {1 \mp \i \over \sqrt{2 |\omega|}} \left(
a_0 + a_2 + \int_0^\infty {a_k dk \over \pi} \right), ~~~\omega
\rightarrow \pm \infty.
\end{eqnarray}
Since the expression in the bracket in this equation is positive, we
will have ${\mathrm Im}~D(\omega) < 0$ for sufficiently large $\omega
> 0$. On the other hand,
\begin{eqnarray} \label{o}
D(\omega) \rightarrow {8 C (-1 \pm \i) \over 3 \sqrt{2 |\omega|}},
~~~\omega \rightarrow \pm 0.
\end{eqnarray}
Therefore, for sufficiently small $\omega > 0$ we must have ${\mathrm
Im}~D(\omega) > 0$. Observe that Eq. (\ref{o}) was obtained for
$\alpha = 0$. When $\alpha$ is small but finite, the two branches in
Eq. (\ref{o}) will actually get connected at ${\mathrm Re}~D(\omega)
\sim - \alpha^{-1/2}$. Thus, the qualitative behavior of $D(\omega)$
should be the one shown in Fig. \ref{cont2}
\begin{figure}
\centerline{\psfig{figure=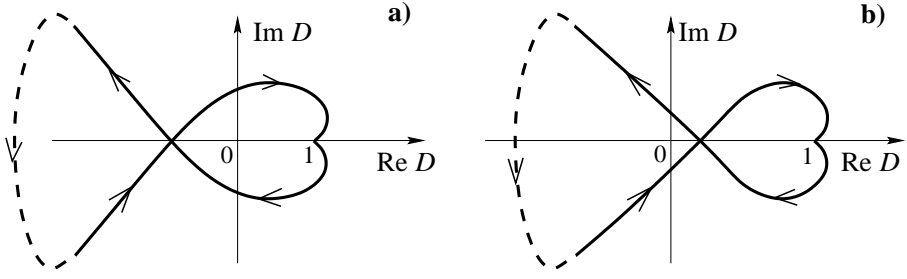,width=12cm,angle=-90}}
\caption{Qualitative form of the function $D(\omega)$ for large values
of $C$ (a) and for small values of $C$ (b).}
\label{cont2}
\end{figure}
Note that the function $D(\omega)$ can be calculated numerically from
Eq. (\ref{dom}) for any value of $C$ and has indeed the form shown in
Fig. (\ref{cont2}).

The number of zeros of $D(\omega)$ in the lower half-plane of the
complex frequency is determined by ${\mathrm
Re}~D(\omega_0)$. According to Eq. (\ref{dom}), if $C$ is sufficiently
small, the first term in Eq. (\ref{omre}) will dominate for $\omega =
\omega_0$, so we will have ${\mathrm Re}~D(\omega_0) > 0$. In this
case the change of the argument of $D(\omega)$ will be $\Delta \arg
D(\omega) = 2 \pi$ [Fig. \ref{cont2}(a)], so we will have $N = 2$ and
therefore an instability. On the other hand, if $C$ is large, we can
neglect $\sqrt{2 \omega}/C$ at $\omega = \omega_0$ in
Eq. (\ref{omre}). From Eq. (\ref{omre}) and the fact that $\omega_0 <
1$ one can then see that ${\mathrm Re}~D(\omega_0) < 0$. In this case
the change of the argument will be $\Delta \arg D(\omega) = - 2 \pi$
[Fig. \ref{cont2}(b)], so the number of zeros in the lower half-plane
is $N = 0$, implying stability. From all this we see that as the value
of $C$ is decreased, at some $C = C_0$ a complex-conjugate pair of
unstable solutions of Eq. (\ref{om1d0}) appears signifying a Hopf
bifurcation. The numerical analysis of Eq. (\ref{dom}) shows that $C_0
\simeq 0.2837$, what corresponds to $\alpha_\omega \simeq 5.15 \ep^2
A^{-4}$, and $\omega_0 \simeq 0.534$, in excellent agreement with the
results of Sec. \ref{stab:ss3}.


\begin{thebibliography}{99}

\bibitem{Prig} G. Nicolis and I. Prigogine, {\em Self-Organization in
Nonequilibrium Systems} (Wiley, New York, 1977).

\bibitem{Vas} V. A. Vasiliev, Y. M. Romanovskii, D. S. Chernavskii,
and V. G.  Yakhno, {\em Autowave Processes in Kinetic Systems} (VEB
Deutscher Verlag der Wissenschaften, Berlin, 1987).

\bibitem{Osc} {\em Oscillations and Traveling Waves in Chemical
Systems}, Ed. by R. J. Field and M. Burger (Wiley, New York, 1985).

\bibitem{Murr} J. D. Murray, {\em Mathematical Biology}
(Springer-Verlag, Berlin, 1989).

\bibitem{Cross} M. C. Cross and P. S. Hohenberg, Rev. Mod. Phys. {\bf
65} (1993) 851.

\bibitem{Mik} A. S. Mikhailov, {\em Foundations of Synergetics}
(Springer-Verlag, Berlin, 1990).

\bibitem{Kapral} R. Kapral and K. Showalter, {\em Chemical Waves and
    Patterns} (Kluwer, Dordrecht, 1995).

\bibitem{KO86} B. S. Kerner and V. V. Osipov, in {\em Nonlinear
Irreversible Processes}, Eds: W.Ebeling and H.Ulbricht (Springer,
Berlin, 1986).

\bibitem{KO89} B. S. Kerner and V. V. Osipov, Sov. Phys. -- Usp. {\bf
32} (1989) 101.

\bibitem{KO90} B. S. Kerner and V. V. Osipov, Sov. Phys. -- Usp. {\bf
33} (1990) 3.

\bibitem{KO94} B. S. Kerner and V. V. Osipov, {\em Autosolitons: a New
Approach to Problem of Self-Organization and Turbulence} (Kluwer,
Dordrecht, 1994).

\bibitem{Semi} {\em Nonlinear Dynamics and Pattern Formation in
Semiconductors and Devices}, Ed. by F. J. Niedernostheide (Springer,
Berlin, 1994).

\bibitem{Bode} M. Bode and H. G. Purwins, Physica D {\bf 86}
(1995) 53.

\bibitem{Gor} M. Gorman, M. el Hamdi, and K. A. Robbins, Combust. Sci.
  Tech. {\bf 98} (1994) 37; {\em ibid.} 71; {\em ibid.} 79.

\bibitem{Lee} K. J. Lee, W. D. McCormick, Q. Ouyong, and
H. L. Swinney, Science {\bf 261} (1993) 192; K. J. Lee,
W. D. McCormick, J. E. Pearson, and H. L. Swinney, Nature {\bf 369}
(1994) 215; K. J. Lee and H. L. Swinney, Phys. Rev. E {\bf 51} (1995)
1899.

\bibitem{KO78} B. S. Kerner and V. V. Osipov, Sov. Phys. -- JETP {\bf
47} (1978) 874; Biophysics (USSR) {\bf 27} (1982) 138; Sov. Phys. --
JETP Lett.  {\bf 41} (1985) 473.

\bibitem{KO79} B. S. Kerner and V. V. Osipov, Sov. Phys. --
Semicond. {\bf 13} (1979) 424; B. S. Kerner and V. V. Osipov,
Sov. Phys. -- Semicond. {\bf 13} (1979) 424; B. S. Kerner and
V. V. Osipov, Sov. Phys. -- Solid State {\bf 21} (1979) 1348.

\bibitem{KO80} B. S. Kerner and V. V. Osipov, Sov. Phys. -- JETP {\bf
52} (1980) 1122; Sov. -- Microelectronics {\bf 10} (1981) 407.

\bibitem{Koga} S. Koga and Y. Kuramoto, Prog. Theor. Phys. {\bf 63}
(1980) 106.

\bibitem{Gol} D. M. Petrich and R. E. Goldstein, Phys. Rev. Lett. {\bf
72} (1994) 1120; R.~E. Goldstein, D.~J. Muraki, and D.~M. Petrich
\newblock {\em Phys. Rev. E}, {\bf 53} (1996) 3933.

\bibitem{Mur96a} C. B. Muratov and V. V. Osipov, Phys. Rev. E {\bf
53} (1996) 3101.

\bibitem{Mur96b} C. B. Muratov and V. V. Osipov, Phys. Rev. E {\bf
54} (1996) 4860.

\bibitem{Hag} A. Hagberg and E. Meron, Phys. Rev. Lett. {\bf 72}
(1994) 2492; A.  Hagberg and E. Meron, Chaos {\bf 4} (1994) 477;
C. Elphick, A. Hagberg and E.  Meron, Phys. Rev. E {\bf 51} (1995)
3052.

\bibitem{thesis} C. B. Muratov, Ph. D. Thesis, Boston University,
1997. 


\bibitem{KO85} B. S. Kerner and V. V. Osipov, Sov. Phys. -- JETP {\bf 62}
(1985) 337.

\bibitem{Pear} J. E. Pearson, Science {\bf 261} (1993) 189.

\bibitem{Ross} P. J. Ortoleva and J. Ross, J. Chem. Phys. {\bf 63}
(1975) 3398; P. Ortoleva and R. Sultan, J. Chem. Phys. {\bf 148}
(1990) 47; . R. G. Casten, H. Cohen, and A. Lagerstrom, Quart.
Appl. Math. {\bf 32} (1975) 365.

\bibitem{KO82} B. S. Kerner and V. V. Osipov, Sov. Phys. -- JETP {\bf
56} (1982) 1275; V. V. Gafiichuk, B. S. Kerner, I. M. Lazurchak and
V. V. Osipov, Mikroelektronika {\bf 15} (1986) 180; V. V. Osipov,
V. V.  Gafiichuk, B. S. Kerner and I. M. Lazurchak, Mikroelektronika
{\bf 16} (1987) 23; V. V. Gafiichuk, V. E. Gashpar, B. S. Kerner and
V. V. Osipov, Sov. Phys. -- Semicond. {\bf 22} (1988) 1298.

\bibitem{KO83} B. S. Kerner and V. V. Osipov, Mikroelektronika {\bf
12} (1983) 512; J. D. Dockery and J. P. Keener, SIAM
J. Appl. Math. {\bf 49} (1989) 539.

\bibitem{Kuzn} E. M. Kuznetsova and V. V. Osipov, Phys. Rev. E {\bf
51} (1995) 148.

\bibitem{Mik94} K. Krischer and A. Mikhailov, Phys. Rev. Lett. {\bf
73} (1994) 3163.

\bibitem{Gaf95} P. Schutz, M. Bode and V. V. Gafiichuk, Phys. Rev. E
{\bf 52} (1995) 4465.

\bibitem{Os96} V. V. Osipov, Physica D {\bf 93} (1996) 143.

\bibitem{GS} P. Gray and S. Scott, Chem. Eng. Sci. {\bf 38} (1983) 29.

\bibitem{GM} A. Gierer and H. Meinhardt, Kybernetik {\bf 12} (1972)
30.

\bibitem{FHN} P. G. FitzHugh, Biophys. J. {\bf 1} (1961) 445;
J. Nagumo, S. Yoshizawa, and S. Arimoto, IEEE Trans. Circuit Theory
{\bf 12} (1965) 400.

\bibitem{RK} J. Rinzel and J. B. Keller, Biophys. J. {\bf 13} (1973)
1313; J. Rinzel and D. Terman, SIAM J. Appl. Math. {\bf 42} (1982)
1111.

\bibitem{Fife} P.~C. Fife.  \newblock {\em Dynamics of internal layers
and diffusive interfaces}.  \newblock (Society for Industrial and
Applied Mathematics, Philadelphia, 1988); G.~Caginalp and P.~C. Fife.
\newblock {Phys. Rev. B} {\bf 33} (1986) 7792; G.~Caginalp.  \newblock
{ Phys. Rev. A} {\bf 39} (1989) 5887; D. W. McLaughlin, D. J. Muraki,
and M. J. Shelly, Physica D {\bf 97} (1996) 471.

\bibitem{Ohta} T. Ohta, M. Mimura and R. Kobayashi, Physica D {\bf
34} (1989) 115.

\bibitem{m:pre96} C. B. Muratov, Phys. Rev. E {\bf 54} (1996) 3369. 

\bibitem{m1:pre97} C. B. Muratov, Phys. Rev. E {\bf 55} (1997) 1463.

\bibitem{Dub} A. L. Dubitskii, B. S. Kerner, and V. V. Osipov,
Sov. Phys. -- Doklady {\bf 34} (1989) 906. 

\bibitem{Os93}  V. V. Osipov, Phys. Rev. E {\bf 48} (1993) 88.

\bibitem{Mur95} V. V. Osipov and C. B. Muratov, Phys. Rev. Lett. {\bf
75} (1995) 388.

\bibitem{Sev} V. V. Osipov and A. V. Severtsev, Phys. Lett. A {\bf
222} (1996) 400; Phys. Lett. A {\bf 227} (1997) 61.

\bibitem{Katz} B. Katz, {\em Nerve, Muscle, and Synapse},
(McGraw-Hill, New-York, 1966).

\bibitem{Wood} P. M. Wood and J. Ross, J. Chem. Phys. {\bf 82} (1985)
1924.

\bibitem{Vin} M. N. Vinoslavskii, Sov. Phys. -- Solid State {\bf 31}
(1989) 1461; M. N. Vinoslavskii, B. S. Kerner, V. V. Osipov and
O. G. Sarbej, J.  Phys. -- Cond. Mat. {\bf 2} (1990) 2863.

\bibitem{Purw} H. Purwins {\em et al.}, Phys. Lett. A {\bf 136} (1989)
480; H.  Willebrandt {\em et al.}, Phys. Lett. A {\bf 149} (1990) 131.

\bibitem{Reyn} W. N. Reynolds, J. E. Pearson, and S. Ponce-Dawson,
Phys. Rev. Lett. {\bf 72} (1994) 2797; Phys. Rev. E {\bf 56} (1997)
185.

\bibitem{kaper} A. Doelman, T. J. Kaper, and P. Zegeling, Nonlinearity
{\bf 10} (1997) 523; A. Doelman, R. A Gardner, and T. J. Kaper,
Tech. Rep. No. 1028, Math. Inst. Univ. Utrecht, 1997.







\bibitem{MS} J. H. Merkin and M. A. Sadiq, IMA J. Appl. Math. {\bf
57}, (1996) 273.

\bibitem{BJ} E. Ben-Jacob {\em et al.}, Physica D {\bf 14} (1985) 348.

\bibitem{Landau} L. D. Landau and E. M. Lifshitz, {\em Course of
Theoretical Physics, Vol. 2}. (Pergamon Press, Oxford, 1965).

\bibitem{nu} Y. Nishiura and D. Ueyama, preprint (1998).

\bibitem{pss} V. Petrov, S. K. Scott, K. Showalter,
Phil. Trans. Roy. Soc. London, {\bf 347A} (1994) 631.

\bibitem{spi}
C. B. Muratov and V. V. Osipov, (submitted to Phys. Rev. Lett.)

\bibitem{peletier} J. K. Hale, L. A. Peletier, and W. C. Troy,
Tech. Rep. No. W98-03, Math. Inst. Univ. Leiden, 1998;
Tech. Rep. No. W98-11, {\em ibid}.

\end{thebibliography}
\end{document}